\documentclass[12pt]{article}
\usepackage{amsfonts}
\usepackage{amsmath, amssymb}
\usepackage{youngtab}
\usepackage{hyperref}
\usepackage[dvips]{color}
\usepackage{psfrag}
\usepackage{feynmp}
\usepackage[pdftex]{graphicx}
\usepackage{braket}
\usepackage{bm}
\usepackage{subfigure}

\usepackage{ulem}

\usepackage{comment}
\includecomment{pdffig}

\allowdisplaybreaks[1]

\Yboxdim{5pt}

\newcommand\non{\nonumber \\}
\newcommand{\bel}{\begin{eqnarray}}
\newcommand{\ee}{\end{eqnarray}}

\def\rem#1{}

\def\lam{{\lambda}}
\def\ep{{\epsilon}}

\def\del{{\partial}}

\def\ep{{\epsilon}}
\def\bep{{\bar{\epsilon}}}

\def\bpsi{{\bar{\psi}}}
\def\bphi{{\bar{\phi}}}

\def\blam{{\bar{\lambda}}}

\newcommand{\ba}{\begin{eqnarray}}
\newcommand{\ea}{\end{eqnarray}}

\makeatletter
    
    \@addtoreset{equation}{section}
  \makeatother

\usepackage{color}

\def\rem#1{}

\renewcommand{\title}[1]{\vbox{\center\LARGE{#1}}\vspace{5mm}}
\renewcommand{\author}[1]{\vbox{\center\large#1}\vspace{5mm}}

\begin{document}
\bibliographystyle{utphys}

\begin{titlepage}
\begin{center}
\vspace{5mm}
\hfill {\tt 
IPMU20-0068
}\\
\vspace{20mm}

\title{
\Large
\bf  
Supersymmetric indices on $I \times T^2$, elliptic genera and  dualities  with boundaries
}
\vspace{7mm}
{
Katsuyuki Sugiyama$^{\dagger,}$\footnote{sugiyama@scphys.kyoto-u.ac.jp } 
 and
Yutaka Yoshida$^{\uplus,}$\footnote{yutaka.yoshida@ipmu.jp} 
\\
}

\vspace{6mm}

\vspace{3mm}
$^{\dagger}${\small {\it Department of Physics, Kyoto University, Kyoto, 606-8502, Japan}} 
\\
$^{\uplus}${\small {\it Kavli IPMU (WPI), UTIAS, The University of Tokyo, Kashiwa, Chiba 277-8583, Japan.}} 
\\

\end{center}

\vspace{7mm}
\abstract{
We study   three dimensional $\mathcal{N} = 2$ supersymmetric  theories on  $I \times M_2$ with 2d $\mathcal{N}=(0,2)$ boundary conditions at the boundaries 
$\partial (I \times M_2)=M_2  \sqcup  M_2$, where $M_2=\mathbb{C}$ or $ T^2$.    
We introduce  supersymmetric indices of  three dimensional $\mathcal{N} = 2$  theories on   $I \times T^2$ that couple  to elliptic genera of  2d $\mathcal{N}=(0,2)$ theories at the two boundaries.
 We evaluate the $I \times T^2$ indices in terms of supersymmetric localization and   study  dualities on the $I \times M_2$. 
We consider  the dimensional reduction of $I \times T^2$ to  $I \times S^1$  and obtain the localization formula of 
2d $\mathcal{N}=(2,2)$ supersymmetric indices on $I \times S^1$. 
We illustrate  computations of open string Witten indices based on gauged linear sigma models. Correlation functions of Wilson loops  on $I \times S^1$ agree with Euler pairings in the geometric phase and also  agree with   cylinder amplitudes for B-type boundary states of  Gepner models  in the Landau-Ginzburg phase. 
}
\vfill

\end{titlepage}

\setcounter{tocdepth}{3}
\tableofcontents
\section{Introduction}

Supersmmetric (SUSY) localization computations 
 provide 
 exact results for partition functions, correlation functions and 
supersymmetric indices in supersymmetric quantum field theories.
After  a seminal work by Pestun \cite{Pestun:2007rz},  many works on supersymmetric localization on manifolds without boundary have appeared. These exact results play 
important roles for understanding various dualities relating different areas of physics and also mathematics. 

In two dimensions, quantum field theories on spacetimes (worldsheets) with boundaries 
have attracted much interest  in the connections with   open strings and D-branes. 
 The  partition function on two dimensional hemisphere ($D^2$) studied in \cite{Honda:2013uca, Hori:2013ika}
  is a typical example of the spacetime with a boundary, and   localization computation  
gives Gamma classes,  central charges of D-branes and period integrals of mirror Calabi--Yau  3-folds in the string theory.
Along with the hemisphere, a basic two dimensional (2d) spacetime with boundaries is the cylinder $I \times S^1$. 
  Supersymmetric indices on $I \times S^1$ with  states at the boundaries are called open string Witten indices that are related to Euler parings for Calabi-Yau 3-folds in string theory. 

When we move to  spacetime dimension higher than two, exact results based on supersymmetric localizations are less known compared with the spacetimes without boundary; 
 3d $\mathcal{N}=2$ gauge theories on $S^1 \times D^2$ \cite{Yoshida:2014ssa}, abelian 3d $\mathcal{N}=4$ gauge theories on a hemisphere \cite{Dedushenko:2017avn}, 4d  $\mathcal{N}=2$ gauge theories on a hemisphere \cite{Gava:2016oep}, 4d $\mathcal{N}=1$ gauge  theories on $D^2 \times T^2$ \cite{Longhi:2019hdh}. 
The  indices on $S^1 \times D^2$ we have evaluated in \cite{Yoshida:2014ssa} are  $S^1$-extensions of 
 partition functions of 2d $\mathcal{N}=(2,2)$ theory on  $D^2$  \cite{Honda:2013uca, Hori:2013ika},  those  have nice properties; first the $K$-theoretic $I$-function \cite{MR1943747} appears in the  index on $S^1 \times D^2$ which is a q-deformation (trigonometric deformation) of the Givental $I$-function  \cite{MR1354600} for the moduli space of  Higgs branch vauca.
For example, see subsequent works for relations between the indices on $S^1 \times D^2$ and the $K$-theoretic $I$-functions  \cite{Jockers:2018sfl, Ueda:2019qhg}. Second, for 3d $\mathcal{N}=4$ gauge theories with an $\mathcal{N}=(2,2)$ boundary condition, the indices on $S^1 \times D^2$ agree with  previous results
 for the  equivariant indices on $S^1 \times \mathbb{C}$ that are equivalent to  Coulomb gas representations of q-deformation of conformal blocks in q-deformed W-algebras  \cite{Aganagic:2013tta,  Aganagic:2014oia},   and also agree with
vertex functions for the $K$-theory of quasimap spaces  of Nakajima quiver varieties \cite{Aganagic:2016jmx}. 

Then  we expect that localization results on other manifolds with boundaries also have  nice properties and useful to understand supersymmetric quantum field theories with boundaries, as in the case of closed manifolds.  To extend the localization calculation to more examples with boundaries,  in this article, we introduce 
new supersymmetric indices   which are regared as   $S^1$-extensions of  the open string Witten indices, i.e., 3d $\mathcal{N}=2$ supersymmetric theories on $I \times T^2$ coupled to 
2d $\mathcal{N}=(0,2)$ boundary theories at the end points of the interval $I$. We formulate 
the supersymmetric localization for the indices on $I \times T^2$ and obtain new exact results.

This article is organized as follows. In section \ref{sec:section2}, we study supersymmetric boundary conditions  on $I \times M_2$, where $M_2=\mathbb{C}$ or $ T^2$ and construct supersymmetric Lagrangians.
 This part is essentially same as the analysis in \cite{Yoshida:2014ssa}, 
where we  constructed   Lagrangians and BPS boundary conditions for 3d $\mathcal{N}=2$ supersymmetric  gauge theories on $S^1 \times D^2$. 
In section \ref{sec:section3}, we define a supersymmetric index on $I \times T^2$ by imposing BPS boundary conditions at the left end and at the right end of  $I$. 
By supersymmetric localization,  we will show that 
 the path integral of the index is reduced to multi-contour integrals called the Jeffrey--Kirwan residue.
In section \ref{sec:section4}, we construct 3d theories in which $I \times T^2$ indices are identical to 2d $\mathcal{N}=(2,2)$ and $\mathcal{N}=(0,4)$ elliptic genera.
  In section \ref{sec:section5}, we study three dimensional dualities; we consider three dimensional IR dualities and put the dual theories on $I \times M_2$. We impose  the boundary conditions that cancel the gauge anomaly between the bulk and the boundary and satisfy the 't Hooft anomaly matching condition. We will show  that the $I \times T^2$ indices match in the dual pairs.  
In section \ref{sec:section6}, we study chiral algebras associated with  simple models on $I \times M_2$;  free chiral multiplets with the Dirichlet and Neumann boundary conditions.
In section \ref{sec:section7}, we perform the dimensional reduction of 3d $\mathcal{N}=2$ theories to 2d $\mathcal{N}=(2,2)$ theories on $I \times S^1$ and obtain the supersymmetric localization formula for $I \times S^1$ indices. We compare $I \times S^1$ indices with 
open string Witten indices in the geometric and  Landau--Ginzburg (LG) phases. 
In the section \ref{sec:summary} we  summarize our results and  comment on future directions.

\section{3d $\mathcal{N}=2$ theories on $I \times M_2$ with 2d $\mathcal{N}=(0,2)$ boundary  conditions}
\label{sec:section2}
In this section we consider 3d $\mathcal{N}=2$ supersymmetric theory on the direct product of one-dimensional interval and two dimensional flat space $I \times M_2$  with $M_2=\mathbb{C}$ or $T^2$
and study  BPS boundary conditions. We couple the 3d $\mathcal{N}=2$ theory to boundary 2d $\mathcal{N}=(0,2)$ theories.
The BPS boundary conditions and the boundary interactions we will consider  are  same as  those for  $S^1 \times D^2$ introduced in \cite{Yoshida:2014ssa},  more precisely those obtained by the flat space limit of  $S^1 \times D^2$. 
We define  coordinates of $I \times \mathbb{C}$ as follows:
\begin{align}
I  \times \mathbb{C}&= \{ (x^1, x^2,x^3)| \, \, x_1 \in [-\pi L, \pi L] , x^2,x^3 \in \mathbb{R}\} \nonumber \\
&=\{ (x^1, w, \bar{w})| \, \, x_1 \in [-\pi L, \pi L] , w=x^2+{\rm i} x^3, \bar{w}=x^2-{\rm i} x^3 \} \, .
\end{align}
A set of coordinates of $T^2$ in $I \times T^2$ is defined by
\begin{align}
T^2&:=\{ (x^2,x^3) | \, x^2+{\rm i} x^3 \sim x^2+{\rm i} x^3+2\pi R \sim x^2+{\rm i} x^3+ 2 \pi R \tau  \}.
\end{align}
Here $\tau=\tau_1+{\rm i} \tau_2$ is the  moduli of the torus $T^2$. 

The  SUSY transformation $\delta$ of 3d $\mathcal{N}=2$ vector and chiral multiplets are defined  by \eqref{eq:SUSYtransvec} and \eqref{eq:SUSYchi1}.
The action of  
 the four supercharges ${\rm Q}_{\alpha}, \bar{\rm Q}_{\alpha}$ $(\alpha=1,2)$ on a field $\mathcal{O}$ is read from the relation
\begin{align}
\delta \mathcal{O}=[\epsilon^{\alpha} {\rm Q}_{\alpha}+\bar{\epsilon}^{\alpha} \bar{\rm Q}_{\alpha}, \mathcal{O}].
\end{align}
Here the contractions of spinors indices are defined below \eqref{eq:SUSYtransvec}.
In this article we choose two component spinors $\epsilon=(\epsilon_1, \epsilon_2)^{T}$ and $\bar{\epsilon}=(\bar{\epsilon}_1,\bar{\epsilon}_2)^T$ in $\delta$  as 
\begin{align}
&\ep
=\ep'  
\left(
\begin{array}{c}
1\\
1
\end{array}
\right)
,
\quad \bep
=\bep' 
\left(
\begin{array}{c}
1 \\
1
\end{array}
\right)
\,,\,\label{eq:spinors}
\end{align}
where $\epsilon^{\prime}$ and $\bar{\epsilon}^{\prime}$ are  Grassmann odd constants. 
Under appropriate  boundary conditions which will be mentioned later, the following two supercharges ${\bf Q}$ and $\bar{\bf Q}$ are preserved. 
\begin{align}
{\bf Q}:={\rm Q}_2-{\rm Q}_1, \quad  \bar{\bf Q}:=\bar{\rm Q}_2-\bar{\rm Q}_1\,.
\label{eq:supercharge}
\end{align}

We will see \eqref{eq:spinors} preserves 2d $\mathcal{N}=(0,2)$ supersymmetry at the boundaries. 
Instead of \eqref{eq:spinors},  if we choose $\epsilon=\epsilon^{\prime }(1,- 1)^{T}$ and $\bar{\epsilon}=\bar{\epsilon}^{\prime }(-1,1)^T$, 2d $\mathcal{N}=(2,0)$ supersymmetry is preserved at the boundaries with appropriate boundary conditions.

\subsection{3d $\mathcal{N}=2$ vector multiplet}
\label{sec:section21}
Let  $G$ be   the gauge group given by a  compact Lie group  
and $\mathfrak{g}$  be the Lie algebra of $G$.
We consider  a $G$ vector multiplet  with BPS boundary conditions at the boundaries $\partial (I \times M_2)= M_ {2,L} \sqcup M_{2, R}$, where $M_{2,R}:=\{ x^1=\pi L \} \times M_2$ and 
$M_{2,L}:=\{ x^1=-\pi L \} \times M_2$. The integrals at the two boundaries  
have  opposite signs:
\begin{equation}
\begin{aligned}
 \int_{M_{2, L} \sqcup M_{2, R}} (\cdots) =-\int_{ \{x^1 = -\pi L \}  \times M_2}  (\cdots)+\int_{ \{x^1 = \pi L \}  \times M_2}  (\cdots)
 \,.
\end{aligned}
\end{equation}

The vector multiplet consists of a $\mathfrak{g}$ valued gauge field $A_{\mu}$, a real scalar $\sigma$, gaugini $\lambda, \bar{\lambda} $, and an auxiliary field $D$.
In our convention, the  covariant derivatives and the field strength of the gauge field are defined by
\bel
&&D_\mu=\del_\mu  +{\rm i}A_\mu \,,\,\,[D_\mu ,D_\nu]={\rm i}F_{\mu\nu}\,,\,\,
F_{\mu\nu}=\del_\mu A_\nu-\del_\nu A_\mu +{\rm i}[A_\mu ,A_\nu ]\, .
\ee
The SUSY transformation of the 3d $\mathcal{N}=2$ vector multiplet is given by
\begin{equation}
\begin{aligned} \label{eq:SUSYtransvec}
&\delta A_{\mu}=\frac{\rm i}{2} (\bar{\ep} \gamma_{\mu} \lambda - \bar{\lambda} \gamma_{\mu} \ep), \\
&\delta \sigma=\frac{1}{2} (\bar{\ep}  \lambda - \bar{\lambda}  \ep), \\
&\delta \lambda=-\frac{1}{2} \gamma^{\mu \nu} F_{\mu \nu} \ep -D \ep+{\rm i} \gamma^{\mu}D_{\mu} \sigma \ep,
\\
&\delta \bar{\lambda}=-\frac{1}{2} \gamma^{\mu \nu} F_{\mu \nu} \bep +D \bep-{\rm i} \gamma^{\mu}D_{\mu} \sigma \bep,
\\
&\delta D=-\frac{\rm i}{2} \bep \gamma^{\mu} D_{\mu} \lambda-\frac{\rm i}{2} D_{\mu} \bar{\lambda} \gamma^{\mu}  \ep
+\frac{\rm i}{2}[\bep \lambda, \sigma]+\frac{\rm i}{2}[\bar{\lambda} \ep , \sigma] \,,
\end{aligned}
\end{equation}
where $\mu=1,2,3$.  $\gamma^{1}, \gamma^{2},\gamma^{3}$ are the Pauli matrices and $\gamma^{\mu \nu}:=\frac{1}{2} [\gamma^{\mu}, \gamma^{\nu}]$. 
Again $\epsilon$ and $\bar{\epsilon}$ are  Grassmann odd two component spinors.
The contractions of spinors are defined by $\epsilon \lambda=\epsilon^{\alpha} \lambda_{\alpha}:=\epsilon^T C \lambda$, $\epsilon \gamma^{\mu} \lambda
=\epsilon^{\alpha} (\gamma^{\mu})_{\alpha}^{\beta} \lambda_{\beta}:=\epsilon^T C \gamma^{\mu} \lambda$,  
where $C:= -{\rm i}\gamma^2$ is a charge conjugation matrix.

A BPS boundary condition at each $x_1=\pm \pi L$ is given by
\begin{align}
\label{eq:vecdbdy}
& \sigma=0, \quad A_1=0, \quad \partial_1 A_2 =0, \quad \partial_1 A_3=0, \quad \partial_1 D=0, \nonumber  \\
&\lambda_1-\lambda_2=0, \quad \bar{\lambda}_1-\bar{\lambda}_2=0,  \quad \partial_1 (\lambda_1+\lambda_2)=0, \quad \partial_1 (\bar{\lambda}_1+\bar{\lambda}_2)=0. 
\end{align}
Here $\lambda_{\alpha}, \bar{\lambda}_{\alpha}$ with $\alpha=1,2$ are the components  of the gaugini  defined by $\lam =(\lam_1,\lam_2)^T$ and $\blam =(\blam_1,\blam_2)^T$.
The SUSY transformations generated by \eqref{eq:spinors} are consistent with the boundary condition \eqref{eq:vecdbdy}.

The restriction of the SUSY transformation \eqref{eq:SUSYtransvec} on a boundary with  \eqref{eq:vecdbdy} 
is written as
\begin{equation}
\begin{aligned} \label{eq:susyvecbdy}
&\delta (A_2+{\rm i} A_3)=0\,,\,\,\\
&\delta (A_2- {\rm i}  A_3)= 2  \bep' \lam_1 +2 \ep' \blam_1 \,,\\
& \delta [\hat{D}+{\rm i} F_{23}]=-2  \bep' (D_2+{\rm i} D_3) \lam_1 \,,\,\, \\
&\delta [\hat{D}-{\rm i} F_{23}]=2 \ep ' (D_2+{\rm i} D_3) \blam_1 \,,\\
&\delta \lam_1 = -(\hat{D}+{\rm i} F_{23})\ep'\,,\,\,\\
&\delta \blam_1 =  (\hat{D}-{\rm i} F_{23})\bep' \, ,
\end{aligned}
\end{equation}
where we defined $\hat{D}:=D- {\rm i} D_1 \sigma $.
From the boundary condition,  $\sigma=0$ and $A_1=0$  at the boundary.
  \eqref{eq:susyvecbdy} is same as the SUSY transformation of the 2d $\mathcal{N}=(0,2)$ $G$ vector multiplet $(A_1,A_2,\lambda_1,\bar{\lambda}_1, \hat{D })$. The boundary condition \eqref{eq:vecdbdy} 
is preserved by the 2d $\mathcal{N}=(0,2)$ SUSY transformation.

In 3d $\mathcal{N}=2$ supersymmetric theories without boundary, the supersymmetric invariant actions of the vector multiplet consist of 
the Chern--Simons term, the super Yang-Mills term and the Fayet–Iliopoulos (FI) term.
 We study the super Yang-Mills term and the FI-term in the presence of boundaries.  
The   3d super Yang-Mills  Lagrangian  is written as a Q-exact form:
\begin{align}
\mathcal{L}_{\text{SYM}}&=-{\bf Q} \cdot  {\bf Q} \cdot 
\mathrm{Tr}\Bigl( \frac{1}{4} \lambda \lambda \Bigr) 
\nonumber \\
&=\frac{1}{2} \mathrm{Tr} \Bigl[ \frac{1}{2}  F^{\mu \nu} F_{\mu \nu}+  D^{\mu} \sigma D_{\mu} \sigma+  D^2
+ {\rm i} \lambda \gamma^{\mu}  D_{\mu} \bar{\lambda}+{\rm i} \lambda [\bar{\lambda},\sigma]   \Bigr].
\label{SYMaction}
\end{align}
Here the action of a supercharge  ${\bf Q} \cdot \mathcal{O}$ is defined by the commutation relation $[{\bf Q}, \mathcal{O}]$ for 
a bosonic field $\mathcal{O}$ and the anti-commutation relation $\{ {\bf Q}, \mathcal{O} \}$ for 
a fermionic field $\mathcal{O}$, respectively.  
 $\mathrm{Tr}$ is a trace taken over $\mathfrak{g}$.
 The super Yang-Mills action $S_{\mathrm{SYM}}$ is defined by 
\begin{align}
S_{\text{SYM}} := \int_{I \times M_2}  \mathcal{L}_{\text{SYM}} .
\label{}
\end{align}
 With the boundary condition \eqref{eq:vecdbdy},  the $ S_{\text{SYM}}$ is invariant by the SUSY transformation \eqref{eq:supercharge}.
 Next we consider the FI-term.  The action for the FI-term is given by
\begin{align}
S_{\text{FI}}=\int_{I \times M_2} \, \mathcal{L}_{\text{FI}}=\int_{I \times M_2}{\rm i} \,  \zeta ( D) \,.
\label{eq:FIterm}
\end{align}
Here $\zeta $ is the FI-parameter. $S_{\text{FI}}$ is invariant by the SUSY transformation \eqref{eq:supercharge}.

\subsection{3d $\mathcal{N}=2$  chiral multiplet}
\label{sec:section22}
Next we consider a chiral multiplet $(\phi, \psi, F)$ in a representation  $\mathbf{R}$ of $G$.  The anti-chiral multiplet $(\bar{\phi}, \bar{\psi}, \bar{F})$ belongs to the complex conjugate representation 
${\bar{\mathbf{R}}}$ of $G$. The  SUSY transformation of the chiral multiplet is given by
\begin{equation}
\begin{aligned} \label{eq:SUSYchi1}
&\delta \phi= \bar{\epsilon} \psi , \\
&\delta \bar{\phi}= \epsilon \bar{\psi} ,\\
&\delta \psi= {\rm i} \gamma^{\mu} \epsilon D_{\mu} \phi+{\rm i} \epsilon \sigma \phi+\bar{\epsilon} F ,\\
&\delta \bar{\psi}= {\rm i} \gamma^{\mu} \bar{\epsilon} D_{\mu} \bar{\phi}+{\rm i} \bar{\phi} \sigma \bar{\epsilon} +\bar{F} \epsilon  ,\\
&\delta F=\epsilon ({\rm i} \gamma^{\mu} D_{\mu} \psi-{\rm i} \sigma \psi -{\rm i}\lambda \phi) ,\\
&\delta \bar{F}=\bar{\epsilon} ({\rm i} \gamma^{\mu} D_{\mu} \bar{\psi}-{\rm i} \bar{\psi} \sigma  +{\rm i} \bar{\phi} \bar{\lambda} ) .
\end{aligned}
\end{equation}
If the chiral multiplet  $(\phi, \psi, F)$ belongs to a representation of a global symmetry group $G_F$,
one can turn on the background gauge fields for the maximal torus of $G_F$.

The kinetic  Lagrangian of the chiral multiplet is  written as a Q-exact form:
\begin{align}
 \mathcal{L}_{\text{chi}}&=  {\bf Q} \cdot {\bf Q} \cdot  \left( \bar{\phi} F \right)   \nonumber \\
&=-\bar{\phi} D^{\mu} D_{\mu}   \phi+\bar{\phi} \sigma^2 \phi+{\rm i} \bar{\phi} D \phi+\bar{F} F 
-{\rm  i}  \bar{\psi} \gamma^{\mu} D_{\mu}  \psi  +{\rm i} \bar{\psi} \sigma \psi    
+{\rm i} \bar{\psi}  \lambda \phi-{\rm i} \bar{\phi} \bar{\lambda}  \psi. 
\label{Lagchiral}  
\end{align}
The kinetic action of the chiral multiplet is defined by
\begin{align}
S_{\text{chi}}= \int_{I \times M_2}  \mathcal{L}_{\text{chi}} \,.
\label{eq:action3dchi}
\end{align}
In a generic choice of  supersymmetric variation parameters $\epsilon=(\epsilon_1,\epsilon_2)^T$ and $\bar{\epsilon}=(\bar{\epsilon}_1,\bar{\epsilon}_2)^T$, 
the  action $S_{\text{chi}}$ is not invariant under the SUSY transformation in the presence of the boundaries. 
In the next two subsections, we will study    two types of BPS boundary conditions for the chiral multiplet; the Dirichlet (denoted by   ${\sf D}$) and the Neumann (denoted by  ${\sf N}$) boundary conditions.  

Other parts constructed by chiral multiplets are the superpotential terms. 
The Lagrangians of the  superpotential terms are given by 
\begin{align}
&\mathcal{L}_{W}=\sum_{i} \frac{ \partial W(\phi)}{\partial \phi_i} F_i -\frac{1}{2}  \sum_{i, j} \psi_i \psi _j  \frac{W(\phi)}{\partial \phi_i \partial \phi_j }, 
\\
&
\mathcal{L}_{\overline{W}}=\sum_{i} \frac{ \partial \overline{W}(\bar{\phi})}{\partial \bar{\phi}_i} \bar{F}_i -\frac{1}{2}  \sum_{i, j} \bar{\psi}_i \bar{\psi} _j  \frac{\overline{W}(\overline{\phi})}{\partial \bar{\phi}_i \partial \bar{\phi}_j } .
\end{align}
Here  $i$ in the sums   \footnote{In this article we use two notations for the representations for the flavor symmetry group.
For example let  $(\phi, \psi, F)$ be a  chiral multiplet in the fundamental representation of $G_F=U(N)$.  $(\phi, \psi, F)$ is also expressed as the collection of  chiral multiplets $(\phi_i, \psi_i, F_i)$ with $i=1,\cdots,N$. 
} 
labels  the chiral multiplets $(\phi_i, \psi_i, F_i)$  in the superpotential $W(\phi)$.
 $\overline{W}(x)$ is the complex conjugate of $W(x)$. 
The actions of the superpotential terms are given by 
\begin{align}
S_{W}=\int_{I \times M_2}  \mathcal{L}_{W}, \quad 
S_{\overline{W}}=\int_{I \times M_2}   \mathcal{L}_{\overline{W}}.
\end{align}
The SUSY transformations of  the superpotential   are written as  the total derivatives:
\begin{align}
\delta \mathcal{L}_W=-{\rm i}  \partial_\mu  \left[ \sum_{i} (\psi_{i} \gamma^\mu \ep) \frac{\partial W}{\partial \phi_{i} } \right]\,,\,\,
\delta \mathcal{L}_{\bar{W}}=-{\rm i} \partial_\mu \left[ \sum_{i} ( \bpsi_{i} \gamma^\mu \bep)\frac{\partial \bar{W}}{\partial \phi_{i}} \right] . 
\end{align}
Then surface terms for the SUSY transformation of the superpotentials  are  given by 
\begin{align}
\delta S_{W}
&= {\rm i}  \ep^{\prime} \sum_{i   } \int_{\partial (I \times M_2)}  (\psi_{1 \, i}-\psi_{2 \, i})   \frac{\partial W}{\partial \phi_{i} } \, ,
\nonumber \\
\delta S_{\bar{W}}
&= {\rm i}  \bar{\ep}^{\prime} \sum_{i   } \int_{\partial (I \times M_2)} (  \bar{\psi}_{1 \, i}-\bar{\psi}_{2 \, i})   \frac{\partial \bar{W}}{\partial \bar{\phi}_i}  \, .
\label{eq:SUSYW}
\end{align}
We will see that the surface terms  \eqref{eq:SUSYW} vanish under the Dirichlet boundary condition. 
In this case the supersymmetry is preserved without adding the boundary degrees of freedom.
 On the other hand, 
the surface terms \eqref{eq:SUSYW} remain under the Neumann boundary condition. 
We have to add boundary degrees of freedom to preserve the supersymmetry.

\subsubsection{Dirichlet boundary condition}
First we consider the Dirichlet boundary condition:
\begin{align}
& \phi=0, \quad \bar{\phi}=0 , \quad \psi_1-\psi_2=0\,,\,\,
\bar{\psi}_1-\bar{\psi}_2=0\,, \nonumber \\
& \partial_1 F=0, \quad \partial_1 \bar{F}=0 , \quad \partial_1 ( \psi_1+\psi_2)=0\,,\,\,
\partial_1 (\bar{\psi}_1+\bar{\psi}_2)=0\,.
\label{eq:Dbdychi}
\end{align}
 \eqref{eq:Dbdychi} is compatible with  
 the SUSY transformation \eqref{eq:SUSYchi1} with the  variation 
parameters defined by 
  \eqref{eq:spinors}. The actions $S_{\text{chi}}$ and $S_{W}$ for the chiral multiplet  are invariant under 
the SUSY transformation with the Dirichlet boundary  condition.

The  restriction of SUSY transformation \eqref{eq:SUSYchi1}  on the boundary with the  Dirichlet boundary condition \eqref{eq:Dbdychi}  is written as
\begin{align}
&\delta \psi_1 = {\rm i} {\ep}^{\prime} \partial_1 \phi + \bep^{\prime} F\,, \quad
\delta F=\ep^{\prime} \left[- 2 {\rm i} \partial_1 \psi_1 + 2   (D_2+{\rm i} D_3) \psi_1\right]  \,\,,\non
&\delta  \bar{\psi}_1 = {\rm i} \bar{\ep}^{\prime} \partial_1 \bar{\phi} + \ep^{\prime} \bar{F}\,, \quad
\delta \bar{F}=\bar{\ep}^{\prime} \left[ -2{\rm i} \partial_1 \bar{\psi}_1 + 2  (D_2+{\rm i} D_3)\bar{\psi}_1 \right]\,.
\label{eq:Dbc}
\end{align}
Note that \eqref{eq:Dbc} is same as the SUSY transformation of the $\mathcal{N}=(0,2)$ fermi multiplet given by \eqref{eq:fermitrans}, if we take $E(\phi^{\prime}):= \phi^{\prime}$ with $\phi^{\prime}=\partial_1 \phi$ and $\psi^{\prime}_{+}:= \partial_1 \psi_1$.

\subsubsection{Neumann boundary condition}
Next we consider  the Neumann boundary condition defined by  
\begin{align}
& \partial_1 \phi=0, \quad F=0, \quad \partial_1 \bar{\phi}=0, \quad \bar{F}=0 ,   \nonumber \\
&\psi_1+ \psi_2=0\,,\,\,
\bar{\psi}_1+ \bar{\psi}_2=0\,,\,\,
\del_1(\psi_1- \psi_2)=0\,,\,\,
\del_1(\bar{\psi}_1- \bar{\psi}_2)=0\,.
\label{eq:Nboundary}
\end{align}
\eqref{eq:Nboundary} is compatible with the SUSY transformation with \eqref{eq:spinors}.
At a boundary, the SUSY  transformation of the chiral multiplet with the Neumann boundary condition gives
\begin{align}
\delta \phi& =2\bep'\psi_1\,,\,\,
\delta \bphi =2 \ep'\bpsi_1\,,\,\,\non
\delta \psi_1&= \ep' (D_2+{\rm i} D_3)\phi\,,\,\,
\delta \bpsi_1= \bep' (D_2+{\rm i} D_3)\bphi\,.
\label{eq:bcchiralSUSY}
\end{align}
The  transformation of $(\phi, \psi_1)$ in \eqref{eq:bcchiralSUSY} is same as the SUSY  transformation of the 2d $\mathcal{N}=(0,2)$ chiral multiplet.

With the Neumann boundary condition \eqref{eq:Nboundary}, the kinetic action of the chiral multiplet $S_{\text{chi}}$  is invariant by the SUSY transformation.
On the other hand, the superpotential term is not invariant under the SUSY transformation. 
The SUSY transformations give the following boundary terms:
\begin{align}
\delta S_{W}
&=2{\rm i  } \ep^{\prime} \int_{ \partial (I \times M_2)  } \sum_{i^{\prime}}    \psi_{1 \, i^{\prime}}   \frac{\partial W}{\partial \phi_{i^{\prime}} } \, , \nonumber \\
\delta S_{\bar{W}}
&=2{\rm i  }  \bar{\ep}^{\prime} \int_{ \partial (I \times M_2)  } \sum_{i^{\prime}}    \bar{\psi}_{1 \, i^{\prime}}  \frac{\partial \bar{W}}{\partial \bar{\phi}_{i^{\prime}} } \,.
\label{eq:surfaceW}
\end{align}
If the surface terms \eqref{eq:surfaceW} are compensated by the SUSY transformation of appropriate 2d $\mathcal{N}=(0,2)$ superpotential terms \cite{Gadde:2013sca},   
this cancellation mechanism is  analogous to the matrix factorization in the 2d $\mathcal{N}=(2,2)$ Landau-Ginzburg models \cite{Kapustin:2002bi} and called the 3d matrix factorization.

In 2d $\mathcal{N}=(2,2)$ gauged linear sigma models (GLSMs) with  boundaries, the surface term of the superpotential is canceled by  brane factors   \cite{Herbst:2008jq}. 
On the other hand, the dimensional reduction of \eqref{eq:surfaceW}  is also  canceled by  the SUSY transformation of the superpotential in  1d $\mathcal{N}=2$ fermi multiplets.
We briefly study these two methods;  brane factors and 1d fermi multiplets  in section \ref{sec:section72}.

\subsection{2d $\mathcal{N}=(0,2)$  theory at boundary and  3d matrix factorization} 
\label{sec:section23}
As we have seen in the previous section, the 3d $\mathcal{N}=2$ theory preserves 
2d $\mathcal{N}=(0,2)$ supersymmetry at the boundaries except for the  superpotential term.  2d $\mathcal{N}=(0,2)$ multiplets are  necessary to   cancel 
 the gauge anomaly  and to cancel the boundary terms for the 3d  superpotential.
 We  explain  2d $\mathcal{N}=(0,2)$ theories at  boundaries and couplings between the 2d theories and the 3d theory.

\subsubsection*{\underline{$\mathcal{N}=(0,2)$ vector multiplet}}
An $\mathcal{N}=(0,2)$ vector multiplet with the gauge group $G^{\prime}$ consists of 
a $\mathfrak{g}^{\prime}$ valued gauge field $A^{\prime}_i$ $(i=2, 3)$, 2d fermions  $\lambda^{\prime}, \bar{\lambda}^{\prime}$ , and an auxiliary field $D^{\prime}$. The action is 
given by
\begin{align}
S_{\text{2d.vec}}&= \int_{M_2}{\cal L}_{\text{2d.vec}}
=- \int_{M_2}  {\bf Q} \cdot \left( \mathrm{Tr} \lam^{\prime}_1 
(\hat{D}-{\rm i} F^{\prime}_{23})  \right) \nonumber \\
&=
\int_{M_2} \mathrm{Tr} \left[
F_{23}^{\prime 2 }+\hat{D}^{\prime 2}+2\lam^{\prime}_1(D_2+{\rm i} D_3)\blam^{\prime}_1 \right] \,.
\label{eq:2dSYM}
\end{align}
Here we defined $F^{\prime}_{23}:=\partial_2 A^{\prime}_3-\partial_3 A^{\prime}_2 +i [A^{\prime}_2, A^{\prime}_3]$. The SUSY transformation  is same as \eqref{eq:susyvecbdy} and 
 ${\bf Q}$ is the restriction of \eqref{eq:supercharge} to \eqref{eq:susyvecbdy}. 
 The FI-term for the 2d $\mathcal{N}=(0,2)$ gauge theory is given by
\begin{align}
S_{\text{2d.FI}}=\int_{M_2}{\cal L}_{\text{2d.FI}} =\int_{M_2} {\rm i} \zeta_{\text{2d}} \left( D^{\prime} \right) \,.
\end{align}

\subsubsection*{\underline{$\mathcal{N}=(0,2)$ chiral multiplet}}
We consider a 2d $\mathcal{N}=(0,2)$ chiral multiplet consisting of $(\phi^{\prime}, \psi^{\prime})$ in a representation of the gauge group $G^{\prime}$.
The SUSY transformation is same as \eqref{eq:bcchiralSUSY}, where $(\phi, \psi_1)$ replaced by $(\phi^{\prime}, \psi^{\prime})$ and the covariant derivative for $G$ is replaced by that for $G^{\prime}$. The Lagrangian of the $\mathcal{N}=(0,2)$ chiral multiplet is written as 
a Q-exact form:
\begin{align}
{\cal L}_{\text{2d.chi}}&=- {\bf Q} \cdot \left( 
\bphi^{\prime} (D_2-{\rm i} D_3)\psi^{\prime} + {\rm i} \bphi^{\prime} \lam^{\prime} \phi^{\prime}
 \right) \non
&
=
- \bphi^{\prime} (D_2-{\rm i} D_3) (D_2+{\rm i} D_3) \phi^{\prime}
 -2\bpsi^{\prime} (D_2-{\rm i} D_3)\psi^{\prime}
 \non
&\qquad  \qquad \qquad -2{\rm i} \bphi^{\prime} \blam_1\psi^{\prime}
+ 2 {\rm i}\bpsi^{\prime} \lam^{\prime}_1 \phi^{\prime} + {\rm i} \bphi^{\prime} (\hat{D}+{\rm i} F^{\prime}_{23}) \phi^{\prime} \,.
\end{align}
The action of the 2d $\mathcal{N}=(0,2)$ chiral multiplet is given by 
\begin{align}
S_{\text{2d.chi}}:= \int_{M_2} {\cal L}_{\text{2d.chi}} \,.
\label{eq:action2dchi}
\end{align}

\subsubsection*{\underline{$\mathcal{N}=(0,2)$ fermi multiplet}}
The SUSY transformation of the 2d $\mathcal{N}=(0,2)$ fermi multiplet $(\psi^{\prime}_-, F^{\prime})$ is given by
\bel
&&\delta \psi^{\prime}_- = {\rm i}  \ep' E+ \bep' F^{\prime}\,,\non
&&\delta F^{\prime}=\ep'\left[-2  {\rm i}   \sum_i \psi^{\prime}_{+ \, i } \partial_{\phi^{\prime}_i} E (\phi^{\prime}) +2 (D_2+{\rm i} D_3)\psi^{\prime}_- \right]\,,\non
&&\delta \bar{\psi}^{\prime} = {\rm i}   \bep' \bar{E}+\ep'\bar{F}^{\prime}\,,\non
&&\delta \bar{F}^{\prime}=\ep'\left[-2  {\rm i}   \sum_i \bpsi^{\prime}_{+ \, i}  \partial_{\bar{\phi}^{\prime}_i} \bar{E} (\bar{\phi}^{\prime})
+2 (D_2+{\rm i} D_3)\bar{\psi}^{\prime}_-\right]\, .
\label{eq:fermitrans}
\ee
Here $(\phi^{\prime}_i, \psi^{\prime}_{+ \, i})$'s are 2d $\mathcal{N}=(0,2)$ chiral multiplets, that can be taken as the boundary values of 3d chiral multiplets with the  Neumann boundary condition.
 $E$ is a function of $\phi_i^{\prime}$'s.
The kinetic term of the fermi multiplet is given by  
\begin{align}
{\cal L}_{\text{fermi}}&={\bf Q} \cdot \left( \bar{\psi}^{\prime}_- F^{\prime}-  {\rm i}   \bar{E} \psi^{\prime}_- \right) \non
&=
-2\bar{\psi}^{\prime}_{-}(D_2+{\rm i} D_3)\psi^{\prime}_{-}
+\bar{F}^{\prime} F^{\prime}+\bar{E}E+2  {\rm i}   \sum_i \bar{\psi}_- \psi_{+ \, i}^{\prime} \partial_{\phi^{\prime}_i}E -2  {\rm i}  \sum_i \bpsi_{+\, i}^{\prime}  \psi^{\prime}_{-} \partial_{\bar{\phi}^{\prime}_i} \bar{E} \,.
\end{align}
The kinetic action of the 2d $\mathcal{N}=(0,2)$ fermi multiplet is given by 
\begin{align}
S_{\text{fermi}}:= \int_{M_2} {\cal L}_{\text{fermi}} \,.
\label{eq:Fermiaction}
\end{align}

The $\mathcal{N}=(0,2)$ superpotential term is constructed by the fermi multiplets $(\psi^{\prime}_{- \, a}, F^{\prime \, a})$'s  and the 2d $\mathcal{N}=(0,2)$ chiral multiplets $(\phi_i^{\prime},\psi^{\prime}_{i +})$'s with functions ${J}^a(\bar{\phi}^{\prime})$'s as
\begin{align}
&
{\cal L}_J=\sum_a \left(  F^{\prime a} J^a-2   \sum_i  \psi^{\prime }_{- \, a} \psi^{\prime}_{ + \, i} \frac{\partial J^a}{\partial \phi_i^{\prime}} \right)
\,,\non
&
{\cal L}_{\bar{J}}=
\sum_a \left(\bar{F}^{\prime a} \bar{J}^a-2  \sum_i\bar{\psi}^{\prime }_{- \, a} \bpsi_{+ \, i }^{\prime} \frac{\partial \bar{J}^a}{\partial \bar{\phi}^{\prime}_i}  \right)
\,.\label{fermi-potential}
\end{align}
The SUSY transformation of the 2d $\mathcal{N}=(0,2)$ superpotential is written as 
\begin{align}
\delta  {\cal L}_J&=2{\rm i} \epsilon^{\prime} \sum_a \left[ -  \sum_i \psi^{\prime}_{+ \, i} \frac{ \partial ( E^a J^a) }{\partial \phi^{\prime}_i} +  (\partial_2 +{\rm i } \partial_3 ) (\psi^{\prime}_{- \, a} J^a )\right] \,,
\end{align}
If $\sum_a E^a J^a=0$, the superpotential preserves the $\mathcal{N}=(0,2)$ supersymmetry. There is another possibility to 
preserve supersymmetry as follows. The surface terms for the 3d $\mathcal{N}=2$ superpotentials and  the SUSY transformation of 2d $\mathcal{N}=(0,2)$ superpotential are
 combined as
\begin{align}
\delta \int_{ I \times M_2}  {\cal L}_W \Big|_{x_1= \pm \pi L}&+ \delta \int_{ \{ \pm \pi L \} \times M_2}  {\cal L}_J
\nonumber \\
 &= 
 \int_{ \{ \pm \pi L \}\times M_2  } 2 {\rm i} \epsilon^{\prime}  \left[ \sum_{i}   \pm  \psi_{1 \, i}   \frac{\partial W}{\partial \phi_{i^{\prime}} } 
-  \psi^{\prime}_{+ \, i} \sum_a  \frac{ \partial ( E^a J^a) }{\partial \phi_i}\right] \,.
\end{align}
Then the boundary terms of the SUSY transformation of the 3d superpotential are canceled by the 2d $\mathcal{N}=(0,2)$ superpotential terms, 
if the 2d chiral multiplets in $E^a$ and $J^a$ take the boundary values of the 3d chiral multiplets with the Neumann  boundary condition;  $\phi^{\prime}_i =\phi_i |_{x^1 = \pm \pi L}, \psi^{\prime}_{- \, i} =\psi_{1 i} |_{x^1 = \pm \pi L}$
 and satisfy the following relations:
\begin{align}
W \Big|_{x^1=  \pm \pi L} =\pm \sum_{a}E^a J^a, \quad \bar{ W} \Big|_{x^1=  \pm \pi L} =\pm \sum_{a}\bar{E}^a \bar{J}^a \,.
\label{eq:3dmatrixfac}
\end{align}

\subsection{Anomaly polynomials  }
\label{sec:section33}

\begin{table}[htb]
\begin{center}
\begin{tabular}{c |  c c  }
			&								$U(1)_y$	&	$U(1)_R$	\\
			\hline
$\phi$			&		$Q$			&	$r$		\\
$\psi$			&	$Q$	&	$r-1$		\\
$F$						&	$Q$				&	$r-2$ \\
\end{tabular} 
\hspace{1.0cm}
\begin{tabular}{c |  c c  }
			&								$U(1)_y$	&	$U(1)_R$	\\
			\hline
$\phi^{\prime}$			&		$Q$			&	$r$		\\
$\psi^{\prime}_+$			&	$Q$	&	$r-1$		\\
\end{tabular} 
\hspace{1.0cm}
\begin{tabular}{c |  c c  }
			&								$U(1)_y$	&	$U(1)_R$	\\
			\hline
$\psi^{\prime}_-$			&		$Q$			&	$r-1$		\\
$F^{\prime}_-$			&	$Q$	&	$r-2$		\\
\end{tabular} 
\caption{Left: The charge assignment for a 3d $\mathcal{N}=2$ chiral multiplet $(\phi,\psi,F)$. 
Middle: The charge assignment for a 2d $\mathcal{N}=(0,2)$ chiral multiplet $(\phi^{\prime},\psi^{\prime}_+)$. Right: The charge assignment for a 2d $\mathcal{N}=(0,2)$ fermi multiplet $(\psi^{\prime}_-, F^{\prime})$.
$U(1)_y$ is a gauge or a flavor symmetry group. $U(1)_R$ is the R-symmetry group.   }
\label{table:chiral1}
\end{center}
\end{table}

 The net contributions  to anomalies  from the fermions in the 3d chiral multiplets
 and the fermions in the 2d  chiral and fermi multiplets  
 are nicely organized as the anomaly polynomials \cite{Dimofte:2017tpi}.
The  3d and 2d  theories have to satisfy the  cancellation of the gauge anomalies at both  left and right boundaries. 
Also the 't Hooft anomalies have to match between the  IR dual theories.
Here let us summarize the contributions to the anomaly polynomials. 
\begin{itemize}
\item 3d chiral multiplet:

A  3d $\mathcal{N}=2$ chiral multiplet with a charge assignment in Table \ref{table:chiral1} contributes to the anomaly polynomial as
\begin{align}
\pm \frac{1}{2} ( Q {\bf y}+(r-1) {\bf r})^2 \,.
\end{align}
Here $+$ (resp. $-$) is taken for the Dirichlet (resp. Neumann) boundary condition for the chiral multiplet.
${\bf y}$ is the field strength for $U(1)_{y}$ and ${\bf r}$ is the field strength for the $U(1)_R$ R-symmetry group.  

The contribution of the 3d chiral multiplet belongs to a representation ${\bf R}$ of $G$: 
\begin{align}
\pm \frac{1}{2} \mathrm{Tr}_{\bf{R}} ({\bf f})^2 \,.
\end{align}
Here $+$ (resp. $-$) is taken for the Dirichlet (Neumann) boundary condition for the chiral multiplet.

\item 2d $\mathcal{N}=(0,2)$ chiral multiplet: 

A  2d $\mathcal{N}=(0,2)$ chiral multiplet with a charge assignment in Table \ref{table:chiral1} contributes to the anomaly polynomial
\begin{align}
-  ( Q {\bf y}+(r-1) {\bf r})^2 \,.
\end{align}
\item 2d $\mathcal{N}=(0,2)$ fermi multiplet:

A 2d $\mathcal{N}=(0,2)$ fermi multiplet with a charge assignment in Table \ref{table:chiral1} contributes to the anomaly polynomial
\begin{align}
 (Q {\bf y}+(r-1) {\bf r})^2 \,.
\end{align}
\end{itemize}

\section{SUSY index of 3d $\mathcal{N}=2$ theory on $I \times T^2$ and localization }
\label{sec:section3}

\subsection{Definition of the index  on $I \times T^2$ and  SUSY localization formula}
In this section we evaluate the SUSY index on $I \times T^2$ in terms of the supersymmetric 
localization method.

For the later convenience, we introduce  the coordinates  $(s,t)$ of $T^2$ defined by 
\begin{align}
x^2=s+\tau_1 t, \quad x^3=t \tau_2\,,
\end{align}
and take a normalization of an integration measure of $I \times T^2$ as
\begin{align}
 \int_{I \times T^2} (\cdots) :=\frac{1}{8 \pi ^3 L R^2} \int_{- \pi L}^{\pi L} dx^1 \int_{0}^{2\pi R } ds \int_{0}^{2\pi R } dt
\,  (\cdots) \,. 
\end{align}

We impose the same boundary conditions at the left and the right boundaries $x^1=\pm \pi L$ and 
 impose the periodic boundary condition along the two-dimensional torus $T^2$. 
The supersymmetric index of the 3d $\mathcal{N}=2$  supersymmetric theory on $I \times T^2$  is defined by 
\begin{align}
Z_{I \times T^2}:&= \mathrm{Tr}_{\mathcal{H}} (-1)^F e^{- 2 \pi   R  H} \prod_i e^{2 \pi {\rm i} F_i z_i}  \nonumber \\
&=\mathrm{Tr}_{\mathcal{H}} (-1)^F e^{ \pi {\rm i} R \tau (P_2-{\rm i}P_3)}  e^{ \pi i R \bar{\tau} (P_2+{\rm i}P_3)}  \prod_i e^{2 \pi {\rm i} F_i z_i}\,.
\label{eq:index}
\end{align}
Here the coordinate $t$ expresses  the time direction and the coordinates $(x^1, s)$ express the   spatial directions. 
$\mathcal{H}$ is the Hilbert space of the states. 
${\rm i} P_i$  $(i =2,3)$ and $H$ are the generators of translations in the directions $x^i$ $(i =2,3)$ and $t$.  $F_i$'s are the generator for the maximal torus of the flavor symmetry group $G_F$. $z_i$ is the fugacity for $F_i$.  
 The $I \times T^2$ index \eqref{eq:index} is a function of the moduli parameter $\tau$ of the torus $T^2$, but independent of 
 the complex conjugate $\bar{\tau}$.  This is because a generator of the translation $P_2 +{\rm i} P_3$ is written  as a Q-exact form:
\begin{align}
\{{\bf Q}, \bar{\bf Q} \}=2{\rm i}(P_2 +{\rm i} P_3) \simeq 4 \partial_{\bar{w}} \,.
\label{eq:QQbar}
\end{align}
Thus we may write the index as
\begin{align}
Z_{I \times T^2}
&=\mathrm{Tr}_{\mathcal{H}} (-1)^F e^{ \pi {\rm i} R \tau (P_2-{\rm i}P_3)} \prod_i e^{2 \pi {\rm i} F_i z_i} \,.
\end{align}
From \eqref{eq:QQbar}, it follows  that  the correlation functions consisting of Q-closed operators are independent of an anti-holomorphic coordinate $\bar{w}$ of  $M_2$. 
In section \ref{sec:section6}, we compute   two point functions of Q-closed operators in the free chiral multiplet and see this property explicitly.

Although it is possible to introduce 2d  vector multiplets  at the boundaries and evaluate the $I \times T^2$ index by localization,  we concentrate on the theories without 2d  vector multiplets  to avoid clutter of the localization computation. 
We summarize  the localization formula for the supersymmetric index $Z_{I \times T^2}  $ of the 3d $\mathcal{N}=2$ theories on $I \times T^2$:
\begin{align}
\label{eq:indexformula}
Z_{I \times T^2}( y ;q ) &= \frac{1}{|W_{ G }|} \sum_{u_{\ast} \in \mathfrak{M}_{\text{sing}}  } 
\mathop{\text{JK-Res}}_{ u = {u}_*}
( { Q}_{\ast}, \eta) 
 Z^{1\text{-loop}}_{I \times T^2}  Z^{1\text{-loop}}_{{T}^2_L}  Z^{1\text{-loop}}_{{T}^2_R}
 \wedge_{a=1}^{\mathrm{rk}(G)} d u^a  \, .
\end{align}
Here  $T^2_R$ (resp. $T^2_L$) is a boundary torus at $x^1=\pi L$ (resp. $x^1=-\pi L$). 
$G$ is the gauge group for the 3d theory on $I \times T^2$. $|W_{G}|$ is the cardinality of the Weyl group of $G$.   
$u$ is  a  flat connection for the maximal torus of $G$.

 $Z^{1\text{-loop}}_{I \times T^2},  Z^{1\text{-loop}}_{{T}^2_L}$ and $  Z^{1\text{-loop}}_{{T}^2_R}$
 are the one-loop determinants of the 3d $\mathcal{N}=2$ theory on $I \times T^2$,  the one-loop determinants of the 2d $\mathcal{N}=(0,2)$ theory on $T^2_L$ and   those of the 2d $\mathcal{N}=(0,2)$ theory on $T^2_R$:
\begin{align}
 Z^{1\text{-loop}}_{I \times T^2}&= Z_{\text{3d.vec}, G} (e^{2 \pi {\rm i} u} ; q) \prod Z_{\text{chi}, {\sf a}, {\bf R}} (e^{2 \pi {\rm i} u}, y ;  q) \,,
\label{eq:loop3d} \\
 Z^{1\text{-loop}}_{T^2_L}&=    \prod Z_{\text{2d.chi}, {\bf R}_L} (e^{2 \pi {\rm i} u}, y ;  q)  \prod Z_{\text{fermi}, {\bf R}^{\prime}_L} (e^{2 \pi {\rm i} u}, y ;  q)
\label{eq:loop2d0}
 \,,\\
 Z^{1\text{-loop}}_{T^2_R}&=    \prod  Z_{\text{2d.chi}, {\bf R}_R} (e^{2 \pi {\rm i} u}, y ;  q)  \prod Z_{\text{fermi}, {\bf R}^{\prime}_R} (e^{2 \pi {\rm i} u}, y ;  q) \,,
\label{eq:loop2d}
\end{align}
where
\begin{align}
Z_{\text{3d.vec}, G}(x; q)&:=\left( \frac{2 \pi \eta (q)^2 }{\rm i} \right)^{\mathrm{rk}(G)} \prod_{\alpha  \in \text{rt}({\mathfrak{g}})}
{\rm i} \frac{\theta_1(x^{\alpha},q)}{\eta(q)} \, ,
\label{eq:loop3dvec} \\
 Z_{\text{chi}, {\sf N},{\bf R}} (x, y ;  q) &= Z_{\text{2d.chi},{\bf R}} (x, y ;  q):=   \prod_{Q \in \text{wt} (\mathbf{R})}  \prod_{Q^F  \in \text{wt} ({\bf F})}
 {\rm i} \frac{ \eta(q)}{\theta_1(x^{Q} y^{Q^F} ,q)}\, , 
\label{eq:onloop2} \\
 Z_{\text{chi}, {\sf D},{\bf R}} (x, y ; q) &= Z_{\text{fermi}, {\bf R}} (x, y ; q):=  \prod_{Q \in \text{wt} (\mathbf{R})}  \prod_{Q^F  \in \text{wt} ({\bf F})}
{\rm i} \frac{ \theta_1(x^{Q} y^{Q^F} ,q)}{\eta(q)} \, .
\label{eq:oneloop}
\end{align}
The 3d $\mathcal{N}=2$ one-loop determinant \eqref{eq:loop3d} consists of a 3d $G$ vector multiplet $Z_{\text{3d.vec}, G}$, a 3d chiral multiplet with the Neumann (denoted by {\sf N}) boundary condition $Z_{\text{chi}, {\sf N},{\bf R}}$ 
and a 3d chiral multiplet with the Dirichlet (denoted by {\sf D}) boundary   condition $Z_{\text{chi}, {\sf D},{\bf R}}$. 
``${\sf a}$" in \eqref{eq:loop3d} belongs to  ${\sf a} \in \{\sf D, N \}$.
The 2d $\mathcal{N}=(0,2)$ one-loop determinants \eqref{eq:loop2d0} and \eqref{eq:loop2d} consist of  the 2d  chiral multiplets $Z_{\text{2d.chi},{\bf R}} (x, y ;  q)$ and the 2d  fermi multiplets $Z_{\text{fermi}, {\bf R}} (x, y ; q)$. The products   are taken over all the multiplets.

As we will see in the next two subsections,  each 3d field is expanded by KK modes along the interval 
and can be expressed as a sum of 2d massive fields and a 2d massless field. 
By setting  the zero mode of the auxiliary field $D=0$,
 massive KK modes are cancelled among bosons and fermions and 
only massless KK modes survive. 
These massless KK modes coincide with 2d $\mathcal{N}=(0,2)$ multiplets on $T^2$ and 
induce one-loop determinants of 2d theories on $T^2$.

$\text{rt}({\mathfrak{g}})$ denotes roots of the Lie algebra $\mathfrak{g}=\mathrm{Lie}(G)$ and $\text{wt} (\mathbf{R})$ denotes  weights of a representation $\mathbf{R}$ of the gauge group.
$\mathbf{F}$ denotes a representation  of the flavor symmetry group $G_F$. 
 $x^Q:=e^{2 \pi {\rm i} \sum_{a=1}^N Q_a u_a}$ and $y^{Q^F }=e^{2 \pi {\rm i} \sum_{i} Q^{F}_i z_i}$. $Q=(Q_1, \cdots,Q_{\mathrm{rk}(G)})$ is a weight of ${\bf R}$.
$Q^F=(Q_1^F, \cdots,Q^F_{\mathrm{rk}(G_F)})$ is a weight of ${\bf F}$. In the path integral formalism, $u$ (resp. $z$) is a flat connection of  the maximal torus $G$ (resp. $G_F$) on the torus $T^2$.

The  theta function $\theta_1 (x,q)$ and the eta function $\eta(q)$  are defined by
\begin{align}
&
\eta (q)=q^{\frac{1}{24}}\prod_{n=1}^\infty (1-q^n)\,,\non
&
\theta_1(x,q)=-{\rm i} q^{\frac{1}{8}} x^{-\frac{1}{2}}
\prod_{n=1}^\infty (1-q^n)(1-x q^{n-1})(1-x^{-1}q^{n})\,,
\end{align}
where  $q=e^{2 \pi {\rm i} \tau}$.

$\mathop{\text{JK-Res}}_{ u = {u}_*} ({Q}_*,{\eta}) $ is the Jeffrey-Kirwan (JK) residue defined as follows.
When the  $\mathrm{rk}(G)$ hyperplanes of  codimension one, called singular hyperplanes ${Q}_i({u}-{u}_*) =\sum_{a=1}^{\text{rk}(G)} {Q}^a_i({u}^a-{u}^a_*)=0$
 $( i=1, \cdots, \mathrm{rk}(G) )$  
intersect at a point $u_*=(u^1_*, \cdots, u^{\text{rk}(G)}_*)$ in the $u$-space, the JK residue at the point $u_*$ is  defined by
\begin{equation}\label{JKnon-deg}
\begin{aligned}
&\mathop{\text{JK-Res}}_{ u = {u}_*} ({Q}_*,{\eta}) \frac{du^1 \wedge\cdots\wedge du^{\text{rk}(G)}} {{Q}_1({u}-{u}_*)\cdots Q_{\text{rk}(G)} ({u}-{u}_*) }
\\
& \qquad \qquad \qquad \qquad
=
\left\{
\begin{array}{cl}
\frac{1}{|\det({Q}_1,\ldots,{Q}_{\text{rk}(G)})|} & \text{ if } {\eta}\in \text{Cone}({Q}_1,\ldots,{Q}_{\text{rk}(G)})\,, \\
0 & \text{ otherwise}\,.
\end{array}
\right. 
\end{aligned}
\end{equation}
Here  $\text{Cone}({Q}_1,\ldots,{Q}_{\text{rk}(G)})= \sum _{i=1}^{\text{rk}(G)} \mathbb{R}_{> 0} Q_i$  is the 
cone spanned by gauge charge vectors;
$Q_i=(Q^1_i, \cdots, Q^{\text{rk}(G)}_i) \in \mathbb{R}^{\mathrm{rk}(G)}$ $(i=1,\cdots, \mathrm{rk}(G))$.
\eqref{JKnon-deg} depends on a set of charges $Q_*=({Q}_1,\ldots,{Q}_{\text{rk}(G)})$ and $\eta$.
 The sum $\sum_{u_{\ast}  }$ runs over all the points $u_{\ast}$, 
where  $N^{\prime}$  singular hyperplanes with $N^{\prime} \ge \text{rk}(G)$ meet at a point and  the condition ${\eta}\in \text{Cone}({Q}_1,\ldots,{Q}_{\text{rk}(G)})$ is satisfied.
 If $N^{\prime}$ singular hyperplanes with $N^{\prime} > \text{rk}(G)$ intersect at a point, we apply the constructive definition  of  the JK residue in \cite{Benini:2013xpa}. 
The condition  $N^{\prime}=\text{rk}(G)$  is satisfied for  the models treated in sections \ref{sec:section3}-\ref{sec:section5}.
   
Note that each  one-loop determinant of the 3d multiplet has the same form as  the one-loop determinant of the vector, chiral, and fermi multiplet in the 2d $\mathcal{N}=(0,2)$ elliptic genus  in the R-sector \cite{Benini:2013xpa}, respectively and the index is independent of the length of the interval $I$. In other words, the fermionic and the bosonic  Kaluza--Klein  modes in the $x^1$-direction cancel out, expect for the lowest modes that form the 2d $\mathcal{N}=(0,2)$ multiplets.

Before we move to  technical details of the localization computation, 
let us  briefly recall  the principle of supersymmetric localization  \cite{Pestun:2007rz}.
 The partition function or the index of the supersymmetric theory in the path integral formalism is expressed as
\begin{align}
Z=\int \mathcal{D} \Psi e^{- S[\Psi]} \, .
\end{align}
 We assume the action $S[\Psi]$ is invariant by 
a fermionic conserved charge (supercharge)  ${\bf Q}$,
where $\Psi$ denotes the component fields of the supermultiplets in the theory.
Without changing the value of the partition function, one can add   one-parameter family of the Q-exact term (action)  $ \frac{1}{g^2} {\bf Q} \cdot V [\Psi]$ to the action.
If  there are more than one Q-exact term,   the action is deformed by a multi-parameter family of  Q-exact terms $\sum_i \frac{1}{g^2_i} {\bf Q} \cdot V_i [\Psi]$.
By taking the weak coupling limit $g^2 \to 0$, the path integral is exactly evaluated in   the one-loop computation of the fluctuations  around the saddle points (zero loci) $\Psi_0$ defined by ${\bf Q} \cdot V [\Psi_0]=0$:
\begin{align}
Z=\lim_{g \to 0} \int \mathcal{D} \Psi e^{- S[\Psi]-\frac{1}{g^2} {\bf Q} \cdot V[\Psi]}
= \int d \Psi_0  e^{- S[\Psi_{0}]} Z_{1\text{-loop} }(\Psi_0)\, .
\end{align}
Here we expanded fields as   $\Psi = \Psi_0+g \tilde{\Psi}$, where $\tilde{\Psi}$ denotes fluctuations  around the saddle point configurations $\Psi_0$.  By integrating out $\tilde{\Psi}$, we obtain
 the one-loop determinant $Z_{1\text{-loop} }$ of ${\bf Q} \cdot V[{\Psi}]$ around the saddle point $\Psi_0$.
When fermion zero-modes exist,  actual localization computation is more involved and 
one has to  treat carefully the zero-mode integral \cite{Benini:2013nda, Benini:2013xpa}. 

In our case, 
the Q-exact terms for the three dimensional 
part are taken as the  super Yang-Mills action \eqref{SYMaction} and the kinetic action of the 3d chiral multiplet \eqref{eq:action3dchi}. The Q-exact terms for the two dimensional 
part are taken as  the  kinetic action of the 2d chiral multiplet \eqref{eq:action2dchi} and the fermi multiplet \eqref{eq:Fermiaction}:  
\begin{align}
\sum_i \frac{1}{g^2_i} {\bf Q} \cdot V_i [\Psi] &= \frac{1}{e^2} S_{\text{SYM}}  +\frac{1}{g^2_1} S_{\text{chi}} 
+\sum_{i=L,R} \left( \frac{1}{g^2_{2.i}} 
S_{\text{2d.chi}, i}+\frac{1}{g^2_{3.i}} S_{\text{fermi},i} \right)\,.
\end{align}
Here $i=L, R$ expresses the boundary theory at $T^2_L, T^2_R$, respectively. 
First we take the limit $g^2_1, g^2_{2.i}, g^2_{3.i}\to 0$ and then we take the limit $e^2 \to 0$. 

The Q-closed actions are the 3d  FI-term, and the 3d and 2d superpotential terms.
 The saddle point configuration  of  the 3d  FI-term is non-zero. On the other hand, 
the saddle point  configuration of superpotential terms is zero . 
The localization formula does not explicitly depend on the  superpotential terms. 
The superpotentials contribute to the localization formula   through the 3d  matrix factorization.

\subsection{Evaluation of the one-loop determinants}
First we evaluate the one-loop determinant for  the 3d $\mathcal{N}=2$ vector multiplet with the boundary condition \eqref{eq:vecdbdy}.
As we have seen in the previous section,   
the super Yang-Mills Lagrangian is written as the SUSY transformation by ${\bf Q}$. We choose it as a Q-exact term. 
The saddle point condition for the vector multiplet, i.e., the zero loci of the  super Yang--Mills Lagrangian  are given by $F_{\mu \nu}=0$ and $D_{\mu} \sigma=0$ and constant values of $\lambda, \bar{\lambda}$.

According to the boundary condition \eqref{eq:vecdbdy}, the saddle point configurations are given by $A_1=0$ and $\sigma=0$
and $F_{23}=0$. Let $\bar{A}$ be the gauge field which satisfies the saddle point condition $F_{23}=0$:
\bel
&&\bar{A}=\bar{A}_2dx^2+\bar{A}_3dx^3=\bar{A}_t dt+\bar{A}_s ds\,
\label{eq:sadllep},
\ee
where $\bar{A}_t$ and $\bar{A}_s$ are constants.
The covariant derivative with the gauge field  \eqref{eq:sadllep} $\bar{D}_{\mu}=\partial_{\mu}+ {\rm i} \bar{A}_{\mu}$ is given by
 \begin{align}
\bar{D}_1&=\partial_1, \non
\bar{D}_2+{\rm i} \bar{D}_3&=\frac{\rm i}{\tau_2}\{(\del_t-\tau\del_s)+{\rm i}(\bar{A}_t-\tau \bar{A}_s)\}
=\frac{\rm i}{\tau_2}\left(\del_t-\tau\del_s+i\frac{u}{R}\right)\,,\,\,\non
\bar{D}_2-{\rm i} \bar{D}_3&=\frac{-{\rm i}}{\tau_2}\{(\del_t-\bar{\tau}\del_s)+{\rm i}(A_t-\bar{\tau} A_s)\}
=\frac{-{\rm i}}{\tau_2}\left(\del_t-\bar{\tau}\del_s+{\rm i}\frac{\bar{u}}{R} \right)\,.
\label{eq:saddleDer}
\end{align}
Here $u$ and $\bar{u}$ are defined by
\begin{align}
2\pi u&:=\oint_t \bar{A}-\tau \oint_s \bar{A}\,, \quad 2\pi \bar{u}:=\oint_t \bar{A}-\bar{\tau} \oint_s \bar{A}\,,
\end{align}
and  $u, \bar{u}$ take  values in a representation of Cartan subalgebra of the gauge group,
\begin{align}
u=\sum_{a=1}^{{\rm rk}(G) } u^a H_a,\, \quad \bar{u}=\sum_{a=1}^{{\rm rk}(G)} \bar{u}^a H_a\,.
\end{align}
The representation is determined by  the matter field on which  the covariant derivative acts.
To make the expressions concise, we use same symbol $\{ H_a \}_{a=1}^{{\rm rk}(G)}$
 to express  a generator of the Cartan subalgebra of $\mathfrak{g}$ and its representation.

In the path integral formalism, a flavor fugacities  correspond to turning on the background gauge  field $A^F_i$ for  the maximal torus of $G_F$. 
Then  the covariant derivative \eqref{eq:saddleDer} is shifted by the background gauge field for the flavor symmetry group:
 \begin{align}
\partial_i+{\rm i} \bar{A}_i \to {\partial}_i+{\rm i} \bar{A}_i +{\rm i} A^F_i \text{ for } i=2, 3\,.  
\end{align}
The fugacity of the flavor symmetry is written in terms of the background gauge field as 
\begin{align}
2\pi z&:=\oint_t {A}^F-\tau \oint_s {A}^F\,, \quad 
2\pi \bar{z}:=\oint_t {A}^F-\bar{\tau} \oint_s {A}^F\,.
\end{align}
where $z=(z_1,\cdots,z_{\mathrm{rk}(G_F)})$.
To make the  equations   concise, we include $A^F_i$ in the definition of $\bar{D}_i$.

We evaluate the one-loop determinants of the fluctuations around the saddle point condition \eqref{eq:sadllep}.
We focus on the mode expansions along the interval $I$ under the boundary condition. 
Then the fluctuations are expanded as 
\begin{align}
\tilde{\sigma}&=\sum_{\ell =1}^{\infty} \sigma^{(\ell)}\sin \frac{\ell x^1}{L}\,,\,\,
\tilde{A}_1=\sum_{\ell =1}^{\infty} A_1^{(\ell)}\sin \frac{\ell x^1}{L}\,,\,\,
\tilde{A}_{i}=\sum_{\ell =0}^{\infty} A_{i}^{(\ell)} \cos \frac{\ell x^1}{L}, \quad (i=2,3), \nonumber \\
\tilde{\lambda}_{\alpha} &= 
 \sum_{\ell =0}^{\infty} \lambda^{(\ell)}_{c}\cos\frac{\ell x^1}{L}
+(-1)^{\alpha-1} \sum_{\ell =1}^{\infty} \lambda^{(\ell)}_{s}\sin\frac{\ell x^1}{L} 
\,,\,\, \nonumber \\
\tilde{\bar{\lambda}}_{\alpha} &=
\sum_{\ell =0}^{\infty} \bar{\lambda}^{(\ell)}_{c}\cos\frac{\ell x^1}{L}
+(-1)^{\alpha-1} \sum_{\ell =1}^{\infty} \bar{\lambda}^{(\ell)}_{s}\sin\frac{\ell x^1}{L} \quad (\alpha=1,2).
\label{eq:modevec}
\end{align}
Fields with tilde $\tilde{}$   express  the  fluctuations around the saddle point configuration.
Each Kaluza-Klein mode  $\Psi^{(\ell)} \in \{ A_{\mu}^{(\ell)}, \sigma^{(\ell)}, {\lambda}^{(\ell)}_{c/s},   \bar{\lambda}^{(\ell)}_{c/s}  \} $  is expanded as
\begin{align}
\Psi^{(\ell) } &= \sum_{\alpha \in \mathrm{rt}(\mathfrak{g}) } \Psi^{(\ell)}_{\alpha} E_{\alpha} \,.
\end{align}
Here $\{ H_a, E_{\alpha} \}$ is the Cartan-Weyl basis for the Lie algebra $\mathfrak{g}$ with 
the normalization $ \mathrm{Tr} (E_{\alpha} E_{\beta} )=\delta_{\alpha+\beta, 0}$.

In \eqref{eq:modevec}, the modes labeled by $\ell$ are functions of the coordinates of the torus $(s, t)$. 
 Since  the fields in the vector multiplet obey the periodic boundary condition, the Kaluza-Klein modes $\Psi^{(\ell)}$'s are 
expanded as
\begin{align}
&\Psi^{(\ell)} (s+2\pi R,t)=\Psi^{(\ell)} (s,t)\,,\,\,
\Psi^{(\ell)} (s,t+2\pi R)=\Psi^{(\ell)} (s,t)\,,\,\,\non
&\rightarrow \Psi^{(\ell)} (s,t)=\sum_{m,n \in \mathbb{Z} }\exp \left({ {\rm i}\frac{m}{R}s+{\rm i}\frac{n}{R}t}\right) \,\Psi_{(\ell,m,n)} \,.
\label{eq:KKmodes}
\end{align}
 To perform the path integral for the gauge field, we introduce a gauge fixing term ${\cal L}_{\text{gf}}$ in the  $R_\xi$-gauge  with $\xi=1$ 
and introduce the Faddeev--Popov ghost $C$ and the anti-ghost $\bar{C}$:
\bel
&&{\cal L}_{\text{gf}}= \frac{1}{2}{\rm Tr}(\bar{D}_{\mu } \tilde{A}^{\mu} )^2\, , \quad 
{\cal L}_{\text{gh}}={\rm Tr} \bar{C}(-\bar{D}_{\mu }\bar{D}^{\mu})C\,.
\ee
We take the Neumann boundary condition for the ghosts at $x^1=\pm \pi L$ and expand them as
\bel
&&
C=\sum_{\ell =0}^{\infty} C^{(\ell)}\cos \frac{\ell x^1}{L}\,,\quad
\bar{C}=\sum_{\ell =0}^{\infty} \bar{C}^{(\ell)}\cos \frac{\ell x^1}{L}\,.
\ee
The bosonic fields are combined into a bilinear form 
\bel
&&
{\frac{1}{\pi L}}\int^{\pi L}_{-\pi L}dx^1\,({\cal L}_{\text{vec}}+{\cal L}_{\text{gf}} )|_{\text{bosonic part}}\non
&&=
{\rm Tr}\left[ \vec{A}^{(0)}{}^T\, {\cal M}_0 \, \vec{A}^{(0)}\right]
+\frac{1}{2}\sum_{\ell =1}^{\infty}
{\rm Tr}\left[ \vec{A}^{(\ell)}{}^T\, {\cal M}_{\ell} \, \vec{A}^{(\ell)}
+
 {\sigma}^{(\ell)}\, \left(\frac{\ell^2}{L^2}-\bar{D}^2_2-\bar{D}^2_3 \right) {\sigma}^{(\ell)}\right]  +{\rm Tr}[D^2]
\,,\non
\ee
with
\begin{align}
&\vec{A}^{(0)}=
\left(
\begin{array}{c}
A^{(0)}_2\\
A^{(0)}_3
\end{array}
\right)\,,\,\,
\vec{A}^{(\ell)}=
\left(
\begin{array}{c}
A^{(\ell)}_1\\
A^{(\ell)}_2\\
A^{(\ell)}_3
\end{array}
\right)\,,\,\,
\non
&
{\cal M}_0=\mathrm{diag} (
-\bar{D}_2^2-\bar{D}_3^2 , -\bar{D}_2^2-\bar{D}_3^2 )
\,,\,\,\non
&
{\cal M}_{\ell}= \mathrm{diag} \left(
\frac{\ell^2}{L^2}-\bar{D}_2^2-\bar{D}_3^2 ,  \frac{\ell^2}{L^2}-\bar{D}_2^2-\bar{D}_3^2
,\frac{\ell^2}{L^2} -\bar{D}_2^2-\bar{D}_3^2 \right)
\,  (\ell =1,2,\cdots )
\,.
\end{align}
Then the bosonic part of the one-loop determinants of the vector multiplet and the gauge fixing term are given by\footnote{As in the case of localization  of the elliptic genera,  we  absorb the factor $(R\tau_2)^{-1}$ in front of $|n-m\tau +\alpha (u)|^2$ by rescaling $L$ and $D$ in the following one-loop computations.}
\begin{align}
&\text{Det}^{-1}\left(-\bar{D}_2^2 - \bar{D}_3^2 \right) \cdot
 \prod_{\ell=0}^{\infty}   \text{Det}^{-2} \left(
\frac{\ell^2}{L^2}-\bar{D}_2^2 - \bar{D}_3^2 
 \right)
\,.
\label{eq:bosloopvec}
\end{align}
Here $\mathrm{Det}$ denotes the functional determinant with respect to derivatives of  the coordinates of  the torus and  $E_{\alpha}$'s.

The fermionic part of the vector multiplet and the ghost action is expanded as
\begin{align}
& \frac{1}{2\pi L}
\int^{\pi L}_{-\pi L}dx^1 \left({\rm i} \tilde{\bar{\lambda}} \gamma^{\mu} \bar{D}_{\mu} \tilde{\lambda}  + \bar{C} (D_{\mu}D^{\mu})C  \right)
=2 \bar{\lambda}_c^{(0)}(\bar{D}_2+{\rm i}  \bar{D}_3)\lambda^{(0)}_c
\non
& \qquad \qquad +\sum_{\ell =1}^{\infty}
\left(
\begin{array}{cc}
\bar{\lambda}_c^{(\ell)} & \bar{\lambda}_s^{(\ell)}
\end{array}
\right)
\left(
\begin{array}{cc}
\bar{D}_2+{\rm i} \bar{D}_3 & -{\rm i}\frac{\ell}{L}\\
{\rm i}\frac{\ell}{L} & \bar{D}_2-{\rm i} \bar{D}_3
\end{array}
\right)
\left(
\begin{array}{c}
{\lambda}_c^{(\ell)} \\
{\lambda}_s^{(\ell)}
\end{array}
\right)
\nonumber \\
&\qquad \qquad  \qquad+\sum_{\ell =0}^{\infty} \bar{C}^{(\ell)}\left(\frac{\ell^2}{L^2}-\bar{D}_2^2-\bar{D}_3^2\right) C^{(\ell)}\,.
\end{align}
Then the fermionic part of the one-loop determinant of the vector multiplet and the ghost action is given by
\begin{align}
&\text{Det} \left(\bar{D}_2+{\rm i} \bar{D}_3 \right) \text{Det}\left(
\bar{D}_2^2 + \bar{D}_3^2 \right)
\prod_{\ell=1}^{\infty}     \text{Det}^{2} \left(\frac{\ell^2}{L^2}-
\bar{D}_2^2 - \bar{D}_3^2 \right)\,.
\label{eq:Fervec}
\end{align}
From  \eqref{eq:bosloopvec} and \eqref{eq:Fervec}, 
 we obtain the one-loop determinant  of the vector multiplet in \eqref{eq:loop3dvec}:
\begin{align}
Z_{\text{3d.vec}, G}(x,  q)&=\text{Det} \left(\bar{D}_2+i \bar{D}_3 \right)
\nonumber \\
& =\prod_{n,m \in \mathbb{Z} \atop (m,n) \neq (0,0)} \left(n + m \tau \right) \cdot
\prod_{n,m \in \mathbb{Z}}\prod_{\alpha \in \text{rt}(\mathfrak{g})} \left(n + m \tau + \alpha (u)\right) \nonumber \\
&=\left( \frac{2 \pi \eta (q)^2 }{\rm i} \right)^{\mathrm{rk}(G)} \prod_{\alpha \in \text{rt}(\mathfrak{g}) } 
{\rm i} \frac{\theta_1(x^{\alpha},q)}{\eta(q)} \,.
\end{align}
 We used the zeta function regularization in the last line.
Next we evaluate the one-loop determinant for the chiral multiplet in a representation ${\bf R}$ of the gauge group $G$ 
and in a representation ${\bf F}$ of a flavor symmetry group $G_F$.

For the Dirichlet boundary condition, the mode expansions of the chiral multiplets along the interval $I$  are expressed as
\begin{align}
&
\phi =\sum_{\ell =1}^{\infty}\phi^{(\ell)}\sin\frac{\ell x^1}{L}\,,\,\,
\bar{\phi} =\sum_{\ell =1}^{\infty}\bar{\phi}^{(\ell)}\sin\frac{\ell x^1}{L}\,,\,\,\non
&
\psi_{\alpha} = 
\sum_{\ell =0}^{\infty} \psi^{(\ell)}_{c}\cos\frac{\ell x^1}{L}
+(-1)^{\alpha-1}\sum_{\ell =1}^{\infty}\psi^{(\ell)}_{s}\sin\frac{\ell x^1}{L}
\,,\,\, \nonumber \\
&
\bar{\psi}_{\alpha} =
\sum_{\ell =0}^{\infty} \bar{\psi}^{(\ell)}_{c}\cos\frac{\ell x^1}{L}
+(-1)^{\alpha-1}\sum_{\ell =1}^{\infty} \bar{\psi}^{(\ell)}_{s}\sin\frac{\ell x^1}{L}, \quad (\alpha=1,2). 
\end{align}
The integration over the interval $I$ gives 
\begin{align}
&\frac{1}{2\pi L}\int^{\pi L}_{-\pi L}dx^1 \, \tilde{\bar{\phi}}(-\bar{D}^{\mu} \bar{D}_{\mu} +{\rm i}D) \tilde{\phi} 
=
\frac{1}{2}\sum_{\ell =1}^{\infty } \bar{\phi}^{(\ell)}\left(-\bar{D}_2^2-\bar{D}_3^2+\frac{\ell^2}{L^2}
+{\rm i}D\right)\phi^{(\ell)}\,,\non
&\frac{1}{2\pi L}\int^{\pi L}_{-\pi L}dx^1\,\bar{\psi}(-{\rm i} \gamma^{\mu} \bar{D}_{\mu} )\psi 
=-2 \bar{\psi}^{(0)}_c(\bar{D}_2+{\rm i} \bar{D}_3)\psi^{(0)}_c \nonumber \\
& \qquad \qquad \qquad  
 \qquad -\sum_{\ell =1}^{\infty}
\left(
\begin{array}{cc}
\bar{\psi}_c^{(\ell)} & \bar{\psi}_s^{(\ell)}
\end{array}
\right)
\left(
\begin{array}{cc}
\bar{D}_2+{\rm i} \bar{D}_3 & -{\rm i}\frac{\ell}{L}\\
{\rm i}\frac{\ell}{L} & \bar{D}_2-{\rm i} \bar{D}_3
\end{array}
\right)
\left(
\begin{array}{c}
{\psi}_c^{(\ell)} \\
{\psi}_s^{(\ell)}
\end{array}
\right)\,.
\label{eq:modeDir}
\end{align}
Here we write the zero-mode of the auxiliary  field $D$  simply by   the same symbol  $D$.

For the Neumann boundary condition, the mode expansions  along the interval $I$ are given by 
\begin{align}
&
\phi =\sum_{\ell =0}^{\infty}\phi^{(\ell)}\cos\frac{\ell x^1}{L}\,,\,\,
\bar{\phi} =\sum_{\ell =0}^{\infty} \bar{\phi}^{(\ell)}\cos\frac{\ell x^1}{L}\,,\,\,\non
&
\psi_{\alpha} =(-1)^{\alpha-1} 
\sum_{\ell =0}^{\infty} \psi^{(\ell)}_{c}\cos\frac{\ell x^1}{L}
+\sum_{\ell =1}^{\infty} \psi^{(\ell)}_{s}\sin\frac{\ell x^1}{L}
\,,\,\,\non
&
\bar{\psi}_{\alpha} =
(-1)^{\alpha-1}\sum_{\ell =0}^{\infty} \bar{\psi}^{(\ell)}_{c}\cos\frac{\ell x^1}{L}
+\sum_{\ell =1}^{\infty} \bar{\psi}^{(\ell)}_{s}\sin\frac{\ell x^1}{L}
\,, \quad (\alpha=1,2).
\end{align}
The mode expansions of the kinetic terms are evaluates as
\begin{align}
&
\frac{1}{2\pi L}\int^{\pi L}_{-\pi L}dx^1\, \tilde{\bar{\phi}}(-\bar{D}^{\mu}\bar{D}_{\mu}+{\rm i}D) \tilde{\phi} \non
&=
\bar{\phi}^{(0)}\left(-\bar{D}_2^2-\bar{D}_3^2+{\rm i}D\right)\phi^{(0)}
+
\frac{1}{2}\sum_{\ell =1}^{\infty}\bar{\phi}^{(\ell)}\left(-\bar{D}_2^2-\bar{D}_3^2+\frac{\ell^2}{L^2}
+{\rm i}D\right)\phi^{(\ell)}\,,\non
&
\frac{1}{2\pi L}\int^{\pi L}_{-\pi L}dx^1 \,\tilde{\bar{\psi}} (-{\rm i}\gamma^{\mu} \bar{D}_{\mu}) \tilde{\psi} \non
&=-2 \bar{\psi}^{(0)}_c(\bar{D}_2-{\rm i} \bar{D}_3)\psi^{(0)}_c
-\sum_{\ell =1}^{\infty}
\left(
\begin{array}{cc}
\bar{\psi}_c^{(\ell)} & \bar{\psi}_s^{(\ell)}
\end{array}
\right)
\left(
\begin{array}{cc}
\bar{D}_2-{\rm i} \bar{D}_3 & -{\rm i}\frac{\ell}{L}\\
{\rm i}\frac{\ell}{L} & \bar{D}_2+{\rm i} \bar{D}_3
\end{array}
\right)
\left(
\begin{array}{c}
{\psi}_c^{(\ell)} \\
{\psi}_s^{(\ell)}
\end{array}
\right)\,.\non
\label{eq:modeNeu}
\end{align}

From \eqref{eq:modeDir} and \eqref{eq:modeNeu}, we obtain
the effect of one-loop determinants with the Dirichlet and the Neumann boundary conditions:
\begin{align}
g_{\text{chi}, {\sf D}} (u, D)&=\mathrm{Det} \left(\bar{D}_2+{\rm i} \bar{D}_3 \right) \prod_{\ell =1}^{\infty}
\frac{\mathrm{Det} \left(\frac{\ell^2}{L^2}-\bar{D}_2^2-\bar{D}_3^2\right)}
{\mathrm{Det} \left(\frac{\ell^2}{L^2}-\bar{D}_2^2-\bar{D}_3^2+{\rm i}Q(D)\right)}
 \nonumber \\
& =\prod_{Q  \in \text{wt} (\mathbf{R})} \prod_{Q^F  \in \text{wt} (\mathbf{F})} \prod_{m,n \in \mathbb{Z}} (n-m\tau +Q (u)+Q^F (z) ) 
\nonumber \\
& \qquad \qquad \times
\prod_{\ell=1}^{\infty}
 \frac{\frac{\ell^2}{L^2}+|n-m\tau +Q (u)+Q^F (z)|^2}{\frac{\ell^2}{L^2}+|n-m\tau +Q (u)+Q^F (z)|^2 +{\rm i} Q (D)} \,,
\label{eq:1loopN}
\\
g_{\text{chi}, {\sf N}}(u, D)&=\frac{\mathrm{Det} \left(\bar{D}_2-{\rm i} \bar{D}_3 \right)}{\mathrm{Det} \left(-\bar{D}_2^2-\bar{D}_3^2+{\rm i} Q(D)\right)}
 \prod_{\ell =1}^{\infty}\frac{\mathrm{Det} \left(-\bar{D}_2^2-\bar{D}_3^2+\frac{\ell^2}{L^2}\right)}
{\mathrm{Det} \left(-\bar{D}_2^2-\bar{D}_3^2+\frac{\ell^2}{L^2}+{\rm i} Q(D)\right)}
 \nonumber \\
&=\prod_{Q \in \text{wt}({\bf R})} \prod_{Q^F  \in \text{wt} (\mathbf{F})} \prod_{m,n \in \mathbb{Z}} \frac{(n-m\bar{\tau} +Q (\bar{u})+Q^F (\bar{z}) )}{|n-m\tau +Q (u)+Q^F (z)|^2 +{\rm i} Q (D)} 
\nonumber \\
& \qquad \times
\prod_{\ell=1}^{\infty}
 \frac{\frac{\ell^2}{L^2}+|n-m\tau +Q (u)+Q^F (z)|^2}{\frac{\ell^2}{L^2}+|n-m\tau +Q (u)+Q^F (z)|^2 +{\rm i} Q (D)} \,.
\label{eq:1loopD}
\end{align}
Here ${\sf D}$ and ${\sf N}$ denote the Dirichlet  and the Neumann boundary condition, respectively.
Again $Q=(Q_1,\cdots, Q_{\mathrm{rk}(G)})$ (resp. $Q^F=(Q^F_1,\cdots, Q^F_{\mathrm{rk}(G_F)})$ ) is 
a weight of a representation of $G$ (resp. $G_F$).  
If  $D=0$, $g_{\text{chi}, {\sf a}}$ for ${\sf a} \in \{{\sf N, D}\}$ becomes  
 the \eqref{eq:onloop2} and \eqref{eq:oneloop}:
\begin{align}
&g_{\text{chi}, {\sf D}} (u, D=0)= Z_{\text{chi}, {\sf D},{\bf R}} (x, y ;  q) =
\prod_{Q  \in \text{wt} (\mathbf{R})} \prod_{Q^F  \in \text{wt} (\mathbf{F})}
 \frac{{\rm i} \theta_1(x^{Q} z^{Q^F} ,q)}{\eta(q)} \, ,  
\label{eq:Dneq1} \\
&g_{\text{chi}, {\sf N}} (u, D=0)= Z_{\text{chi}, {\sf N},{\bf R}} (x, y ;  q) =\prod_{Q  \in \text{wt} (\mathbf{R})} \prod_{Q^F  \in \text{wt} (\mathbf{F})}
 \frac{{\rm i} \eta(q)}{\theta_1(x^{Q} y^{Q^F} ,q)} \,.
\label{eq:Dneq2}
\end{align}
Here we used the  zeta function regularization.

In  similar way, the one-loop determinants of the 2d $\mathcal{N}=(0,2)$ multiplets
 are computed as 
\begin{align}
g_{\text{fermi}} (u)&=\prod_{Q  \in \text{wt} (\mathbf{R})} \prod_{Q^F  \in \text{wt} (\mathbf{F})} \prod_{m,n \in \mathbb{Z}} (n-m\tau +Q (u)+Q^F (z) ) \,,
\nonumber \\
g_{\text{2d.chi}}(u, D)
&=\prod_{Q \in \text{wt}({\bf R})} \prod_{Q^F  \in \text{wt} (\mathbf{F})} \prod_{m,n \in \mathbb{Z}} \frac{(n-m\bar{\tau} +Q (\bar{u})+Q^F (\bar{z}) )}{|n-m\tau +Q (u)+Q^F (z)|^2 +{\rm i} Q (D)} \,.
\end{align}
Here ${\bf R}$ and ${\bf F}$ are  representations of  the gauge and flavor symmetry groups of the 2d multiplets on  the boundary torus, respectively.
Note that $g_{\text{fermi}} (u)$ and $g_{\text{2d.chi}}(u, D=0)$ are given by \eqref{eq:Dneq1} and \eqref{eq:Dneq2} with the zeta function regularization.

Next we perform the integral over the zero-modes of the gaugini $\lambda, \bar{\lambda}$ and the auxiliary field $D$, which  imposes  $D=0$ and gives \eqref{eq:Dneq1} and \eqref{eq:Dneq2}.

\subsection{Integration over  zero modes}
In the previous subsection,  we have evaluated the one-loop determinant for the 
fluctuation around the saddle point locus.  When the gaugino zero-modes exist, the path integral over the gaugino zero-modes contributes to the supersymmetric localization procedure. In this subsection,  we will perform the integration over the gaugino zero-modes. 
 The gaugino zero modes   $\lambda_{c, (0,0,0)}, \bar{\lambda}_{c, (0,0,0)}$ are expanded as
\begin{align}
\lambda_{c, (0,0,0)}=\sum_{a=1}^{\mathrm{rk}(G)} \lambda^a_{({\bm 0})}  H_a, \quad \bar{\lambda}_{c, (0,0,0)}=\sum_{a=1}^{\mathrm{rk}(G)} \bar{\lambda}^a_{({\bm 0})}  H_a.
\end{align}
Then the  following combination of the Yukawa couplings saturates the integral over the gaugino zero-modes:
\begin{align}
&\int \prod_{a=1}^{\mathrm{rk}(G)} d \lambda_{({\bm 0})}^a d \bar{\lambda}^a_{({\bm 0})} \mathcal{D} \bar{\tilde{\phi}}
\mathcal{D} \tilde{\phi}  \mathcal{D} \bar{\tilde{\psi}}  \mathcal{D} \tilde{\psi}  
e^{ \int_{I \times T^2} [ \tilde{\bar{\phi}}(\bar{D}^{\mu} \bar{D}_{\mu} -{\rm i}D) \tilde{\phi} +\tilde{\bar{\psi}} (-i\gamma^{\mu} \bar{D}_{\mu}) \tilde{\psi}]}
\nonumber \\
&  \qquad \qquad \times  \frac{1}{(\mathrm{rk}(G) !)^2}  \left( \int_{I \times T^2} {\rm i} \tilde{\bar{\psi}} \lambda_{0,0,0} \tilde{\phi} \right)^{\mathrm{rk}(G)} \left( \int_{I \times T^2} {\rm i} \tilde{\bar{\phi}} \bar{\lambda}_{0,0,0} \tilde{\psi} \right)^{\mathrm{rk}(G)}   \nonumber \\
&  \qquad \qquad =  \mathrm{det} h_{\sf a} (u, D)
g_{\text{chi}, {\sf a} } (u,D )\,. 
\label{eq:intgaugino}
\end{align}
Here ${\sf a} \in \{ {\sf D, N} \}$ denotes the Dirichlet boundary condition ({\sf D}) or the Neumann boundary condition ({\sf N}).  $h^{a b}_{\sf a}$ with $a,b=1,\cdots, \mathrm{rk}(G)$ is defined by   
\begin{align}
h^{a b}_{\sf D}&= \sum_{Q^F \in \text{wt}(\mathfrak{g})} \sum_{\ell =1}^{\infty} \sum_{m,n \in \mathbb{Z}}
\frac{2 Q^a Q^b \left(n -m \tau + Q (u)+Q^F (z) \right)  }{    |n -m \tau +Q (u)+Q^F (z)|^2 +\frac{\ell^2}{L^2}   } 
\nonumber \\ & \qquad \qquad \times 
 \frac{1}{ |n -m \tau+ Q (u)+Q^F (z)|^2 +\frac{\ell^2}{L^2}  + {\rm i} Q (D) }
\, , \nonumber \\
h^{a b}_{\sf N}&= \sum_{Q^F \in \text{wt}(\mathfrak{g})} \sum_{\ell =0}^{\infty} \sum_{m,n \in \mathbb{Z}}
\frac{2 Q^a Q^b  \left(n -m \tau + Q (u)+Q^F (z) \right)  }{   |n -m \tau +Q (u)+Q^F (z)|^2 +\frac{\ell^2}{L^2} } \nonumber \\
& \qquad \times 
\frac{1}{ |n -m \tau+ Q (u)+Q^F (z)|^2 +\frac{\ell^2}{L^2}  + {\rm i} Q_i (D)} \,,
\end{align}
 
  $h^{a b}_{\sf a}$ satisfies the following relations: 
\begin{align}
\frac{\partial g_{\sf a}(u,D)}{\partial \bar{u}_a}&=-{\rm i}  h^{ab}_{\sf a} (u, D)  D_b g_{\sf a}(u,D), 
\label{eq:pro1}\\
\frac{\partial h^{ab}_{\sf a}(u,D)}{\partial \bar{u}_c}&=\frac{\partial h^{c a}_{\sf a}(u,D)}{\partial \bar{u}_b}=\frac{\partial h^{b c }_{\sf a}(u,D)}{\partial \bar{u}_a}\,.
\label{eq:pro2} 
\quad 
\end{align}
 The evaluation of  integrals over the 3d gaugino zero-modes with   the Yukawa couplings including  boundary 2d fields are parallel to \eqref{eq:intgaugino}:
\begin{align}
h^{a b}_{\text{2d.chi}}&=\sum_{Q^F \in \text{wt}(\mathfrak{g})} \sum_{m,n \in \mathbb{Z}}
 \frac{2 Q^a Q^b}{\left(  |n -m \tau+ Q (u)+Q^F (z)|^2   + {\rm i} Q (D) \right)
\left(  n -m \bar{\tau}+ Q (\bar{u})+Q^F (\bar{z})\right) } \,. 
\end{align}

After performing the path integral for the fluctuations and the gaugino zero-modes, we obtain the following result:
\begin{align}
Z_{I \times T^2}&= \frac{c}{|W_G|}\lim_{e \to 0 \atop \varepsilon \to 0} \int_{\mathfrak{M} \backslash \Delta_{\varepsilon} } d^{\mathrm{rk}(G)} u \, \, d^{\mathrm{rk}(G)} \bar{u}  \int_{  \mathbb{R}^{\mathrm{rk}(G)} }  d^{\mathrm{rk}(G)} D 
\nonumber \\  & \qquad \qquad \times  \det  h (u,\bar{u}, D)  g (u,\bar{u},  D) \exp \left[-   \frac{1}{2e^2}\mathrm{Tr}(D^2) -{\rm i} \zeta (D) \right]\,,
\label{eq:zeromodeint}
\end{align}
with
\begin{align}
\det h &= \prod {\rm det }  h_{\sf  a} \prod {\rm det } h_{{\rm 2d}.\text{chi}} \,, \nonumber \\
g &=Z_{\text{3d.vec}, G}   \prod g_{\text{chi}, {\sf a}} \prod g_{\text{2d.chi}} \, \prod g_{\text{fermi}} \,.
\end{align}
Here the products run over the 3d chiral multiplets,  the 2d chiral and the fermi multiplets. 
$c$ is an overall constant. For elliptic genera,  the overall constant is taken to reproduce the free field computation of the elliptic genera \cite{Witten:1993jg, Kawai:1993jk, Kawai:1994np}. 
Since the indices on $I \times T^2$ do not depend on the length of $I$, we take the same  normalization as  the elliptic genera; $c=(4 \pi^2 {\rm i} )^{-\mathrm{rk}(G)}$.

$\mathfrak{M}$ is the space of flat connections $u$ and $\bar{u}$. 
 $\Delta_{\varepsilon}$ is the union of the  $\varepsilon$-neighborhood around the singular loci of the one-loop determinant
 defined as follows. 
First we define $H_i$ called  a singular hyperplane associated with  the $i$-th  3d chiral multiplet with the Neumann boundary condition or the  $i$-th 2d $\mathcal{N}=(0,2)$ chiral multiplet by
\begin{align}
H_i:=\{ u=(u_1,\cdots,u_{\mathrm{rk}(G)}) | u_i \in \mathbb{C}/\mathbb{Z}+\tau \mathbb{Z}, \, Q_i (u)+Q_i^F(z) =0 \} .
\end{align}
Then  $ \Delta_{\varepsilon} (H_i) $ is  the $\varepsilon$-neighborhood of the singular hyperplane $H_i$:
\begin{align}
\Delta_{\varepsilon}(H_i):=\{ u=(u_1,\cdots,u_{\mathrm{rk}(G)}) |  \, \, |Q_i (u)+Q_i^F(z)| \le \varepsilon \} \,.
\end{align}
$ \Delta_{\varepsilon}$ is the union of $ \Delta_{\varepsilon} (H_i) $ defined by
\begin{align}
\Delta_{\varepsilon}=\bigcup_{i } \Delta_{\varepsilon} (H_i) \,,
\end{align}
 where index $i$ runs over all the singular hyperplanes in the theory. For the higher rank gauge theories, the relations \eqref{eq:pro1} and \eqref{eq:pro2} satisfy the same  properties in order to perform the integration over  $\bar{u}$ and $D$  in \cite{Benini:2013xpa}.  
By  repeating  the argument  in \cite{Benini:2013xpa}, we obtain   the expression \eqref{eq:indexformula} after some tedious computations. For simplicity  we shall consider $U(1)$ gauge theories and evaluate explicitly the integrals of $\bar{u}$ and $D$. In this case, \eqref{eq:zeromodeint} with $\mathrm{rk}(G)=1$ is written as 
\begin{align}
Z_{I \times T^2}&= \lim_{e \to 0 \atop \varepsilon \to 0} \int_{\mathfrak{M} \backslash \Delta_{\varepsilon} } \frac{d u d \bar{u}}{2 \pi} \int_{  \mathbb{R}} 
 \frac{d D}{2 \pi {\rm i}} 
    h (u, D)  g (u,  D) \exp \left[-   \frac{1}{2e^2}D^2 -{\rm i} \zeta D \right]  \,.
\label{eq:rankoneZ}
\end{align}
For the rank one gauge theories, we omitted the labels for the Cartan part of the gauge group as $D=D^1$, $u=u^1$, $h=h^{11}$,  and so on.

\begin{pdffig}
\begin{figure}[thb]
\centering
\subfigure[]{\label{fig:Gamma1}
\includegraphics[height=4cm]{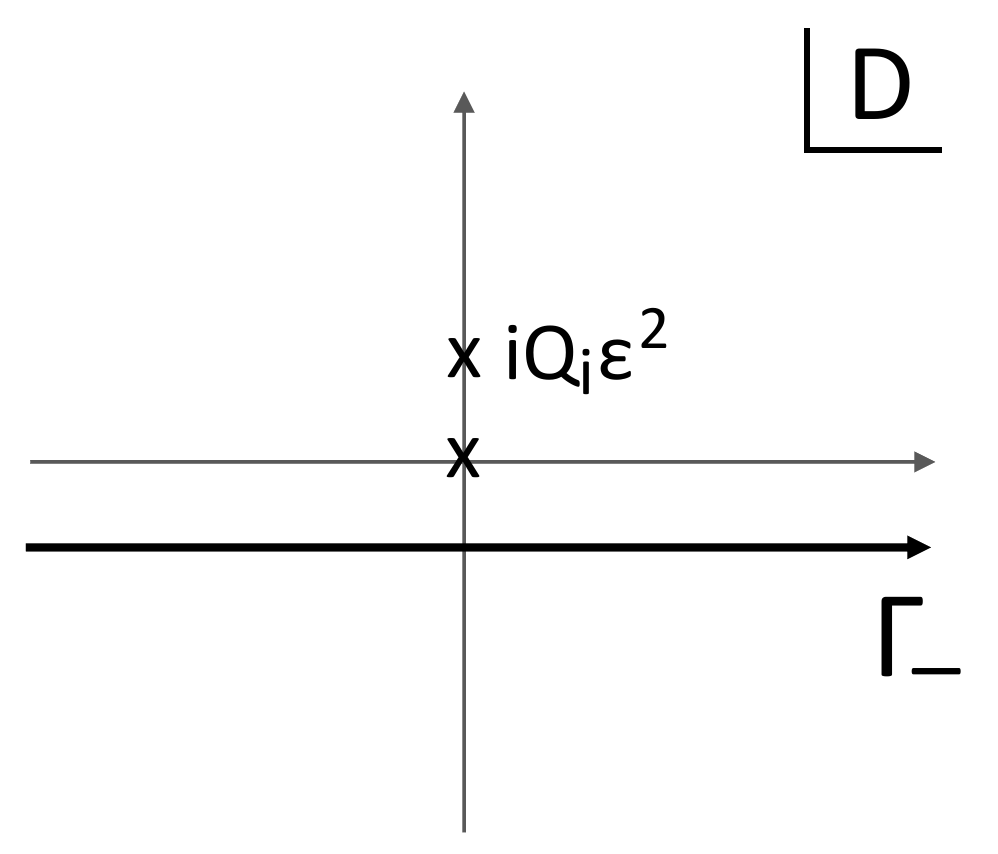}
}
\subfigure[]{\label{fig:Gamma2}
\includegraphics[height=4cm]{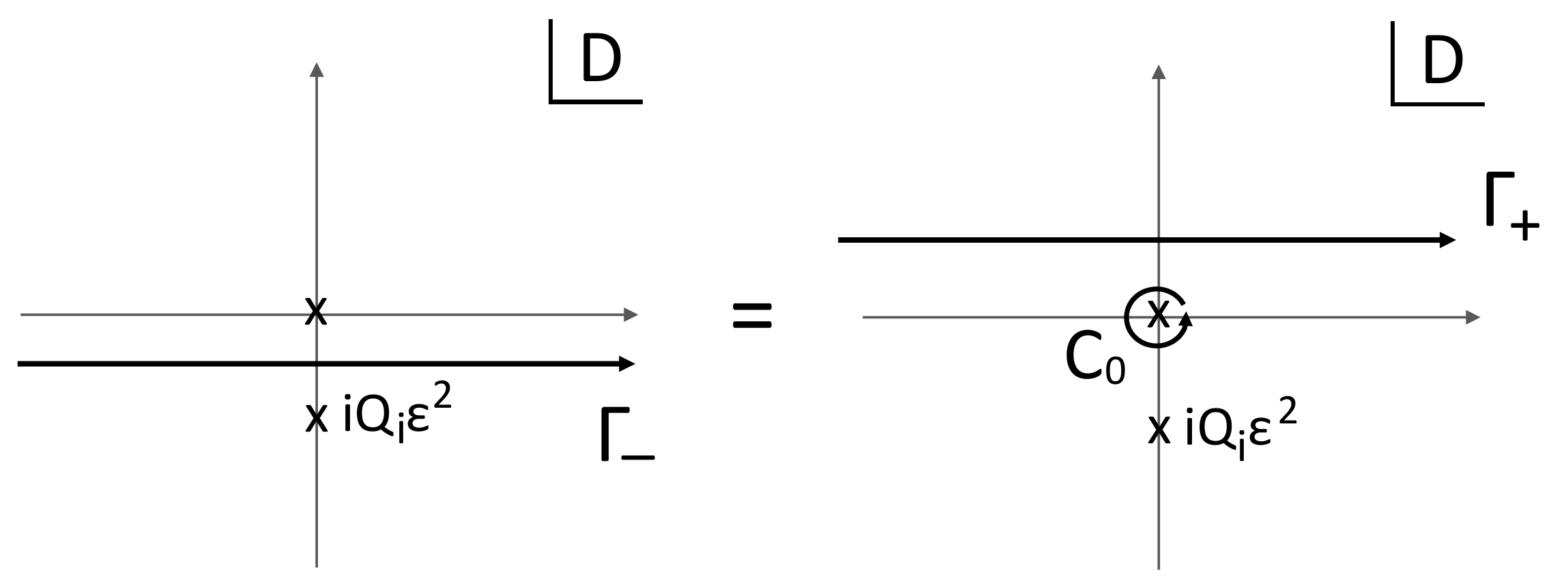}}
\caption{(a): The integration contour $\Gamma_-$ is specified  by the solid black arrow. Any pole arising from  $\Delta_{\varepsilon}^{+}$ does not 
hit $\Gamma_-$ in the limit $ \varepsilon \to 0$. From the disjointness of $\Delta^+_{\varepsilon}$ and 
$\Delta^-_{\varepsilon}$, any pole approaches to $\Gamma_-$ from the negative imaginary axis in the limit $\varepsilon \to 0$.
 (b): The decomposition of $\Gamma_-$ to $C_0+\Gamma_+$. Any pole arising from  $\Delta_{\varepsilon}^{-}$ does not 
hit $\Gamma_+$ in the limit $ \varepsilon \to 0$. }
\label{fig:Dcontour}
\end{figure}
\end{pdffig}

When $u$ locates on a center of  the tube $|n -m \tau+ Q_i u+Q^F_i(z)| = \varepsilon $, the zero-mode of the auxiliary field $D$ has a pole at $D= {\rm i}  Q_i \varepsilon^2 $. 
The contour of $D$ 
 can be deformed away from the origin of the imaginary axis if the contour does not hit the  pole specified as
\begin{align}
 |n -m \tau+ Q_i u+Q^F_i(z)|^2   + {\rm i} Q_i D=0 \,.
\label{eq:poleD}
\end{align} 
We define such deformed integration contours $\Gamma_{\pm}$ by $\Gamma_{\pm}:= \mathbb{R}\pm {\rm i} \delta  $ with $0< \delta < \varepsilon$. 
First we take $\Gamma_{-}$.  Eq.\eqref{eq:pro1}  for the rank one gauge theory is given by
\begin{align}
\frac{\partial g (u,D)}{\partial \bar{u}}=-{\rm i }D   h (u, D)   g(u,D) \,.
\end{align}
From this relation we can rewrite the integral \eqref{eq:rankoneZ} as
\begin{align}
Z_{I \times T^2}&=-\lim_{e \to 0 \atop \varepsilon \to 0}   \int_{ \Gamma_-} 
 \frac{d D}{2 \pi {\rm i} D} \int_{\mathfrak{M} \backslash \Delta_{\varepsilon} } \frac{d u d \bar{u}}{ 2 \pi {\rm i}} \frac{\partial g (u,D)}{\partial \bar{u}}
    \exp \left[-   \frac{1}{2e^2}D^2 -{\rm i} \zeta D \right]  \nonumber \\
&= -\lim_{e \to 0 \atop \varepsilon \to 0}   \int_{ \Gamma^-} 
 \frac{d D}{2 \pi  {\rm i} D} 
    \exp \left[-   \frac{1}{2e^2}D^2 -{\rm i} \zeta D \right]  \oint_{\partial(\mathfrak{M} \backslash \Delta_{\varepsilon}) =\partial \Delta_{\varepsilon} } \frac{d u}{2 \pi {\rm i} }   g (u,D) \,.
\label{eq:rankonD}
\end{align}
Here we assume that the $ \Delta_{\varepsilon}$ in the rank one gauge theory is decomposed to the disjoint union: 
\begin{align}
\Delta_{\varepsilon}=\Delta_{\varepsilon}^{+} \sqcup \Delta_{\varepsilon}^{-} \,,
\label{eq:projective} 
\end{align}
where   $\Delta_{\varepsilon}^{+}$ is the union of  the $\varepsilon$-neighborhoods around the singular hyperplanes (=points) $Q_i u+Q_i^F(z) =0$  with $ Q_i>0$ and 
$\Delta_{\varepsilon}^{-}$ is the union of the $\varepsilon$-neighborhoods around the singular hyperplanes  $Q_i u+Q_i^F(z) =0$  with $ Q_i<0$. 
If the condition \eqref{eq:projective} is satisfied, the singular hyperplane arrangements are  called projective.
For the higher rank gauge theories,  the singular hyperplane arrangements mean that weights $Q=(Q_1,\cdots,Q_{\mathrm{rk}(G)})$ for gauge representations at each singular point $u_*$ are contained in 
a half space of $\mathbb{R}^{\mathrm{rk}(G)}$. In this article we assume  ``projective'' condition is satisfied.
 
Since the pole $D= \mathrm{i} Q_i \varepsilon^2$ with $Q_i > 0$ does not hit the integration contour $\Gamma_-$  in the limit $\varepsilon \to 0$ as depicted by (a) in Figure \ref{fig:Dcontour} and  the integrand is  bounded, the contribution from a boundary $\partial \Delta^+_{\varepsilon}$ in \eqref{eq:rankonD} vanishes: 
\begin{align}
 \lim_{e \to 0 \atop \varepsilon \to 0}   \int_{\Gamma_- } 
 \frac{d D}{{\rm i} D} 
    \exp \left[-   \frac{1}{2e^2}D^2 -{\rm i} \zeta D \right]  \oint_{ \partial \Delta^+_{\varepsilon} } \frac{d u}{2 \pi {\rm i}}  \,  g (u,D)=0\,.
\label{eq:vanishes}
\end{align}
Next we will see the contribution from $\Delta^-_{\varepsilon}$  is written as  the contour integral  on $\partial \Delta^-_{\varepsilon}$.  
As depicted  by  Figure \ref {fig:Gamma2}, we decompose the integration contour $\Gamma_-$ as
\begin{align}
\Gamma_-=C_0+\Gamma_+\,.
\end{align}
Here $C_0$ is a small circle around the origin of the $D$-plane and  the index is expressed as
\begin{align}
Z_{I \times T^2}&= - \lim_{e \to 0 \atop \varepsilon \to 0}   \int_{ \Gamma^- } 
 \frac{d D}{2 \pi  {\rm i} D} 
    \exp \left[-   \frac{1}{2e^2}D^2 -{\rm i} \zeta D \right]  \oint_{ \partial \Delta^-_{\varepsilon} } \frac{d u}{2 \pi {\rm i}} g (u,D)
\nonumber \\
&=- \lim_{e \to 0 \atop \varepsilon \to 0}   \oint_{C_0} 
 \frac{d D}{2 \pi  {\rm i} D} 
    \exp \left[-   \frac{1}{2e^2}D^2 -{\rm i} \zeta D \right]  \oint_{ \partial \Delta^-_{\varepsilon} } \frac{d u}{2 \pi {\rm i}} g (u,D)
\nonumber \\
& \qquad - \lim_{e \to 0 \atop \varepsilon \to 0}   \int_{\Gamma_+} 
 \frac{d D}{2 \pi {\rm i} D} 
    \exp \left[-   \frac{1}{2e^2}D^2 -{\rm i} \zeta D \right]  \oint_{ \partial \Delta^-_{\varepsilon} } \frac{d u}{2 \pi {\rm i}}   g (u,D) \, .
\label{eq:contour2}
\end{align}
The last term in \eqref{eq:contour2} vanishes due to a similar reason of \eqref{eq:vanishes}.
The residue at $C_0$  gives an expression:
\begin{align}
Z_{I \times T^2}= - \int_{ \partial \Delta^-_{\varepsilon} } \frac{d u}{2 \pi {\rm i}}   g (u,D=0).
\label{eq:rank1JK1}
\end{align}
If we choose $\Gamma_+$, the same argument gives 
\begin{align}
Z_{I \times T^2}=  \int_{ \partial \Delta^+_{\varepsilon} } \frac{d u}{2 \pi {\rm i}}  g (u,D=0).
\label{eq:rank1JK2}
\end{align}
\eqref{eq:rank1JK1} and \eqref{eq:rank1JK2} are  the  localization formula \eqref{eq:indexformula} for the   $G=U(1)$ gauge theories. The sign of $\delta$ corresponds to $\eta$ in the JK residues.

\section{2d $\mathcal{N}=(2,2)$ and $\mathcal{N}=(0,4)$ theories from  3d $\mathcal{N}=4$ theories }
\label{sec:section4}
We study the relations between $I \times T^2$ indices for 3d $\mathcal{N}=4$ theories and 2d elliptic genera. A similar construction of 2d 
$\mathcal{N}=(2,2)$ and $\mathcal{N}=(0,4)$ theories  based on 4d $\mathcal{N}=2$ theories  on $T^2 \times S^2$ was studied in \cite{Honda:2015yha}.

The 3d $\mathcal{N}=4$ $G$ vector multiplet is decomposed to 3d $\mathcal{N}=2$ $G$ vector multiplet and a chiral multiplet $\varphi$ in the adjoint representation.
The charge assignments for 3d $\mathcal{N}=2$ chiral multiplets $q, \tilde{q}$ in the 3d $\mathcal{N}=4$ multiplet 
are depicted  in Table \ref{tabel:glsm1}.
\begin{table}[h]
\begin{center}
\begin{tabular}{c | c c c  }
			&	$G$						&	$U(1)_y$		\\
			\hline
$\varphi$		&	\text{adj}												&	$-r_1-r_2$	\\		
$q$			&	${\bf R}$									&	$r_1$					\\
$\widetilde{q}$	&	$\overline{\bf R}$									&	$r_2$			
\end{tabular} 
\caption{The charge assignments of 3d $\mathcal{N}=2$ multiplets in the 3d $\mathcal{N}=4$ multiplets. $\varphi$ denotes the chiral multiplet in the $\mathcal{N}=4$ vector multiplet. 
$q$ and $\tilde{q}$ denote the chiral multiplets in the 3d $\mathcal{N}=4$ hypermultiplet. $U(1)_y$ is a flavor symmetry. 
}
\label{tabel:glsm1}
\end{center}
\end{table}

 We impose the Dirichlet boundary condition for the adjoint chiral multiplets $\varphi$ in the $\mathcal{N}=4$ vector multiplet and
impose the Dirichlet boundary conditions for both $q$ and $ \tilde{q}$, or  the Neumann boundary conditions for both $q$ and $\tilde{q}$.
We take flavor charges as $r_1=r_2=\frac{1}{2}$.
Then the one-loop determinants of the 3d $\mathcal{N}=4$ multiplets are given by
\begin{align}
 Z^{\mathcal{N}=4}_{\text{vec}, {\sf D}} (x ,y; q) &= Z^{\mathcal{N}=2}_{\text{vec}} Z^{\mathcal{N}=2}_{\text{chi}, {\sf D}} \nonumber \\
& =\left( 2 \pi {\rm i}  \eta (q)  \theta_1 (q, y^{-1}) \right)^{\text{rk}(G)}  \prod_{\alpha \in \text{rt}(\mathfrak{g})}  \frac{\theta_1(x^{\alpha},q) \theta_1(x^{\alpha} y^{-1},q)}{\eta(q)^2}
\label{eq:04vec}\,,
 \\
 Z^{\mathcal{N}=4}_{\text{hyp}, {({\sf D},{\sf D})}} (x ,y, q) &=Z^{\mathcal{N}=2}_{\text{chi}, {\sf D}} Z^{\mathcal{N}=2}_{\text{chi}, {\sf D}} 
= 
\prod_{Q \in \text{wt} ({\bf R})} 
 \frac{\theta_1(x^{Q}  y^{\frac{1}{2}}  ,q) \theta_1(x^{-Q} y^{\frac{1}{2}}  ,q)}{-\eta (q)^2}
 \label{eq:04Fermi} \,,
 \\
 Z^{\mathcal{N}=4}_{\text{hyp}, {({\sf N},{\sf N})}} (x ,y, q) &=Z^{\mathcal{N}=2}_{\text{chi}, {\sf N}}
Z^{\mathcal{N}=2}_{\text{chi}, {\sf N}} 
= \prod_{Q \in \text{wt} (\mathbf{R})} 
 \frac{-\eta (q)^2}{\theta_1(x^{Q}  y^{\frac{1}{2}}  ,q) \theta_1(x^{-Q} y^{\frac{1}{2}}  ,q) }
\label{eq:04hyp} \,.
\end{align}
Here ${\sf D}$ and ${\sf N}$ denote the boundary conditions for a 3d $\mathcal{N}=2$ chiral multiplet.
\eqref{eq:04vec}, \eqref{eq:04Fermi} and \eqref{eq:04hyp} agree with the one-loop determinants of the  vector, the long fermi, and the hypermultiplet in 
  the 2d $\mathcal{N}=(0,4)$ elliptic genus, respectively.

Next  we choose the Neumann boundary condition for the adjoint chiral multiplet in the 3d $\mathcal{N}=4$ vector multiplet. 
  We choose the Neumann (resp. Dirichlet) boundary condition for a chiral multiplet in the representation ${\bf R}$ (resp. $\overline{\bf R}$) in the hypermultiplet. 
Then the 3d $\mathcal{N}=4$ vector multiplet preserves $\mathcal{N}=(2,2)$ supersymmetry at the boundaries.  
According to \eqref{eq:loop3dvec}-\eqref{eq:oneloop}, the one-loop determinant of the 3d $\mathcal{N}=4$ vector multiplet 
 $Z^{\mathcal{N}=4}_{\text{vec}}$ and the hypermultiplet $Z^{\mathcal{N}=4}_{\text{hyp}}$ in a representation  
$\mathbf{R} \oplus \overline{\mathbf{R}}$ are given by
\begin{align}
 Z^{\mathcal{N}=4}_{\text{vec}, {\sf N}} (x ; q) &
 =\left( \frac{2 \pi \eta (q)^3 }{\theta_1 ( y^{-1}, q)} \right)^{\text{rk}(G)} \prod_{\alpha \in \text{rt}(\mathfrak{g})} 
 \frac{\theta_1(x^{\alpha},q)}{\theta_1(x^{\alpha} y^{-1},q)}\, , \nonumber \\
 Z^{\mathcal{N}=4}_{\text{hyp}, ({\sf N, D})} (x ; z, q) &
=\prod_{Q  \in \text{wt}(\mathbf{R})} 
 \frac{\theta_1(x^{Q} y^{\frac{r}{2}-1} ,q)}{\theta_1(x^{Q}  y^{\frac{r}{2}}  ,q)} \, .
\label{eq:22genus}
\end{align}
Here we choose  flavor charges as $r_1=r$ and $r_2=1-r$.
\eqref{eq:22genus} agrees with the one-loop determinants of  the vector multiplet and  
the chiral multiplet in the representation ${\bf R}$  for the 2d $\mathcal{N}=(2,2)$ elliptic genus. 

\subsection{2d $\mathcal{N}=(2,2)$ $U(N)$ gauge theory from 3d $\mathcal{N}=4$ gauge theory}
As an example of 2d $\mathcal{N}=(2,2)$ theory, we take a  $G=U(N_c)$ gauge theory with ${\bf R}={\Box}^{\oplus N_f}$  in Table \ref{tabel:glsm1}, where $\Box$ denotes 
the fundamental representation of $U(N_c)$.  We assume $N_c \le N_f$.
The $I \times T^2$ index for the 3d $\mathcal{N}=4$ $U(N_c)$ gauge theory   with ${\bf R}={\Box}^{\oplus N_f}$  is given by
\begin{align}
&Z^{\mathcal{N}=4}_{I \times T^2}( y,  z,q; N_c, N_f )
= \frac{1}{N_c!} \sum_{u^*} \mathop{\text{JK-Res}}_{ u = {u}_*} ({Q}_*,{\eta})
 Z^{\mathcal{N}=4}_{\text{vec}, {\sf N}}  Z^{\mathcal{N}=4}_{\text{hyp}, ({\sf N, D})}  \wedge_{a=1}^N d u^a\nonumber \\
& \quad =\left( \frac{\eta (q)^3 }{\theta_1 ( y^{-1}, q)} \right)^{N_c}  \sum_{1 \le i_1 < \cdots < i_{N_c} \le N_f  }\oint_{x_a=z_{i_a}} \prod_{a=1}^{N_c} \frac{d x_a}{2 \pi i x_a}   \nonumber \\
&  \qquad \qquad \qquad \qquad \times \prod_{1 \le a \neq b \le N_c}\frac{\theta_1(x_a x_b^{-1},q)}{\theta_1(x_a x_b^{-1} y^{-1},q)} \cdot \prod_{a=1}^{N_c} \prod_{i=1}^{N_f}
 \frac{\theta_1(x_{a} y^{-1} z^{-1}_i ,q)}{\theta_1(x_{a}  z^{-1}_i  ,q)} \, .
 \label{eq:22genusUN}
\end{align}
\eqref{eq:22genusUN} is same as the elliptic genus for the 2d $\mathcal{N}=(2,2)$ $U(N_c)$ gauge theory with 
$N_f$ chiral multiplets in the fundamental representation of $U(N_c)$ in \cite{Benini:2013xpa}.
Here we have taken  $\eta=(1,1\cdots,1)  \in \mathbb{R}^{N_c} $. The JK residue  is evaluated as  
\begin{align}
Z^{\mathcal{N}=4}_{I \times T^2}( y,  z,q; N_c, N_f ) &=\sum_{\mathcal{I}\subset \{1,2,\cdots, N_f \} } \prod_{a \in \mathcal{I} } \prod_{b \in \{1,\cdots, N_f \} \backslash \mathcal{I} }   \frac{\theta_1(z_a z^{-1}_b y ,q)}{\theta_1(z_a z^{-1}_b ,q)} \,,
\end{align}
where $\mathcal{I}:=\{i_1,\cdots, i_{N_c} \}$ with $1 \le i_1 < i_2< \cdots <i_{N_c} \le N_f$. The sum $\sum_{\mathcal{I}\subset \{1,2,\cdots, N_f \} }$ runs over all the possible configurations of $\mathcal{I}$ in $\{1,\cdots, N_f \}$.
Note that $Z^{\mathcal{N}=4}_{I \times T^2}(  N_c, N_f )$ 
 satisfies the following relation:
\begin{align}
Z^{\mathcal{N}=4}_{I \times T^2}( y,  z,q; N_c, N_f )  &=
Z^{\mathcal{N}=4}_{I \times T^2}( y,  z,q; N_f-N_c, N_f ). 
\end{align}
A pair of $U(N_c)$ and $U(N_f-N_c)$ gauge theories is  known as   a  Seiberg-like duality in two dimensions \cite{Hanany:1997vm}, where the Higgs branch is  the Grassmann manifold $\mathrm{Gr}(N_c,N_f) \simeq \mathrm{Gr}(N_f-N_c.N_f)$ in positive FI-parameter  regions. The flavor symmetry $U(1)_y$ is broken to $\mathbb{Z}_{N_f}$ due to the  anomaly.

\subsection{Mirror of 3d $\mathcal{N}=8$ Super Yang--Mills and M-strings}

\begin{table}[htb]
\begin{center}
\begin{tabular}{c | c  c c c | c }
			&	$U(N)_{\text{gauge}}$				 & $U(1)_{\epsilon_1}$ 		&	$U(1)_{\epsilon_2}$	&	$U(1)_{y}$ & b.c.	\\
			 \hline
$\varphi$			&	${\bf adj}$					 &	$1$			&	$1$			&	$0$	 & ${\sf D}$	\\
$B_1$			&	${\bf adj}$					 &	$1$			&	$0$			&	$0$	 & ${\sf N}$	\\
$B_2$	&	${\bf adj}$			 &	$0$						&	$1$	&	$0$	 &  ${\sf N}$	\\
$I$			&	${\Box}$					 &	$\frac{1}{2}$			&	$\frac{1}{2}$			&	$0$	 & ${\sf N}$	\\
$J$	&	$\overline{\Box}$			 &	$\frac{1}{2}$						&	$\frac{1}{2}$	&	$0$	 &  ${\sf N}$	\\
\hline
$\psi^{\prime}_{ - \, L}$ & ${\Box}$&    $0$ & $0$& $1$ & - \\
$\psi^{\prime}_{ - \, R}$ & $\overline{\Box}$ & $0$ & $0$& $1$ & - \\
\end{tabular} 
\caption{The charge assignments and the boundary conditions. $\varphi$ denotes the adjoint chiral multiplet in the 3d $\mathcal{N}=4$ vector multiplet. $(B_1,B_2)$ is an adjoint 
hypermultiplet and $(I,J)$ is a fundamental hypermultiplet. ${\bf adj}$ is the adjoint representation. $\Box$ (resp. $\overline{\Box}$) denotes the fundamental (resp. anti-fundamental) representation. The b.c. represents the boundary condition. The flavor symmetry group $U(1)_y$ exists for $L=0$. }
\label{tabel:ADHM}
\end{center}
\end{table} 
As an example of 3d $\mathcal{N}=4$ theory on $I \times T^2$  leading to a 2d $\mathcal{N}=(0,4)$ elliptic genus, we consider 
 the 3d $\mathcal{N}=4$ $U(N)$ gauge theory with an adjoint hypermultiplet $(B_1,B_2)$ and a fundamental hypermultiplet $(I,J)$. The moduli space of  Higgs branch vacua is the ADHM 
moduli space of the $N$-instantons in the $U(1)$ gauge theory.   
This theory is known as the mirror dual of the 3d $\mathcal{N}=8$ super Yang-Mills theory,
which flows to the same IR fixed point of the $U(N)_{1} \times U(N)_{-1}$ ABJM model  \cite{Aharony:2008ug} describing the world volume theory on $N$-stacks of M2-branes on $\mathbb{C}^4$.

We impose the boundary conditions  specified in Table \ref{tabel:ADHM}.
They preserve the supersymmetry of the  superpotential term:
\begin{align}
W =\mathrm{tr} \varphi ([B_1,B_2]+IJ).
\end{align}
Under the boundary condition in Table \ref{tabel:ADHM},  we find that the one-loop determinant of 
each 3d $\mathcal{N}=4$ multiplet agrees with that of  the 2d $\mathcal{N}=(0,4)$ multiplet.
Since 3d multiplets induce the gauge anomalies,
 we have to introduce fermi multiplets $\psi^{\prime}_{- \, L}$ at $x^1=-\pi L$ and $\psi^{\prime}_{- \, R}$ at $x^1=\pi L$ in Table \ref{tabel:ADHM} to cancel the  gauge anomaly. 
In the  limit  $L \to 0$, $\psi^{\prime}_{-\,L}$ and $\psi^{\prime}_{-\,R}$ live on the same spacetime and form a long fermi multiplet.  
An extra  $U(1)_y$ flavor symmetry  appears  in the limit $L=0$. The charge assignments for the $U(1)_y$ symmetry are depicted in Table \ref{tabel:ADHM}.

We shall compute  the $I \times T^2$ index
\begin{align}
Z^{\text{ADHM}}_{I \times T^2} &= \frac{\eta(q)^{N}}{N!}
\oint \prod_{a=1}^N {d u^a} 
\frac{ \prod_{1 \le a \neq b \le N}   \theta_1(x_a x^{-1}_b ,q)  \prod_{a,  b =1}^N   \theta_1(x_a x^{-1}_b q_1 q_2,q)} 
{\prod_{a, b=1}^{N} \theta_1(x_a x^{-1}_b q_1 ,q)\theta_1(x_a x^{-1}_b q_2,q)}
\nonumber \\ 
&\qquad \times \prod_{a=1}^{N} 
 \frac{\theta_1(x_a    y ,q)\theta_1(x_a^{-1}    y ,q)}{\theta_1(x_a    (q_1 q_2)^{\frac{1}{2}} ,q)\theta_1(x^{-1}_a    (q_1q_2)^{\frac{1}{2}} ,q)} \,,
\label{eq:parM}
\end{align}
where $x_a=e^{2 \pi {\rm i} u^a}$. The fugacities $q_i=e^{2 \pi {\rm i} \epsilon_i}$ with $i=1,2$  correspond to the $\Omega$-background parameters. We included a formal fugacity $y$ for $U(1)_y$ in \eqref{eq:parM} to compare with 
the M-string partition function.  
The JK residue computations are   same as  those of Nekrasov's  $N$-instanton partition functions in \cite{Hwang:2014uwa}. Then we obtain the result:
\begin{align}
Z^{\text{ADHM}}_{I \times T^2} &= 
\sum_{Y: |Y|=N} \prod_{(i,j) \in Y} 
 \frac{\theta_1( q_1^{i-\frac{1}{2}} q_2^{j-\frac{1}{2}} y ,q)\theta_1 (q_1^{-i+\frac{1}{2}} q_2^{-j+\frac{1}{2}}   y ,q)}{\theta_1(   q_1^{-\lambda^T_j+i} q_2^{\lambda_i-j+1} ,q)\theta_1( q_1^{\lambda^T_j-i} q_2^{-\lambda_i-j} ,q)} .
\label{eq:Mstring}
\end{align}
Here the sum is taken over the Young diagrams $Y=(\lambda_1,\lambda_2, \cdots)$ with $\lambda_1 \ge \lambda_2 \ge  \cdots  \ge 0$ and  the  number of boxes of $Y$ is $|Y|=N$.
$Y^T=(\lambda^T_1,\lambda^T_2 ,\cdots)$ is the  transpose of $Y$.  \eqref{eq:parM} and  \eqref{eq:Mstring} reproduce
 the elliptic genus of M-strings  suspended between 2 M5-branes on the
single center Taub-NUT space \cite{Haghighat:2013gba}, except for the fugacity $y$. In our case $y=1$ for $L >0$.

\section{Three dimensional dualities on $I \times M_2$}
\label{sec:section5}

\subsection{3d $\mathcal{N}=2$   SQED and XYZ model}
\begin{table}[h]
\begin{center}
\begin{tabular}{c | c c c | c }
			&	$U(1)_{\text{gauge}}$						&	$U(1)_y$	&	$U(1)_R$ & b.c.	\\
			 \hline
$\phi$			&	$1$									&	$1$			&	$0$	 & ${\sf N}$	\\
$\widetilde{\phi}$	&	$-1$									&	$1$	&	$0$	 &  ${\sf N}$	\\
\hline
$\psi^{\prime}_{ - R}, \psi^{\prime}_{- \, L}$ & $1$& $0$& $0$ & - \\
\end{tabular} 
\caption{ The charge assignments and boundary conditions for the SQED and the fermi multiplets. $\phi$ and $\tilde{\phi}$ denote scalars in the chiral multiplets. 
$\psi^{\prime}_{- R}$ and $\psi^{\prime}_{- L}$ denote the boundary fermions at  $x^1= \pi L$ and $x^1= -\pi L$, respectively. 
}
\label{tabel:SQED}
\end{center}
\end{table}

\begin{table}[h]
\begin{center}
\begin{tabular}{c |  c c | c }
									&	$U(1)_y$	&	$U(1)_R$& b.c. \\
			 \hline
$\phi_X$													&	$-1$			&	$1$	 & ${\sf D}$ \\
$\phi_Y$										&	$-1$	&	$1$	 & ${\sf D}$	\\
$\phi_Z$ & $2$& $0$ & ${\sf N}$ \\
\end{tabular} 
\caption{ The charge assignments and boundary conditions for the  XYZ model. $\phi_X, \phi_Y,\phi_ Z$ express the 
scalars in the three chiral multiplets $X, Y$ and $Z$.
}
\label{tabel:XYZ}
\end{center}
\end{table}

We consider a simple 3d $\mathcal{N}=2$ mirror symmetry; the 3d $\mathcal{N}=2$ one-flavor SQED and the XYZ model \cite{deBoer:1997kr, Aharony:1997bx}. 
The charge assignments of the SQED and the XYZ model are listed in Table \ref{tabel:XYZ}. 
We put these theories on the interval and study a duality with the boundaries based on  indices and anomaly matching. The Neumann boundary conditions  for two chiral multiplets $(q, \tilde{q})$ in the SQED are specified by $({\sf N,N})$.
 In  the XYZ model, the Dirichlet boundary conditions for  chiral multiplets $X,Y$ and the Neumann boundary condition for a chiral multiplet $Z$  are specified by $({\sf D,D,N})$.
To cancel the gauge  anomaly in the SQED,
 we add one fermi multiplet which couples to 3d $U(1)$ gauge field
 on the left and the right boundary, respectively.
At these boundaries $x^1=\pm \pi L$, the  anomaly polynomials are evaluated as
\begin{align} 
{\bf I}_{\text{SQED}}&=-\frac{1}{2} ({\bf f} +{\bf y}-{\bf r})^2-\frac{1}{2} (-{\bf f} +{\bf y}-{\bf r})^2+\frac{1}{2} {\bf r}^2\,,
\nonumber  \\
{\bf I}_{\text{XYZ}} &=\frac{1}{2} {\bf y}^2+\frac{1}{2}{\bf y}^2-\frac{1}{2} (2{\bf y}-{\bf r})^2 \, ,\nonumber \\
{\bf I}_{\text{fermi}}&={\bf f}^2 \,,
\end{align}
where ${\bf I}_{\text{SQED}}$, ${\bf I}_{\text{XYZ}} $ and ${\bf I}_{\text{fermi}}$ are the anomaly polynomials for the SQED, the XYZ model and the fermi multiplet, respectively.
Then  we find that the anomaly polynomials match: 
\begin{align} 
{\bf I}_{\text{SQED}}+{\bf I}_{\text{Fermi}}={\bf I}_{\text{XYZ}} .
\end{align}
The $I \times T^2$ index for the SQED with two boundary fermi multiplets is written as 
\begin{align} 
Z^{\text{SQED}}_{I \times T^2, ({\sf N,N})} &= -\eta (q)^2    \oint_{x=y^{-1}} \frac{d x}{2 \pi {\rm i} x} \frac{ \theta_1 (x , q)^2}{\theta_1 (x y , q) \theta_1 (x^{-1} y, q)}  =\left({\rm i}\frac{ \theta_1 (y^{-1} , q)^2  }{\eta (q)} \right)^2 \frac{{\rm i}\eta(q)} {\theta_1 ( y^2, q)} \nonumber \\
&=Z^{\text{XYZ}}_{I \times T^2, ({\sf D,D,N}) } \,,
\end{align}
where we have chosen   $\eta >0$ in the  JK residue evaluation. When $\eta <0$  the residue is evaluated at $x=y$, which gives the same result.
Therefore  we find that the $I \times T^2$ index of the SQED agrees  with  the $I \times T^2$ index of the XYZ model.

\subsection{Aharony duality of  $U(N)$ gauge theory with $N$-flavors}
\begin{table}[htb]
\begin{center}
\begin{tabular}{c | c c c c c | c }
			&	$U(N)_{\text{gauge}}$				&$SU(N)_y$ & $SU(N)_{\tilde{y}}$ 		&	$U(1)_a$	&	$U(1)_R$ & b. c	\\
			 \hline
$\phi$			&	${\Box}$					& $\overline{\Box}$ &	${\bf 1}$			&	$1$			&	$0$	 & ${\sf N}$	\\
$\tilde{\phi}$	&	$\overline{\Box}$			& ${\bf 1}$ &	${\Box}$						&	$1$	&	$0$	 &  ${\sf N}$	\\
\hline
$\psi^{\prime}_{ - R}, \psi^{\prime}_{- \, L}$ & ${\bf det}$&  ${\bm 1}$ & ${\bm 1}$ & $0$& $0$ & - \\
\end{tabular} 
\caption{ The charge assignments and boundary conditions for  the chiral multiplets in the $U(N)$ gauge theory and  the boundary fermi multiplets. $\phi$ and $\tilde{\phi}$ denote scalars in the chiral multiplets. 
$SU(N)_y \times SU(N)_{\tilde{y}} \times U(1)_a$ is the set of  flavor symmetry groups.
$\psi^{\prime}_{- R}$ and $\psi^{\prime}_{- L}$ denote the fermions at  $x^1= \pi L$ and $x^1= -\pi L$, respectively. ${\bf det}$ is the determinant representation. ${\bf 1}$
 is the trivial representation.}
\label{tabel:Ahar}
\end{center}
\end{table}

\begin{table}[htb]
\begin{center}
\begin{tabular}{c |  c c c c | c }
							&$SU(N)_y$ & $SU(N)_{\tilde{y}}$ 		&	$U(1)_a$	&	$U(1)_R$ & b. c\\
			 \hline
$q$								& ${\bf 1}$ &	${\bm 1}$			&	$-N$			&	$1$	 & ${\sf D}$	\\
$\tilde{q}$			& ${\bf 1}$ &	${\bm 1}$						&	$-N$	&	$1$	 &  ${\sf D}$	\\
$M$ &   $\overline{\Box}$ & ${ \Box}$ & $2$& $0$ & ${\sf N}$  \\
\end{tabular} 
\caption{ The charge assignments and boundary conditions for scalars $q, \tilde{q}$ and $M$ in the three chiral multiplets . 
}
\label{tabel:dualAhar}
\end{center}
\end{table}
Mirror symmetry for one flavor $\mathcal{N}=2$ SQED is generalized to  an Aharony duality  \cite{Aharony:1997gp} for the $U(N)$ gauge theory with $N$ fundamental and anti-fundamental chiral multiplets. The dual theory  consists of  chiral multiplets $M$, $q, \tilde{q}$ with the superpotential $W=\det(M) \tilde{q}  q$. The charge assignments of two theories are listed in Table \ref{tabel:Ahar} and Table \ref{tabel:dualAhar}.
The anomaly polynomials ${\bf I}_{U(N)+N\text{-flavors}}$, ${\bf I}_{\det (M) \tilde{q}  q}$, ${\bf I}_{\text{fermi}}$ of the $U(N)$ gauge theory, the dual theory with $W=\det(M) \tilde{q}  q$, and the boundary fermions  
are given by
\begin{align} 
{\bf I}_{U(N)+N\text{-flavors}}&=N \mathrm{Tr} \, ({\bf f}^2)-(\mathrm{Tr} {\bf f})^2+ \frac{N^2}{2} {\bf r}^2 \nonumber \\
&\qquad -\frac{N}{2}  \mathrm{Tr} \, ({\bf f}^2)-\frac{N}{2}  \mathrm{Tr} \, ({\bf y})^2+ N (\mathrm{Tr} {\bf f})({\bf a}-{\bf r })-\frac{N^2}{2} ({\bf a}-{\bf r })^2
\nonumber  \\
&\qquad -\frac{N}{2}  \mathrm{Tr} \, ({\bf f}^2)-\frac{N}{2}  \mathrm{Tr} \, (\tilde{\bf y})^2+ N (\mathrm{Tr} {\bf f})({\bf a}-{\bf r })-\frac{N^2}{2} ({\bf a}-{\bf r })^2, 
\nonumber \\
{\bf I}_{\det (M) \tilde{q}  q} &=
-\frac{N}{2}  \mathrm{Tr} \, ({\bf y})^2-\frac{N}{2}  \mathrm{Tr} \, (\tilde{\bf y})^2-\frac{N^2}{2}(2 {\bf a}- {\bf r})^2 \nonumber \\
&\qquad \qquad +\frac{1}{2} (-N {\bf a} )^2+\frac{1}{2} (-N {\bf a})^2  \, ,\nonumber \\
{\bf I}_{\text{fermi}}&=(\mathrm{Tr} \, {\bf f})^2 \,.
\end{align}
The anomaly polynomials satisfy a matching condition:
\begin{align} 
{\bf I}_{U(N)+N\text{-flavors}}+{\bf I}_{\text{fermi}}
 &=
{\bf I}_{\det (M) \tilde{q}  q} \, .
\end{align}

The $I \times T^2$ index for the dual theory is given by 
\begin{align} 
Z^{W=\det (M) \tilde{q}  q}_{I \times T^2, ({\sf N, N, D}) }
&=
\left( {\rm i}\frac{  \theta_1 (a^{-N}  )}{\eta(q)} \right)^2  \prod_{i,j=1}^N  {\rm i} \frac{\eta(q)}{   \theta_1 (a^2 y_i^{-1}   \tilde{y}_j, q)}.
\end{align}
Here $y_i$'s and $\tilde{y}_i$'s $(i=1,\cdots, N)$ with $\prod_{i=1}^N y_i=1$ and  $\prod_{i=1}^N \tilde{y}_i=1$ are fugacties for the $SU(N)_y$ and the $SU(N)_{\tilde{y}}$, respectively.
On the other hand,  the index for the $U(N)$ gauge theory  is given by 
\begin{align} 
Z^{{U(N)+N\text{-flavors}}}_{I \times T^2, ({\sf N, N}) }&=\frac{1}{N!} \left( \frac{\eta (q) }{{\rm i}} \right)^{2N}\sum_{j=1}^N \sum_{k_j=1}^N \oint_{x_i=a^{-1} y_{k_i}}  \prod_{i=1}^N \frac{d x_i}{2 \pi {\rm i} x_i}    
\prod_{1 \le i \neq j \le N} {\rm i} \frac{\theta_1 (x_ i x_j^{-1} , q)}{\eta(q)}
\nonumber \\
& \quad \quad \times 
  \left( {\rm i} \frac{ \theta_1 (\prod_{i=1}^N x_i, q )}{\eta(q)} \right)^2  \prod_{i,j=1}^N  \frac{({\rm i} \eta(q))^2}{\theta_1 (x_i a  y^{-1}_j, q)  \theta_1 (x_i^{-1} a  \tilde{y}_j, q)} \nonumber \\
&=\left( {\rm i}\frac{  \theta_1 (a^{-N}  )}{\eta(q)} \right)^2  \prod_{i,j=1}^N  {\rm i} \frac{\eta(q)}{   \theta_1 (a^2 y_i^{-1}   \tilde{y}_j, q)} \,,
\end{align}
 where we have chosen $\eta=(1,\cdots,1)$ in the JK residue formula. If we choose $\eta=(-1,\cdots,-1)$, we obtain the same result.
Thus we have agreement between the indices of two theories.

\subsection{Triality on the interval  from  3d Seiberg-like dualities}
\begin{pdffig}
\begin{figure}[thb]
\centering
\subfigure[]{\label{fig:trial1}
\includegraphics[height=3cm]{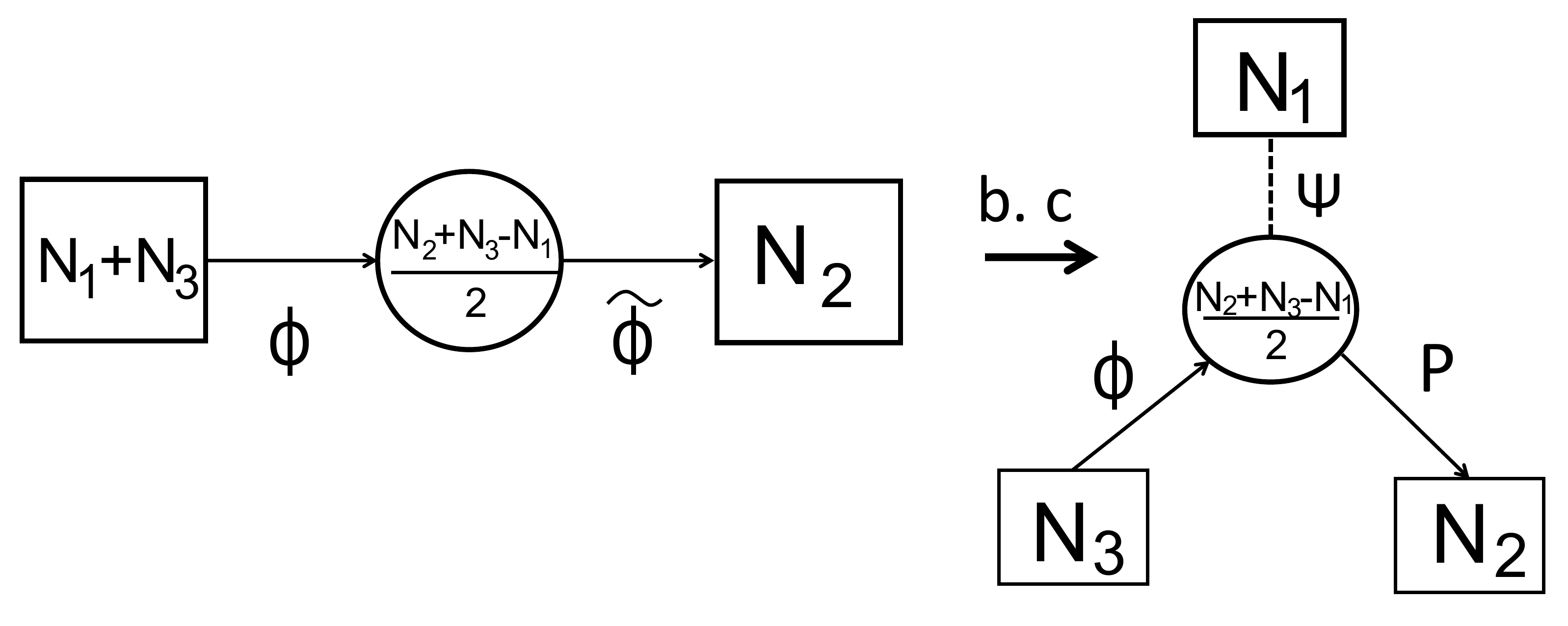}
}
\subfigure[]{\label{fig:trial2}
\includegraphics[height=3cm]{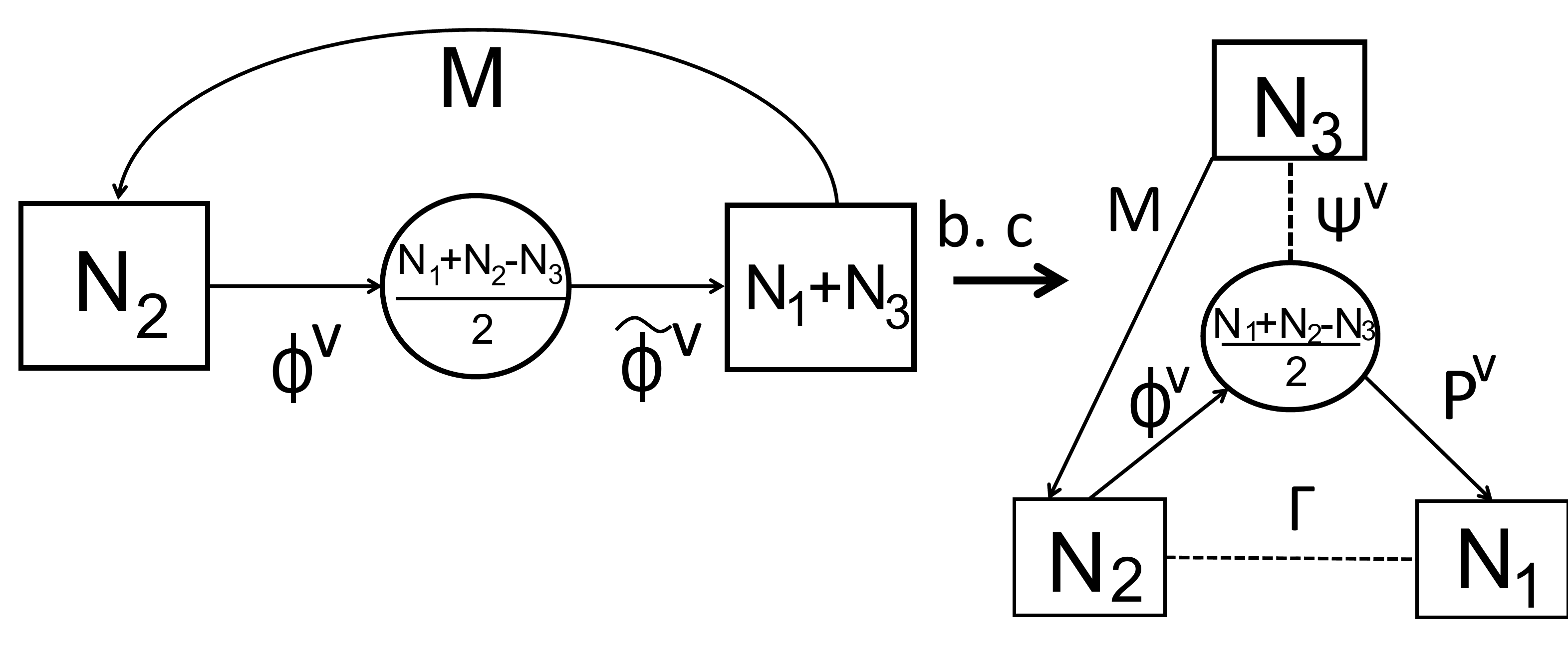}}
\caption{(a): The decomposition of 3d multiplets $\{\Phi_i \}_{i=1}^{N_1+N_3} \mapsto (\Phi, \Psi)$ and $\{ \tilde{\Phi} \}_{i=1}^{N_2} \mapsto P$ in the theory ${\bf A}$ by the boundary conditions. In the right figure of (a),
 the dashed line denotes the $N_1$ chiral multiplets with the Dirichlet boundary condition $\Psi$. The solid arrows denote the chiral multiplets with the Neumann boundary condition $\Phi, P$. 
(b): The decomposition of 3d multiplets $\{ \Phi^{\vee}_{i} \}_{i=1}^{N_2} \mapsto  \Phi^{\vee}$, $\{ \tilde{\Phi}^{\vee}_i \}_{i=1}^{N_1+N_3} \mapsto (P^{\vee}, \Psi^{\vee})$ and $M_{i k} \mapsto (M, \Gamma)$ in the theory ${\bf A}^{\vee}$.}
\label{fig:decompose}
\end{figure}
\end{pdffig}

\begin{table}[htb]
\begin{center}
\begin{tabular}{c | c  c c  c | c }
			&	$U( \frac{N_2+N_3-N_1}{2})_{\text{gauge}}$			& $SU(N_3)_y$	 & $SU(N_1)_{y^{\prime}}$ 		&	$S U(N_2)_{\tilde{y}}$	 & b.c	\\
			 \hline
$\Phi=\{ \Phi_i \}_{i=1}^{N_3}$			&	$\Box$					 &	$\overline{\Box}$	&	${\bf 1}$	&	${\bf 1}$				 & ${\sf N}$	\\
$P:=\{\tilde{\Phi}_i\}_{i=1}^{N_2}$			&	$\overline{\Box}$					 &	${\bf 1}$			&	${\bf 1}$	&	${\Box}$	&	 ${\sf N}$	\\
$\Psi:=\{ \Phi_i \}_{i=N_3+1}^{N_1+N_3} $	&	${\Box}$			 &	${\bf 1}$						&	$\overline{\Box}$	&{\bf 1} &  ${\sf D}$	\\
\hline
$\Omega_{i=1,2}$ &	${\bf det}$					 &	${\bf 1}$			& {\bf 1}&	${\bf 1}$			 & - \\
\end{tabular} 
\caption{The charge assignments and  the boundary conditions in the theory ${\bf A}$. The subscripts of the groups correspond to fugacities for these symmetries in the $I \times T^2$ index.
 $\Omega_{i=1,2}$'s are  fermi multiplets at the left and the right boundaries introduced to cancel the gauge anomaly.}
\label{table:trial1}
\end{center}
\end{table}

\begin{table}[htb]
\begin{center}
\begin{tabular}{c | c  c c c  | c }
			&	$U( \frac{N_1+N_2-N_3}{2})_{\text{gauge}}$			& $SU(N_3)_y$	 & $SU(N_1)_{y^{\prime}}$ 		&	$S U(N_2)_{\tilde{y}}$	 & b.c	\\
			 \hline
$\Phi^{\vee}=\{ \Phi^{\vee}_i \}_{i=1}^{N_2}$			&	$\Box$					 &	${\bf 1}$	&	${\bf 1}$	&	$\overline{\Box}$				 & ${\sf N}$	\\
$P^{\vee}:=\{ \tilde{\Phi}^{\vee}_{i} \}_{i=1}^{N_1}$			&	$\overline{\Box}$					 &		${\bf 1}$		&	${\Box}$	&	${\bf 1}$		 & ${\sf N}$	\\
$\Psi:=\{ {\Phi}^{\vee}_i \}_{i=N_1+1,\cdots,N_1+N_3} $	&	$\overline{\Box}$			 &					${\Box}$		&	${\bf 1}$	&{\bf 1}&	  ${\sf D}$	\\
$M:=\{M_{k i}\}_{(k, i)=(1,1)}^{(N_2,N_1)} $			&	${\bf 1}$					 &	$\overline{\Box }$			& ${\bf 1 }$&	${\Box}$				 & ${\sf N}$	\\
$\Gamma=\{M_{k i}\}_{(k,i)=(1,N_1+1)}^{(N_2,N_1+N_3)} $			&	${\bf 1}$					 &	${\bf 1}$			& $\overline{\Box }$&	${\Box}$				 & ${\sf D}$	\\
\hline
$\Omega^{\vee}_{i=1,2}$ &	${\bf det}$					 &	${\bf 1}$			& {\bf 1}&	${\bf 1}$				 & - \\
\end{tabular} 
\caption{The charge assignments and  the boundary conditions for the theory ${\bf A}^{\vee}$ with $N_c=\frac{N_2+N_3-N_1}{2}$, $N_f=N_1+N_3$ and $N_a=N_2$. $\Omega^{\vee}_{i=1,2}$'s are  fermi multiplets at the left and  the right boundaries introduced to cancel the gauge anomaly.  }
\label{table:trial2}
\end{center}
\end{table}

In this section, we start from two pairs of Seiberg-like dualities and construct three theories on the interval.  We will see the three theories  satisfy the 't Hooft anomaly matchings and the $I \times T^2$ indices agree one another. Our construction is analogous to  the relation  \cite{Tachikawa2013} between  the 2d $\mathcal{N}=(0,2)$ triality  \cite{Gadde:2013lxa} and  the twisted compactification on $S^2$ with fluxes of  
4d $\mathcal{N}=1$  Seiberg dualities \cite{Seiberg:1994pq}. For example, see   computations of $T^2 \times S^2$ indices in \cite{Honda:2015yha}.

Let us consider a 3d Seiberg-like dual pair  \cite{Benini:2011mf}:
\begin{align}
&G=U(N_c)+ \Phi_{i=1,\cdots N_f} \text{ and }  \tilde{\Phi}_{k=1,\cdots N_a}, \nonumber  \\
&G=U(N_a-N_c)+\Phi^{\vee}_{k=1,\cdots N_a},  \tilde{\Phi}^{\vee}_{i=1,\cdots N_f} \text{ and } M_{i=1,\cdots,N_f \,, k=1,\cdots,N_a}\,. 
\label{eq:Seiberglike}
\end{align}
Here $G$ denotes the gauge group and  we assume $N_f \le N_a$. 
In the $U(N_c)$ gauge theory,   $\Phi_i$ with  $i=1,\cdots,N_f$ (resp.  $\tilde{\Phi}_k$ with $k=1,\cdots,N_a$)   represent  chiral multiplets in the fundamental (resp. anti-fundamental) representation of $U(N_c)$.
In the dual $U(N_a-N_c)$ gauge theory ,   $\Phi^{\vee}_k$ with   $k=1,\cdots,N_a$ (resp.  $\tilde{\Phi}^{\vee}_i$ with   $i=1,\cdots,N_f$) correspond to  chiral multiplets in 
the fundamental (resp. anti-fundamental) representation of $U(N_a-N_c)$.
$M_{i  k}$ with  $k=1,\cdots,N_a$ and  $i=1,\cdots,N_f$ are mesons and the dual theory has a superpotential $W= \sum_{i,k}\tilde{\Phi}^{\vee}_i M_{i k} \Phi^{\vee}_k$.

Now we take $N_c=\frac{N_2+N_3-N_1}{2}$, $N_f=N_1+N_3$, and  $N_a=N_2$ and impose the boundary conditions depicted in Table \ref{table:trial1} and in Table \ref{table:trial2}. 
 We call  these two theories the ``theory  ${\bf A}$'' and the ``theory $ {\bf A}^{\vee}$'': 
\begin{align}
&\text{Theory } {\bf A}: G=U(\frac{N_2+N_3-N_1}{2})+ \Phi_{i=1,\cdots N_1+N_3} \text{ and }  \tilde{\Phi}_{k=1,\cdots N_2}\,,
 \label{eq:Seiberglike0}  \\
&\text{Theory } {\bf A}^{\vee}: G=U(\frac{N_1+N_2-N_3}{2})+\Phi^{\vee}_{k=1,\cdots N_2},  \tilde{\Phi}^{\vee}_{i=1,\cdots N_1+N_3} \text{ and } M_{i=1,\cdots,N_1+N_3 \,, k=1,\cdots,N_2}\,. 
\nonumber 
\end{align}
Under the boundary conditions,  the  quiver diagram of the theory ${\bf A}$ on the interval is depicted by the right quiver  in Figure \ref{fig:trial1}. 
To cancel  the gauge anomaly of the theory ${\bf A}$,  we introduce a fermi multiplet $\Omega_1$ at $x^1= - \pi L$ and another fermi multiplet $\Omega_2$ at $x^1=\pi L$ in the  determinant representation. 
 The boundary conditions  in  the  theory ${\bf A}^{\vee}$ on the interval  is depicted in Table \ref{table:trial2}.
The  quiver diagram of the theory ${\bf A}^{\vee}$ on the interval is  depicted by the right quiver  in Figure \ref{fig:trial2}. 
To cancel  gauge anomaly of the theory ${\bf A}^{\vee}$, 
 we introduce a fermi multiplet $\Omega^{\vee}_1$ at the left boundary and another fermi multiplet $\Omega^{\vee}_2$ at the right boundary in the  determinant representation.
The anomaly polynomials of the theories ${\bf A}$, ${\bf A}^{\vee}$ and the boundary fermi multiplets are given by
\begin{align}
{\bf I}_{\text{Theory} {\bf A}}&= \left(\frac{N_2+N_3-N_1}{2} \right)  \mathrm{Tr} ({\bf f}^2) - (\mathrm{Tr} {\bf f})^2 +\frac{1}{2} \left(\frac{N_2+N_3-N_1}{2}  \right)^2 {\bf r}^2  \quad (\text{vector}) \nonumber \\
&-\frac{1}{2} \left( N_3 \mathrm{Tr} ({\bf f}^2)  +\frac{N_2+N_3-N_1}{2} \left[ \mathrm{Tr} ( {\bf y})^2 +N_3  (-\left(\frac{N_3+N_1-N_2}{N_1+N_2+N_3} -1\right) {\bf r})^2 \right] \right)  \quad (\Phi) \nonumber \\
&-\frac{1}{2} \left(N_2 \mathrm{Tr} ({\bf f}^2)  
+\frac{N_2+N_3-N_1}{2} \left[ \mathrm{Tr} ( \tilde{\bf y})^2 +N_2(-\left(\frac{N_1+N_3-N_2}{N_1+N_2+N_3} -1\right) {\bf r})^2 \right] \right) \quad (P) \nonumber \\
&+\frac{1}{2} \left(N_1 \mathrm{Tr} ({\bf f}^2)  
+\frac{N_2+N_3-N_1}{2}\mathrm{Tr} ( {\bf y}^{\prime})^2 \right) \quad (\Psi), 
 \\
{\bf I}_{\text{Theory} {\bf A}^{\vee}}&=  \left(\frac{N_1+N_2-N_3}{2} \right)  \mathrm{Tr} ({\bf f}^2) - (\mathrm{Tr} {\bf f})^2 +\frac{1}{2} \left(\frac{N_1+N_2-N_3}{2}  \right)^2 {\bf r}^2  \quad (\text{vector}) \nonumber \\
&-\frac{1}{2} \left( N_2 \mathrm{Tr} ({\bf f}^2)  +\frac{N_1+N_2-N_3}{2} \left[ \mathrm{Tr} ( \tilde{\bf y})^2 +N_2  (\left(\frac{N_2+N_3-N_1}{N_1+N_2+N_3} -1\right) {\bf r})^2 \right] \right)  \quad (\Phi^{\vee}) \nonumber \\
&-\frac{1}{2} \left(N_1 \mathrm{Tr} ({\bf f}^2)  
+\frac{N_1+N_2-N_3}{2} \left[ \mathrm{Tr} ( {\bf y}^{\prime})^2 +N_1(\left(\frac{N_1+N_3-N_2}{N_1+N_2+N_3} -1\right) {\bf r})^2 \right] \right) \quad (P^{\vee}) \nonumber \\
&+\frac{1}{2} \left(N_3 \mathrm{Tr} ({\bf f}^2)  
+\frac{N_1+N_2-N_3}{2}\mathrm{Tr} ( {\bf y})^2 \right) \quad (\Psi^{\vee})  \nonumber \\
& -\frac{1}{2} \left( N_2 \mathrm{Tr} ({\bf y})^2+N_3 \mathrm{Tr} (\tilde{\bf y})^2 
+N_2 N_3 (\left(\frac{N_2+N_3-N_1}{N_1+N_2+N_3} \right) {\bf r})^2\right) \quad (M)
\nonumber \\
& +\frac{1}{2} \left( N_2 \mathrm{Tr} ({\bf y}^{\prime})^2+N_1 \mathrm{Tr} (\tilde{\bf y})^2 
+N_1 N_2 (\left(\frac{N_1+N_2-N_3}{N_1+N_2+N_3} \right) {\bf r})^2\right) \quad  (\Gamma)\,,
\nonumber \\
{\bf I}_{\Omega_i}&={\bf I}_{\Omega^{\vee}_i}=(\mathrm{Tr} {\bf f})^2\,.
 \end{align}
Here each line corresponds to the anomaly contribution from a multiplet specified by the $(\cdot)$.
Then the anomaly polynomials satisfy a relation:
\begin{align}
{\bf I}_{\text{Theory} {\bf A}}+{\bf I}_{\Omega_i}={\bf I}_{\text{Theory} {\bf A}^{\vee}}+{\bf I}_{\Omega^{\vee}_i} \,.
 \end{align}

Next we compare the  $I \times T^2$ indices of two theories. 
From the localization formula, we have the $I \times T^2$ indices of two theories $Z^{\text{Theory} {\bf A}}_{I \times T^2}$ and $Z^{\text{Theory} {\bf A}^{\vee}}_{I \times T^2}$:
\begin{align}
&Z^{\text{Theory} {\bf A}}_{I \times T^2}(y, \tilde{y},y^{\prime}) 
 =\frac{({\rm i}\eta(q))^{(N_2+N_3-N_1)}}{((N_2+N_3-N_1)/2)!}   
 \oint \prod_{a=1}^{\frac{N_2+N_3-N_1}{2}} \frac{d x_a}{2 \pi {\rm i} x_a}    
 \, \,  {\rm i}^2 \frac{\theta_1(  \prod_{a=1}^{\frac{N_2+N_3-N_1}{2}}  x_a )^2}{\eta (q)^2} 
 \nonumber \\ 
& \quad \times 
\prod_{a \neq b}^{\frac{N_2+N_3-N_1}{2}} {\rm i} \frac{\theta_1(x^{-1}_a x_b )}{\eta (q)} \prod_{a=1}^{\frac{N_2+N_3-N_1}{2}} \prod_{i=1}^{N_3} {\rm i} \frac{\eta (q)}{  \theta_1(x_a y^{-1}_i) } 
\prod_{j=1}^ {N_2} {\rm i}\frac{\eta (q)}{ \theta_1(x^{-1}_a \tilde{y}_i) } 
   \prod_{i=1}^{N_1}  {\rm i} \frac{\theta_1(x^{-1}_a  y^{\prime}_i) }{\eta(q)} \,.
\end{align}
If we choose $\eta$  as $(-1,\cdots,-1)$ in the JK residue operation, the index is  given by residues at $x_a=\tilde{y}_{i_a}$: 
\begin{align} 
Z^{\text{Theory} {\bf A}}_{I \times T^2}(y, \tilde{y},y^{\prime})= 
\sum_{\tilde{\mathcal{I}} \subset \{1,\cdots, N_2 \}} 
& \left(  {\rm i} \frac{\theta_1(  \prod_{a \in \tilde{\mathcal{I}}}  \tilde{y}_a )}{\eta (q)} \right)^2
\prod_{a \in \tilde{\mathcal{I}}} \prod_{i=1}^{N_3} {\rm i} \frac{\eta (q)}{  \theta_1(\tilde{y}_a y^{-1}_i) } \nonumber \\
&  \times
\prod_{ j \in \{1,\cdots, N_2 \} \backslash \tilde{\mathcal{I}} } {\rm i}\frac{\eta (q)}{ \theta_1( \tilde{y}^{-1}_a \tilde{y}_i) } 
   \prod_{i=1}^{N_1}  {\rm i} \frac{\theta_1( \tilde{y}^{-1}_a  y^{\prime}_i) }{\eta(q)} \,,
\label{eq:trialell1}
\end{align}
where we take $\tilde{{\mathcal{I}}}=\{i_1,\cdots,i_{\frac{N_2+N_3-N_1}{2}} \}$ with $1 \le i_1 < i_2 <\cdots < i_{\frac{N_2+N_3-N_1}{2}} \le N_2$. 
The sum $\sum_{\tilde{\mathcal{I}}} $ runs over all the 
possible $\tilde{\mathcal{I}}$ in $\{1,\cdots, N_2 \}$. 

If  we choose $\eta$ as $(1,\cdots,1)$ in the JK residue formula, the index is  expressed by  residues at $x_a={y}_{i_a}$: 
\begin{align} 
Z^{\text{Theory} {\bf A}}_{I \times T^2}  (y, \tilde{y},y^{\prime})= 
\sum_{{\mathcal{I}} \subset \{1,\cdots, N_3 \}} 
& 
 \left(  {\rm i} \frac{\theta_1(  \prod_{a \in \mathcal{I}}  {y}_a )}{\eta (q)} \right)^2
\prod_{a \in \mathcal{I}} \prod_{ j \in \{1,\cdots, N_3 \} \backslash \mathcal{I} } {\rm i} \frac{\eta (q)}{  \theta_1({y}_a y^{-1}_i) } 
\nonumber \\
& \times \prod_{i=1}^{N_2} {\rm i}\frac{\eta (q)}{ \theta_1( {y}^{-1}_a \tilde{y}_i) } 
   \prod_{i=1}^{N_1}  {\rm i} \frac{\theta_1( {y}^{-1}_a  y^{\prime}_i) }{\eta(q)} \,,
\label{eq:trialell11}
\end{align}
where we take ${\mathcal{I}}=\{i_1,\cdots,i_{\frac{N_2+N_3-N_1}{2}} \}$ with $1 \le i_1 < i_2 <\cdots < i_{\frac{N_2+N_3-N_1}{2}} \le N_3$. 
The sum $\sum_{{\mathcal{I}}}$ runs over all the 
possible $\mathcal{I}$ in $\{1,\cdots, N_3 \}$. Since the moduli space $\mathfrak{M}$ is compact,  \eqref{eq:trialell1} agrees with \eqref{eq:trialell11}. 
 Next we will show the matching of the indices between theories ${\bf A}$ and ${\bf A}^{\vee}$. 

The $I \times T^2$ index for the dual theory ${\bf A}^{\vee}$ is given by
\begin{align}
&Z^{\text{Theory} {\bf A}^{\vee}}_{I \times T^2}(y, \tilde{y},y^{\prime})  =\frac{({\rm i}\eta(q))^{(N_1+N_2-N_3)}}{((N_1+N_2-N_3)/2)!}   
\prod_{i=1}^{N_2} \prod_{j=1}^{N_3}  {\rm i} \frac{ \eta (q)  }{\theta_1( \tilde{y}_i {y}^{ -1}_j)}
 \prod_{j=1}^{N_1}  {\rm i} \frac{\theta_1( \tilde{y}_i y^{\prime -1}_j )  }{\eta (q)}
 \nonumber \\
& \qquad \qquad \times
\sum_{\tilde{y}_{i_a}} \oint_{x_a=\tilde{y}_{i_a}} \prod_{a=1}^{\frac{N_1+N_2-N_3}{2}} \frac{d x_{a}}{2 \pi {\rm i} x_a}  
 \prod_{a \neq b}^{\frac{N_1+N_2-N_3}{2}} {\rm i} \frac{\theta_1(x^{-1}_a x_b )}{\eta (q)}  
 \,\,  {\rm i}^2 \frac{\theta_1(  \prod_{a=1}^{\frac{N_1+N_2-N_3}{2}}  x_a )^2}{\eta (q)^2} 
\nonumber \\
&\qquad \qquad \times 
\prod_{a=1}^{\frac{N_1+N_2-N_3}{2}} \prod_{i=1}^{N_2} {\rm i} \frac{\eta (q)}{  \theta_1(x_a \tilde{y}^{-1}_i) } 
\prod_{j=1}^ {N_1} {\rm i}\frac{\eta (q)}{ \theta_1(x^{-1}_a {y}^{\prime}_i) } 
   \prod_{i=1}^{N_3}  {\rm i} \frac{\theta_1(x^{-1}_a  y_i) }{\eta(q)} \nonumber \\
&\qquad =   
\prod_{i=1}^{N_2} \prod_{j=1}^{N_3}  {\rm i} \frac{ \eta (q)  }{\theta_1( \tilde{y}_i {y}^{ -1}_j)}
 \prod_{j=1}^{N_1}  {\rm i} \frac{\theta_1( \tilde{y}_i y^{\prime -1}_j )  }{\eta (q)} 
\sum_{\mathcal{I}^{\prime} \subset \{1,\cdots, N_2\}}
  \left(  {\rm i} \frac{\theta_1(  \prod_{a \in \mathcal{I}^{\prime}}  \tilde{y}_a )}{\eta (q)} \right)^2 
\nonumber \\
& \qquad  \quad \times 
\prod_{a \in \mathcal{I}^{\prime}} \prod_{i \in \{1,\cdots, N_2  \}  \backslash  \mathcal{I}^{\prime} } {\rm i} \frac{\eta (q)}{  \theta_1(\tilde{y}_a \tilde{y}^{-1}_i) } 
\prod_{j=1}^ {N_1} {\rm i}\frac{\eta (q)}{ \theta_1(\tilde{y}^{-1}_a {y}^{\prime}_i) } 
   \prod_{i=1}^{N_3}  {\rm i} \frac{\theta_1( \tilde{y}^{-1}_a  y_i) }{\eta(q)}\,.
\label{eq:trialell2}
\end{align}
Here we have chosen $\eta$  as $(1,\cdots,1)$ in the JK residue operations.
 ${\mathcal{I}^{\prime}}=\{i_1,\cdots,i_{\frac{N_1+N_2-N_3}{2}} \}$ with $1 \le i_1 < i_2 <\cdots < i_{\frac{N_1+N_2-N_3}{2}} \le N_2$.  The sum $\sum_{{\mathcal{I}^{\prime}}}$ runs over all the 
possible $\mathcal{I}^{\prime}$ in $\{1,\cdots, N_2 \}$.

For an arbitrary ${\mathcal{I}^{\prime}}$ in the sum $\sum_{\mathcal{I}^{\prime}}$, there exists a unique $\tilde{\mathcal{I}}$ in the sum $\sum_{\tilde{\mathcal{I}}}$
 such that $\tilde{\mathcal{I}} =\{1,\cdots, N_2  \}  \backslash  \mathcal{I}^{\prime}$.
Then we have the following identities:
\begin{align}
 \prod_{i=1}^{N_2} \prod_{j=1}^{N_3}  {\rm i} \frac{ \eta (q)  }{\theta_1( \tilde{y}_i {y}^{ -1}_j)} \cdot
\prod_{a \in \mathcal{I}^{\prime}}   \prod_{i=1}^{N_3}  {\rm i} \frac{\theta_1( \tilde{y}^{-1}_a  y_i) }{\eta(q)}
&=\prod_{a \in \tilde{\mathcal{I}}} \prod_{i=1}^{N_3} {\rm i} \frac{\eta (q)}{  \theta_1(\tilde{y}_a y^{-1}_i) } \,,
\nonumber \\
\prod_{a \in \mathcal{I}^{\prime}} \prod_{i \in \{1,\cdots, N_2  \}  \backslash  \mathcal{I}^{\prime} } {\rm i} \frac{\eta (q)}{  \theta_1(\tilde{y}_a \tilde{y}^{-1}_i) } 
&=\prod_{ a \in  \tilde{\mathcal{I}} } \prod_{ j \in \{1,\cdots, N_2 \} \backslash \tilde{\mathcal{I}} } -{\rm i}\frac{\eta (q)}{ \theta_1( \tilde{y}^{-1}_a \tilde{y}_i) } \,,\nonumber \\
\prod_{i=1}^{N_2}  
 \prod_{j=1}^{N_1}   {\rm i} \frac{\theta_1( \tilde{y}_i y^{\prime -1}_j )  }{\eta (q)}  
\cdot \prod_{a \in \mathcal{I}^{\prime}} \prod_{j=1}^ {N_1} {\rm i}\frac{\eta (q)}{ \theta_1(\tilde{y}^{-1}_a {y}^{\prime}_i) } 
&=  \prod_{ a \in  \tilde{\mathcal{I}} }  \prod_{i=1}^{N_1}  {\rm i} \frac{\theta_1( \tilde{y}^{-1}_a  y^{\prime}_i) }{\eta(q)}\,, \nonumber \\
\theta_1(\prod_{a \in \mathcal{I}^{\prime}}  \tilde{y}_a )&=-\theta_1(\prod_{a \in \tilde{\mathcal{I}}}  \tilde{y}_a )\,.
\label{eq:thetaid}
\end{align}
Applying these identities to \eqref{eq:trialell1} and \eqref{eq:trialell2}, we obtain the agreement of the $I \times T^2$ indices between the theory ${\bf A}$ and the theory ${\bf A}^{\vee}$:
\begin{align}
Z^{\text{Theory} {\bf A}}_{I \times T^2}(y, \tilde{y},y^{\prime}) 
=Z^{\text{Theory} {\bf A}^{\vee}}_{I \times T^2}(y, \tilde{y},y^{\prime}) \,.
\end{align}

To obtain the third theory in the triality, 
we take $N_c=\frac{N_2+N_3-N_1}{2}$, $N_f=N_3$ and $N_a=N_1+N_2$  in \eqref{eq:Seiberglike} with $N_3 >N_1+N_2$.
We call the $U(\frac{N_2+N_3-N_1}{2})$ gauge theory and its Seiberg-like dual as a ``theory ${\bf B}$'' and a `` theory ${\bf B}^{\vee}$'':
\begin{align}
&\text{Theory } {\bf B}: U \left(\frac{N_2+N_3-N_1}{2} \right)+ \Phi^{\prime}_{i=1,\cdots N_3} \text{ and }  \tilde{\Phi}^{\prime}_{k=1,\cdots N_1+N_2} \,,
\label{eq:Seiberglike3} \\
&\text{Theory } {\bf B}^{\vee}: U\left( \frac{N_1+N_3-N_2}{2} \right)+\Phi^{ \prime \vee}_{k=1,\cdots N_1+N_2},  \tilde{\Phi}^{ \prime \vee}_{i=1,\cdots N_3} \text{ and } 
M^{\prime}_{i=1,\cdots,N_3 \,, k=1,\cdots,N_1+N_2}\,. \nonumber 
\end{align}

\begin{table}[thb]
\begin{center}
\begin{tabular}{c | c  c c  c | c }
			&	$U( \frac{N_2+N_3-N_1}{2})_{\text{gauge}}$			& $SU(N_3)_y$	 & $SU(N_1)_{y^{\prime}}$ 		&	$S U(N_2)_{\tilde{y}}$	 & b.c	\\
			 \hline
$\Phi=\{ \Phi^{\prime}_i \}_{i=1}^{N_3}$			&	$\Box$					 &	$\overline{\Box}$	&	${\bf 1}$	&	${\bf 1}$				 & ${\sf N}$	\\
$P:=\{ \tilde{\Phi}^{\prime}_i \}_{i=1}^{N_2}$			&	$\overline{\Box}$					 &	${\bf 1}$			&	${\bf 1}$	&	${\Box}$		 & ${\sf N}$	\\
$\Psi:=\{ \tilde{\Phi}^{\prime}_i \}_{i=N_2+1}^{N_1+N_2} $	&	${\Box}$			 &	${\bf 1}$						&	$\overline{\Box}$	&{\bf 1}&	  ${\sf D}$	\\
\hline
${\Omega}_{i=1,2}$ &	${\bf det}$					 &	${\bf 1}$			& {\bf 1}&	${\bf 1}$				 & - \\
\end{tabular} 
\caption{The charge assignments and  the boundary conditions for the theory ${\bf B}$.  $\Omega_{i=1,2}$'s are  fermi multiplets at the left and the right boundaries introduced to cancel the gauge anomaly.
  The theory ${\bf B}$ on the interval is identical to the theory ${\bf A}$ on the interval.}
\label{table:trial3}
\end{center}
\end{table}

\begin{table}[htb]
\begin{center}
\begin{tabular}{c | c  c c c  | c }
			&	$U( \frac{N_1+N_3-N_2}{2})_{\text{gauge}}$			& $SU(N_3)_y$	 & $SU(N_1)_{y^{\prime}}$ 		&	$S U(N_2)_{\tilde{y}}$	 & b.c	\\
			 \hline
$\Phi^{\prime }:=\{ \Phi^{\prime  \vee}_i \}_{i=1}^{N_1}$			&	$\Box$					 &	${\bf 1}$	&	$\overline{\Box}$	&	${\bf 1}$				 & ${\sf N}$	\\
$P^{\prime  }:=\{ \tilde{\Phi}^{ \prime \vee}_{i} \}_{i=1}^{N_3}$			&	$\overline{\Box}$					 &		${\Box}$		&	${\bf 1}$	&	${\bf 1}$		 & ${\sf N}$	\\
$\Psi^{\prime}:=\{ {\Phi}^{\prime \vee}_i \}_{i=N_1+1,\cdots,N_1+N_2} $	&	${\Box}$			 &					${\bf 1}$		&	${\bf 1}$	&$\overline{\Box}$&	  ${\sf D}$	\\
$M^{\prime}:=\{M^{\prime}_{k i}\}_{(k, i)=(1,1)}^{(N_3,N_2)} $			&	${\bf 1}$					 &	$\overline{\Box }$			& ${\bf 1 }$&	${\Box}$				 & ${\sf N}$	\\
$\Gamma^{\prime}=\{M^{\prime}_{k i}\}_{(k,i)=(1,N_2+1)}^{(N_3,N_1+N_2)} $			&	${\bf 1}$					 &	$\overline{\Box }$			& ${\Box}$& ${\bf 1}$					 & ${\sf D}$	\\
\hline
$\Omega^{\prime}_{i=1,2}$ &	${\bf det}$					 &	${\bf 1}$			& {\bf 1}&	${\bf 1}$				 & - \\
\end{tabular} 
\caption{The charge assignments and  the boundary conditions for the theory ${\bf B}^{\vee}$.  $\Omega^{\prime}_{i=1,2}$'s are  fermi multiplets at the left and the right boundaries introduced to cancel the gauge anomaly.}
\label{table:trial4}
\end{center}
\end{table}

\begin{pdffig}
\begin{figure}[thb]
\centering
\subfigure[]{\label{fig:trial322}
\includegraphics[height=3cm]{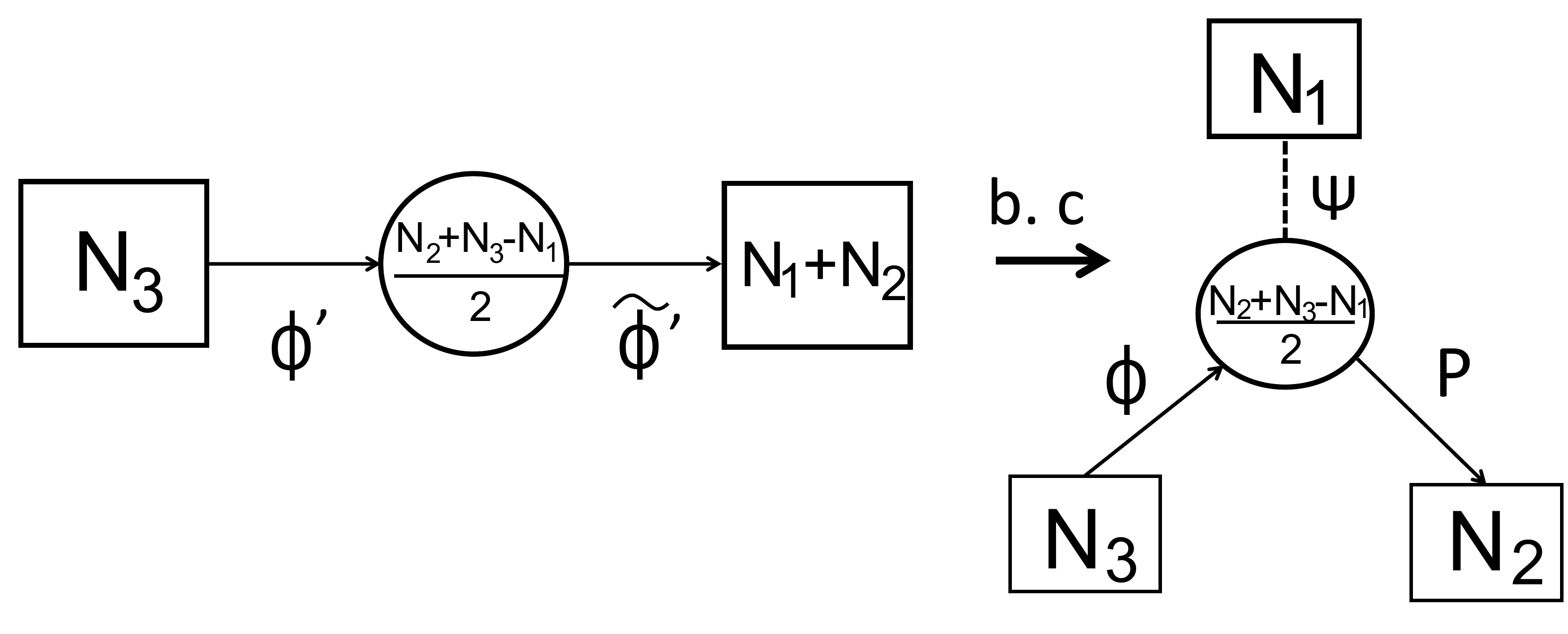}}
\subfigure[]{\label{fig:trial422}
\includegraphics[height=3cm]{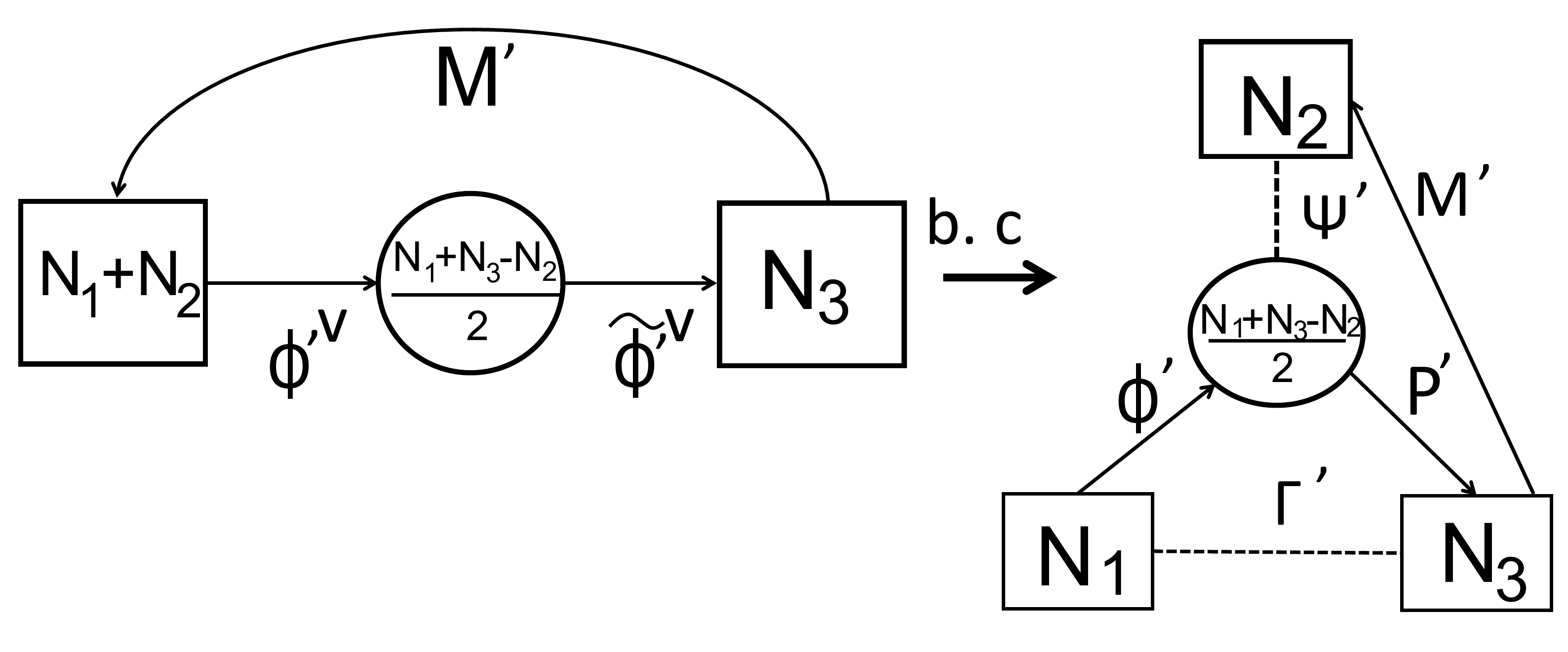}}
\caption{The decomposition of 3d multiplets $\{\tilde{\Phi}^{\prime}_i \}_{i=1}^{N_1+N_2} \mapsto (P, \Psi)$ and $\{ {\Phi}^{\prime}_i \}_{i=1}^{N_3} \mapsto \Phi$ in the theory ${\bf B}$ by the boundary conditions. In the right figure in (a),
 the dashed line denotes the chiral multiplet with the Dirichlet boundary condition $\Psi$. The solid arrows denote the chiral multiplets with the Neumann boundary condition $\Phi, P$. 
(b): The decomposition of 3d multiplets $\{ \tilde{\Phi}^{\prime \vee }_{i} \}_{i=1}^{N_3} \mapsto  P^{\prime}$, $\{ {\Phi}^{\prime \vee}_i \}_{i=1}^{N_1+N_2} \mapsto ( \Phi^{\prime}, \Psi^{\prime})$ and $M^{\prime}_{i k} \mapsto (M^{\prime}, \Gamma^{\prime})$ in the theory ${\bf B}^{\vee}$.}
\label{fig:decompose322}
\end{figure}
\end{pdffig}

We put the theory ${\bf B}$ on the interval and impose the boundary conditions depicted  by 
Table \ref{table:trial3}. The boundary conditions for the theory ${\bf B}^{\vee}$ are depicted  by Table \ref{table:trial4}. The quiver diagrams for the theories ${\bf B}$ and ${\bf B}^{\vee}$ are depicted  by 
Figure \ref{fig:decompose322}.

Under the boundary conditions in Table \ref{table:trial1} and Table \ref{table:trial3}, we find that the theory ${\bf A}$ and the theory ${\bf B}$ are identical. On the 
other hand,  the theory ${\bf B}^{\vee}$ is distinct from the theory ${\bf A}$ and the theory ${\bf A}^{\vee}$. In a similar way of ${\bf I}_{\text{Theory} {\bf A}^{\vee}}$, we can write down   the anomaly polynomial ${\bf I}_{\text{Theory} {\bf B}^{\vee}}$ of the theory ${\bf B}^{\vee}$ which matches with ${\bf I}_{\text{Theory} {\bf A}}$ and ${\bf I}_{\text{Theory} {\bf A}^{\vee}}$:
\begin{align}
{\bf I}_{\text{Theory} {\bf A}}+{\bf I}_{\Omega_i}={\bf I}_{\text{Theory} {\bf A}^{\vee}}+{\bf I}_{\Omega^{\vee}_i}={\bf I}_{\text{Theory} {\bf B}^{\vee}}+{\bf I}_{\Omega^{\prime}_i} \,.
 \end{align}

Next we evaluate the $I \times T^2$ index of  the theory ${\bf B}^{\vee}$:
\begin{align}
&Z^{\text{Theory} {\bf B}^{\vee}}_{I \times T^2}(y, \tilde{y},y^{\prime})  =\frac{({\rm i}\eta(q))^{(N_1+N_3-N_2)}}{(N_1+N_3-N_2)!}   
\prod_{i=1}^{N_3} \prod_{k=1}^{N_2}  {\rm i} \frac{ \eta (q)  }{\theta_1( \tilde{y}_k {y}^{ -1}_i)}
 \prod_{j=1}^{N_1}  {\rm i} \frac{\theta_1( {y}^{-1}_i y^{\prime }_j )  }{\eta (q)}
 \nonumber \\
& \qquad \qquad \times
\sum_{{y}_{i_a}} \oint_{x_a={y}_{i_a}} \prod_{a=1}^{\frac{N_1+N_3-N_2}{2}} \frac{d x_{a}}{2 \pi {\rm i} x_a}  
 \prod_{a \neq b}^{\frac{N_1+N_3-N_2}{2}} {\rm i} \frac{\theta_1(x^{-1}_a x_b )}{\eta (q)}  
 \,\,  \cdot \left( {\rm i} \frac{\theta_1(  \prod_{a=1}^{\frac{N_1+N_3-N_2}{2}}  x_a )}{\eta (q)} \right)^2 
\nonumber \\
&\qquad \qquad \times 
\prod_{a=1}^{\frac{N_1+N_3-N_2}{2}} \prod_{i=1}^{N_1} {\rm i} \frac{\eta (q)}{  \theta_1(x_a {y}^{\prime -1}_i) } 
\prod_{j=1}^ {N_3} {\rm i}\frac{\eta (q)}{ \theta_1(x^{-1}_a {y}_i) } 
   \prod_{i=1}^{N_2}  {\rm i} \frac{\theta_1(x^{-1}_a  \tilde{y}_i) }{\eta(q)} \nonumber \\
&\qquad =   
\prod_{i=1}^{N_3} \prod_{k=1}^{N_2}  {\rm i} \frac{ \eta (q)  }{\theta_1( \tilde{y}_k {y}^{ -1}_i)}
 \prod_{j=1}^{N_1}  {\rm i} \frac{\theta_1( {y}^{-1}_i y^{\prime }_j )  }{\eta (q)}
 \sum_{\mathcal{I}^{\prime \prime} \subset \{1,\cdots, N_3\}} \left({\rm i} \frac{\theta_1(  \prod_{a \in \mathcal{I}^{\prime \prime}}  y_a )}{\eta (q)} \right)^2 
\nonumber \\
& \qquad  \quad \times 
\prod_{a \in \mathcal{I}^{\prime \prime}} \prod_{i=1}^{N_1} {\rm i} \frac{\eta (q)}{  \theta_1( y_a {y}^{\prime -1}_i) } 
\prod_{i \in \{1,\cdots, N_3  \}  \backslash  \mathcal{I}^{\prime \prime} } {\rm i}\frac{\eta (q)}{ \theta_1({y}^{-1}_a {y}_i) } 
   \prod_{i=1}^{N_2}  {\rm i} \frac{\theta_1( y^{-1}_a  \tilde{y}_i) }{\eta(q)}\,.
\label{eq:trialell3}
\end{align}

 Here we have chosen $\eta $  as $(-1,\cdots,-1)$ in the JK residue operations.  We also define
 $\mathcal{I}^{\prime \prime}=\{i_1,\cdots,i_{\frac{N_1+N_3-N_2}{2}} \}$ with $1 \le i_1 < i_2 <\cdots < i_{\frac{N_1+N_3-N_2}{2}} \le N_3$.  The sum $\sum_{\mathcal{I}^{\prime \prime}}$ runs over all the 
possible $\mathcal{I}^{\prime \prime}$ in $\{1,\cdots, N_3 \}$. Since  there are similar identifies of \eqref{eq:thetaid}, the $I \times T^2$ indices for theories ${\bf A}$ and ${\bf B}^{\vee}$ agree each other.
 Thus we have shown that the equality of $I \times T^2$ indices between theories ${\bf A}$, ${\bf A}^{\vee}$ and ${\bf B}^{\vee}$  specified by quiver diagrams in Figure \ref{fig:decompose3}:
\begin{align}
Z^{\text{Theory} {\bf A}^{\vee}}_{I \times T^2}(y, \tilde{y},y^{\prime})=Z^{\text{Theory} {\bf A}}_{I \times T^2}(y, \tilde{y},y^{\prime}) 
=Z^{\text{Theory} {\bf B}^{\vee}}_{I \times T^2}(y, \tilde{y},y^{\prime}) \,.
\end{align}

\begin{pdffig}
\begin{figure}[thb]
\centering
\subfigure[]{\label{fig:trial5}
\includegraphics[height=4.5cm]{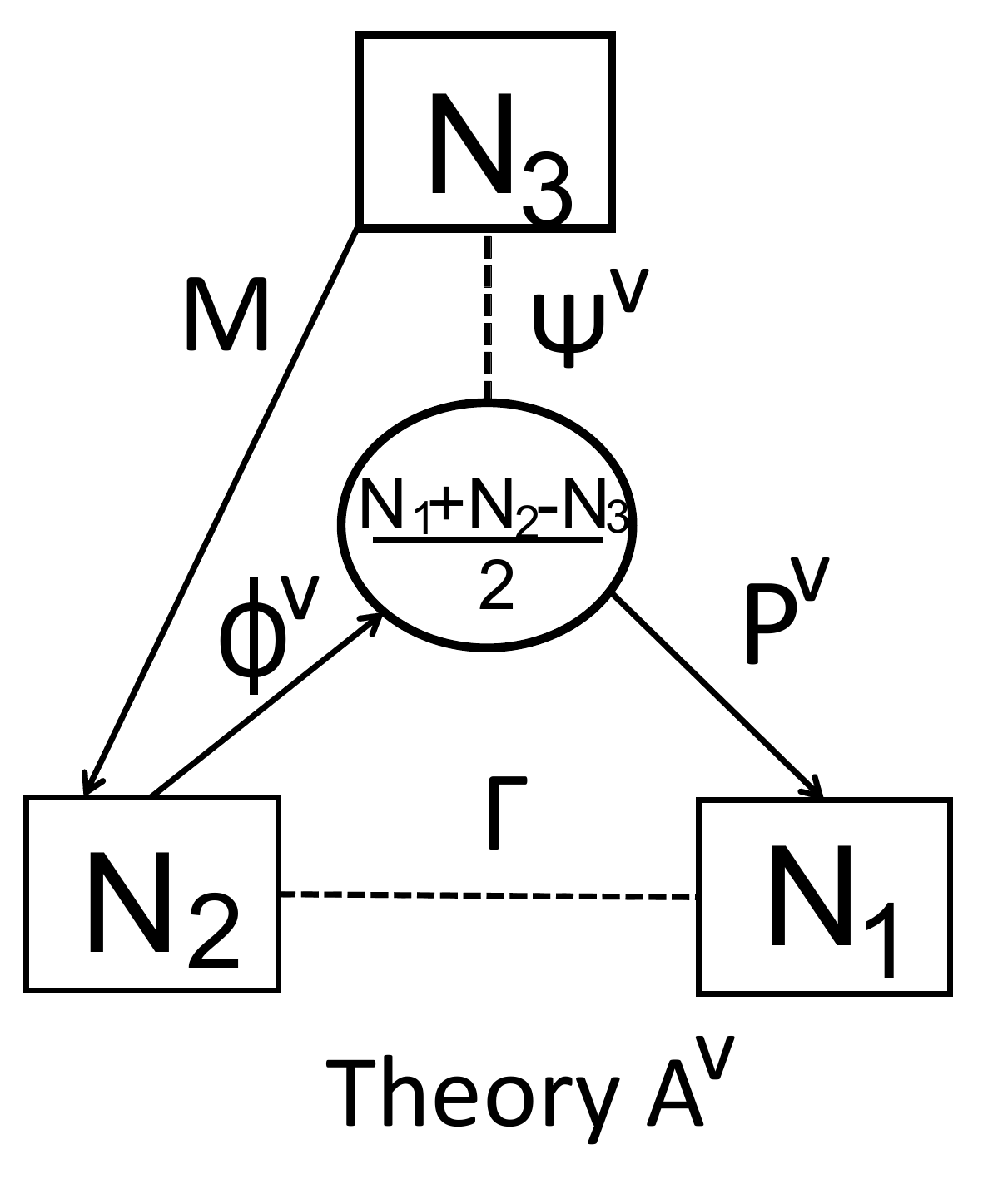}}
\subfigure[]{\label{fig:trial6}
\includegraphics[height=4.5cm]{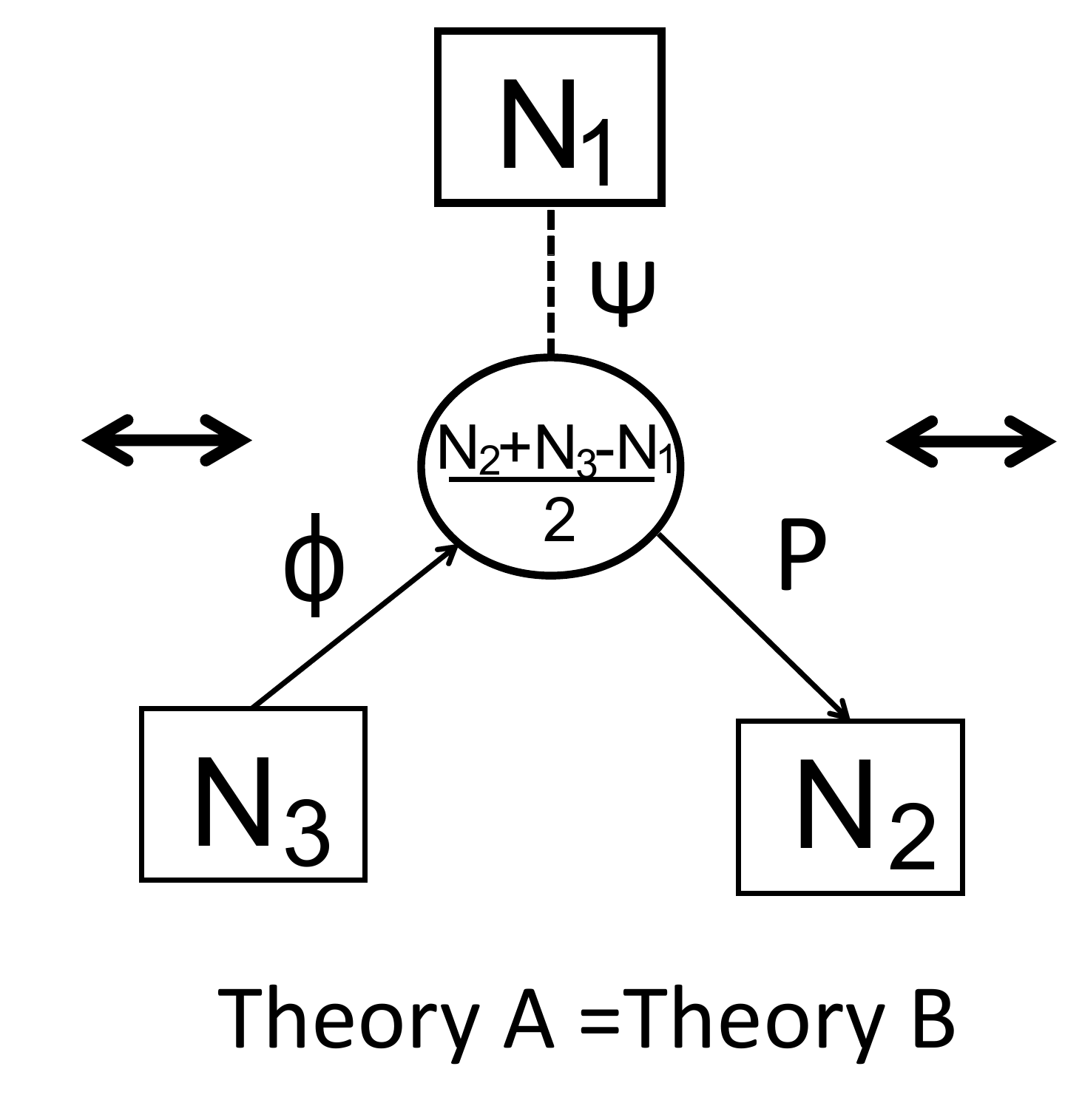}}
\subfigure[]{\label{fig:trial7}
\includegraphics[height=4.5cm]{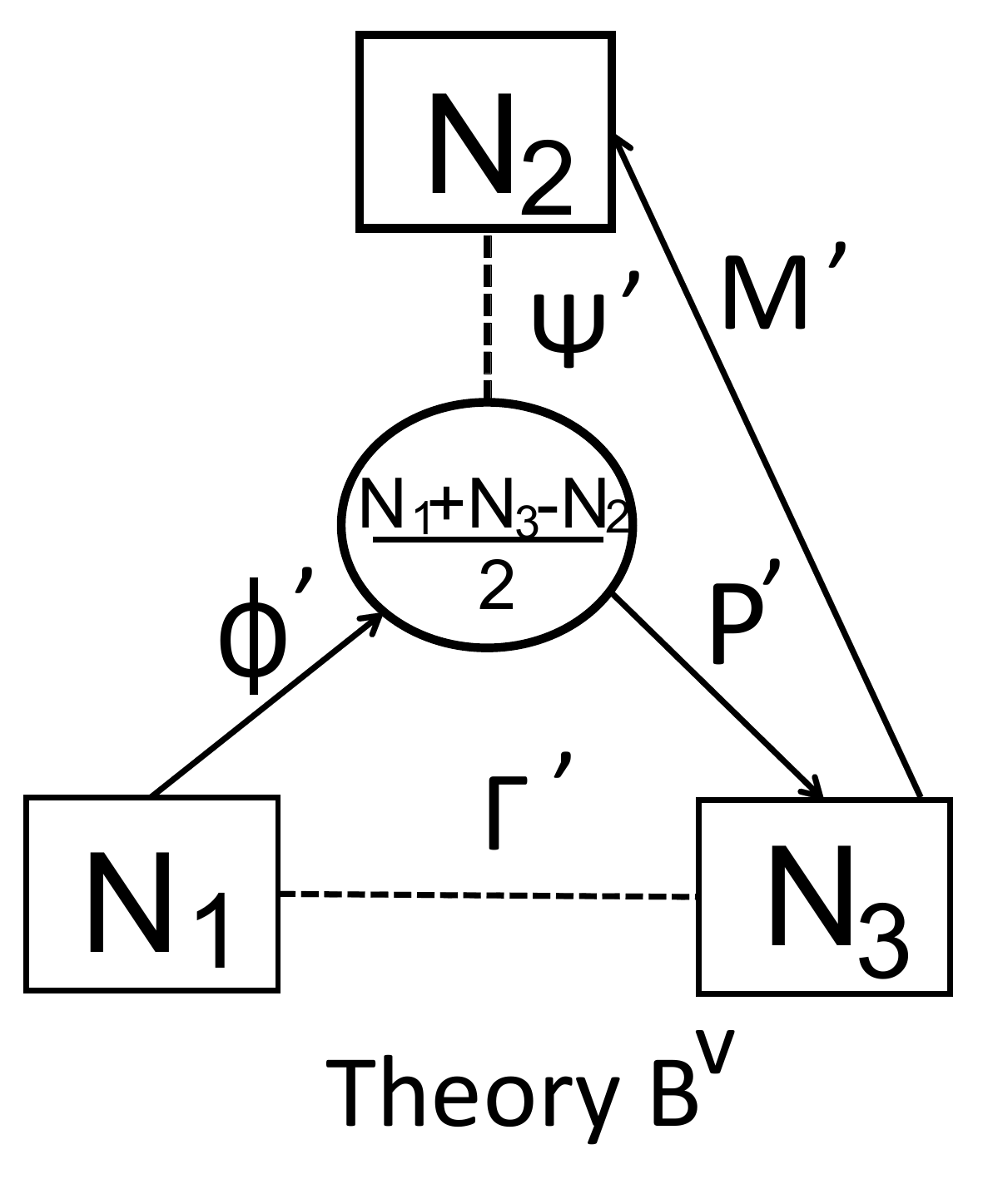}}
\caption{The quiver diagrams  for  the triality of 3d gauge theories with boundaries. (a) and (b) are obtained by a pair of Seiberg-like duality; theory ${\bf A}$ and theory ${\bf A}^{\vee}$.
(b) and (c) are obtained by a  pair of  Seiberg-like duality; ${\bf B}$ and ${\bf B}^{\vee}$. With the boundary conditions,  the theory ${\bf A}$ and  the theory ${\bf B}$ have the same matter content.
In the quiver diagrams, we suppressed  the boundary fermi multiplets $\Omega_i$ in (b), $\Omega^{\vee}_i$ in (a), $\Omega^{\prime}_i $  in (c) with $i=1,2 $.}
\label{fig:decompose3}
\end{figure}
\end{pdffig}

\section{3d theory on the interval and  $\beta \gamma$, $b c$ systems}
\label{sec:section6}
Recently chiral algebras associated with  3d $\mathcal{N}=2$ theories on a 3d half space $\mathbb{R}_{\le 0} \times \mathbb{C}$ were studied in \cite{Costello:2020ndc}. 
Although the general rules for chiral algebras associated with the 3d theories on $I \times \mathbb{C}$ are not  studied yet,  
the author of \cite{Costello:2020ndc} considered the simplest model on $I \times \mathbb{C}$, namely, a free chiral multiplet and  
 expected that  the chiral algebra for the free chiral multiplet on the interval with the Neumann (resp. Dirichlet ) boundary condition  is realized as the $\beta \gamma$-system (resp. the $b c$-system).
We study the relation among free chirals on $I \times M_2$,  $\beta \gamma$ and $b c$-systems.

The $I \times T^2$ indices for the 3d free chiral multiplet with the Dirichlet (${\sf D}$) and with 
the Neumann (${\sf N}$) boundary conditions  are given by

\begin{align}\label{eq:Freechiral}
&Z^{\text{free chiral}}_{I \times T^2, {\sf N}}= \frac{{\rm i} \eta(q) }{\theta_1( x ,q)},  \quad 
Z^{\text{free chiral}}_{I \times T^2, {\sf D}}=
 \frac{{\rm i} \theta_1( x ,q)}{\eta(q)}\, .
\end{align}
Here $x$ is the fugacity of the $U(1)$ flavor symmetry and the fields in the 3d chiral multiplet have  the charge +1. 
We find that $Z^{\text{free chiral}}_{I \times T^2, {\sf N}}$ is nothing but  the  character of the $\beta \gamma$-system with  anti-periodic boundary conditions and 
the weight  $(1, 0)$ for $(\beta,\gamma) $. On the other hand, $Z^{\text{free chiral}}_{I \times T^2, {\sf D}}$ agrees with the  character of the $b c$-system with  anti-periodic boundary conditions and the
weight  $(1, 0)$ for $(b, c)$ . Thus we have the agreement between the characters of $bc$-system and the supersymmetric indices.
 Note that \eqref{eq:Freechiral} is same as the 2d $\mathcal{N}=(0,2)$ elliptic genera for 
the free chiral multiplet and  the free fermi multiplet, respectively.  This  simplest case corresponds to   realizations of  2d $\mathcal{N}=(0,2)$ elliptic genera 
based on  the $\beta \gamma$-system and the $bc$-system  in \cite{Kawai:1994np}.

Next we study properties of  Q-closed operators. 
Since  we have chosen a supercharge ${\bf Q}:= {\rm Q}_2-{\rm Q}_1$ to define the index and  to perform the supersymmetric localization, 
 the index   is expected to  count the  Q-closed operators  modulo the Q-exact operators and the boundary conditions.
The SUSY transformation of the 3d chiral multiplet  by ${\bf Q}:={\rm Q}_2-{\rm Q}_1$  is written as
\begin{align} 
&{\bf Q} \cdot \phi= 0 \,, \quad
{\bf Q} \cdot \bar{\phi}=  \bar{\psi}_2-\bar{\psi}_1 , 
\nonumber \\ 
&{\bf Q} \cdot (\bar{\psi}_2 \pm \bar{\psi}_1)= 0, \quad
{\bf Q} \cdot ({\psi}_1-{\psi}_2)=2 (\partial_2 +{\rm i} \partial_3) \phi, \quad
{\bf Q} \cdot ({\psi}_1+{\psi}_2)=2 {\rm i} \partial_1 \phi.
\end{align}
Here $F, \bar{F}$ are set to zero by the equations of motion. Then  we find that $( \partial_2-{\rm i} \partial_3)^n {\phi} \simeq \partial^n_w {\phi}  $ with $n \ge 0$ are Q-closed operators.

\subsubsection*{\underline{Neumann boundary condition}}
For  the Neumann boundary condition \eqref{eq:Nboundary} , operators $\bar{\psi}_1+ \bar{\psi}_2$ and ${\psi}_1+ {\psi}_2$ are set to zero by the boundary condition. 
Another Q-closed operator is the first descendant of $\bar{\psi}_1+\bar{\psi}_2$ \cite{Costello:2020ndc}:
\begin{align} 
{\bf Q} \cdot \int_I  ( \partial_2-{\rm i} \partial_3) \bar{\phi}&=-\int_I ( \partial_2-{\rm i} \partial_3)(\bar{\psi}_1-\bar{\psi}_2) \nonumber \\
&=-{\rm i} \int_I  \partial_{1} (\bar{\psi}_1+\bar{\psi}_2)=0.
\end{align}
Here we used the equations of motion of $\bar{\psi}_1, \bar{\psi}_2$ and  the boundary condition $\bar{\psi}_1+\bar{\psi}_2=0$. 
Then $( \partial_2-{\rm i} \partial_3)^n \int_I \bar{\phi} \simeq ( \partial_2-{\rm i} \partial_3)^n  \bar{\phi}^{(0)}$ with $n \ge 1$ are Q-closed operators. 
The counting of the  Q-closed operators $ \partial_{{w}}^{n+1} \int_I \bar{\phi} $ and $ \partial_{{w}}^{n} {\phi}$ with $n\ge 0$ 
is consistent with  $Z^{\text{free chiral}}_{I \times T^2, {\sf N}}$ up to the zero point energy.

Since we expect that the Q-closed operators with the Neumann boundary condition are associated with the $\beta \gamma$-system,
the correlation functions of  these Q-closed operators  should be consistent with the OPE of the $\beta\gamma$-system.
On $I \times \mathbb{C}$, the normalized  two point functions of $\int_I \partial_{w}\bar{\phi}  $ and $\phi$  are 
\begin{align} 
&\langle  \int_I \partial_{w}\bar{\phi} (w) \cdot  \phi (x^1, 0)  \rangle_{I \times \mathbb{C}} = 
\langle   \partial_{w}\bar{\phi}^{(0)} (w) \cdot  \phi^{(0)} ( { 0})  \rangle_{\mathbb{C}}=\frac{1}{2 \pi w}\,,
\nonumber \\
&\langle  \int_I \partial_{w}\bar{\phi} (w) \cdot  \int_I \partial_{w}\bar{\phi} (w) \rangle_{I \times \mathbb{C}} =\langle   \partial_{w}\bar{\phi}^{(0)} (w) \cdot \partial_{w} \bar{\phi}^{(0)} ( { 0})  \rangle_{\mathbb{C}}= 0\,,
\nonumber \\
&\langle {\phi} (x^1,w)   {\phi} (0) \rangle_{I \times \mathbb{C}} =\langle   {\phi}^{(0)} (w) \phi^{(0)} ( { 0})  \rangle_{\mathbb{C}}= 0 \,.
\label{eq:freechiral}
\end{align}
Here $\langle \cdots \rangle_{\mathbb{C}}$ means the correlation functions of a 2d $\mathcal{N}=(0,2)$ free chiral multiplet with a lowest component scalar $\phi^{(0)}$  on $\mathbb{C}$.
We find that these correlation functions of Q-closed operators are independent of coordinates $\bar{w}$ and $x^1$.  The correlators in \eqref{eq:freechiral} are independent of $\bar{w}$, because the translation along $\bar{w}$ is expressed as  the anti-commutator of {\bf Q} and $\bar{\bf Q}$.
Two point functions \eqref{eq:freechiral} match with   
  the OPEs of $\beta\gamma$ system \footnote{The factor $\frac{1}{2 \pi}$ in \eqref{eq:freechiral} can be absorbed to the normalization of the action of the chiral multiplet.} :
\begin{align} 
\beta (w) \gamma (0) \sim \frac{1}{w}, \quad \beta (w) \beta (0) \sim 0, \quad \gamma (w) \gamma (0) \sim 0.
\end{align}

 \subsubsection*{\underline{The Dirichlet boundary condition}}
In this case $\bar{\psi}_1+\bar{\psi}_2$ is a Q-closed operator. 
Meanwhile  for the Dirichlet boundary condition \eqref{eq:Dbdychi},  $\phi$ is set to zero.
Another Q-closed operator is obtained by the descent equation:
\begin{align} 
{\bf Q} \cdot \int_I ( {\psi}_1+{\psi}_2)  &=2 {\rm i}\int_I \partial_1 \phi =0,
\end{align}
where we used the boundary condition $\phi=0$ at $x^1=\pm \pi L$. 
In a similar way, $ \partial^n_w \int_I ( {\psi}_1+{\psi}_2)$ is Q-closed.

Again we compute  two point functions on $I \times \mathbb{C}$  and relate them to the $b c$-system.
 Two point functions of Q-closed operators are given by 
\begin{align} 
&\langle  \int_I ({\psi}_1(w)+{\psi}_2 (w))  \cdot (\bar{\psi}_1(0)+\bar{\psi}_2(0))  \rangle_{I \times \mathbb{C}} 
=
4 \langle   \bar{\psi}^{(0)}_c(w)  \cdot \psi^{(0)}_c(0)  \rangle_{ \mathbb{C}}
=\frac{1}{\pi w}\,, \nonumber \\
&\langle  \int_I  ({\psi}_1(w)+{\psi}_2 (w))  \cdot \int_I  ({\psi}_1(0)+{\psi}_2 (0))    \rangle_{I \times \mathbb{C}} =0
 \,, \nonumber \\
& \langle  (\bar{\psi}_1(x^1,w)+\bar{\psi}_2(x^1,w))   (\bar{\psi}_1(0)+\bar{\psi}_2(0))   \rangle_{I \times \mathbb{C}}= 0
 \,.
\end{align}
Here $\langle \cdots \rangle_{\mathbb{C}}$ means the correlation functions in a 2d $\mathcal{N}=(0,2)$ free fermi multiplet with the fermion $\psi^{(0)}_c$ on $\mathbb{C}$.
These correlation functions are consistent with the OPEs of the $bc$ system:
\begin{align} 
b (w) c (0) \sim \frac{1}{w}, \quad b (w) b (0) \sim 0, \quad c (w) c (0) \sim 0 \,.
\end{align}

\section{Dimensional reduction  and 2d $\mathcal{N}=(2,2)$ theories on $I \times S^1$}
\label{sec:section7}
In section \ref{sec:section3}, we have studied the supersymmetric localization computation of the indices on $I \times T^2$. In this section
we perform  the dimensional reduction in the $x^3$-direction and study localization formula of 
supersymmetric indices for 2d $\mathcal{N}=(2,2)$ theories on $I \times S^1$.  The detailed analysis for the $I \times S^1$ indices   is left in  our upcoming future work \cite{Sugiyama2020}.

\subsection{SUSY localization formula  for 2d $\mathcal{N}=(2,2)$ theories on $I \times S^1$ }
First we define the coordinates of $I \times S^1$ as
\begin{align}
I \times S^1= \{ (x^1,x^2) | -\pi L \le  x^1 \le \pi L, \, \, x^2 \sim  x^2 + 2 \pi R \}.
\end{align}
 
In the dimensional reduction, the SUSY transformation, the Lagrangians and boundary conditions at $x^{1}=\pm \pi L$ for the 2d theory
  are naturally originated from  the 3d theory  in sections \ref{sec:section21} and section \ref{sec:section22} with the replacement: 
\begin{align}
&A_{3}(x_1,x_2,x_3) \mapsto \sigma^{\prime }(x_1,x_2), \quad  D_3 \Psi (x^1,x^2,x^3) \mapsto {\rm i} \sigma^{\prime} \Psi (x^1,x^2) \, , \nonumber \\
&\Psi (x^1,x^2, x^3)  \mapsto \Psi (x^1, x^2)  \, .
\end{align}
Here $\sigma^{\prime}$ is an adjoint scalar in the 2d $\mathcal{N}=(2,2)$ vector multiplet and $\Psi(x^1,x^2, x^3)$ and $\Psi(x^1,x^2)$ are  fields in the 3d $\mathcal{N}=2$ and the  2d $\mathcal{N}=(2,2)$ theories.
The supersymmetric quantum mechanics at the boundaries $\partial (I \times S^1) =S^1_L \sqcup S^1_R$ are given by
the dimensional reduction of the 2d $\mathcal{N}=(0,2)$ theories  in section \ref{sec:section23} to the 1d $\mathcal{N}=2$ theories.  
In three dimensions, the surface terms for the bulk 3d superpotential are compensated by the SUSY transformation of 
  the boundary 2d $\mathcal{N}=(0,2)$ superpotentials \eqref{eq:3dmatrixfac}. In the same way as \eqref{eq:3dmatrixfac}, the surface terms of  the bulk 2d superpotential on  $I \times S^1$ are canceled by the SUSY transformation of the superpotentials of the 1d Fermi multiplets. 

In 2d $\mathcal{N}=(2,2)$ GLSMs with boundaries, 
there is another choice of boundary interaction called a Chan--Paton factor or a brane factor that cancels the surface term of the superpotential \cite{Herbst:2008jq}:
\begin{align}
\mathcal{W}_{\mathcal{V}}=\mathrm{Str}_{\mathcal{V}} P \exp  \left( \pm  {\rm i} \oint dx^2  \mathcal{A}_2 \right) \,,
\label{eq:branefac}
\end{align}
with
\begin{align}
 \mathcal{A}_2 =\rho_* (A_2+{\rm i}\sigma^{\prime})+\frac{\rm i}{2} \{{\sf Q}, \bar{\sf Q} \}
-\frac{1}{2}\sum_{i} (\psi_{1,i}-\psi_{2, i})\frac{\partial {\sf Q}}{\partial \phi_i}
+\frac{1}{2}\sum_{i}  (\bar{\psi}_{1,i}-\bar{\psi}_{2 , i})\frac{\partial \bar{\sf Q}}{\partial \bar{\phi}_i}\,.
\end{align}
Here $\mathcal{V}=\mathcal{V}_{\text{even}}\oplus \mathcal{V}_{\text{odd}}$ is a $\mathbb{Z}_2$-graded vector space,  called a Chan-Paton vector space. $\rho_*$ is a map from the Lie algebra of 
the gauge and the flavor symmetry groups to $\mathcal{V}$.

 ${\sf Q} \in \mathrm{End} (\mathcal{V}) $ is called  a matrix factorization or a tachyon profile. ${\psi}_{1, i}, {\psi}_{2, i}$ are 
the first and second components of the fermion $\psi_i=({\psi}_{1, i}, {\psi}_{2, i})^{T}$ in the $i$-th 2d  $\mathcal{N}=(2,2)$ chiral multiplet. 
The subscript $i$ in the sum labels  the chiral multiplets in the tachyon profile.
The surface term of the  bulk 2d $\mathcal{N}=(2,2)$ superpotential $W$, i.e., the dimensional reduction of \eqref{eq:SUSYW}  is canceled by the SUSY transformation of a Chan--Paton factor $\mathcal{W}_{\mathcal{V}}$, if  tachyon profiles  satisfy the following relations:
\begin{align}
 {\sf Q}^2=W \, {\rm id}_{\mathcal{V}}, \quad  \bar{\sf Q}^2=\bar{W} \, {\rm id}_{\mathcal{V}} \,.
\label{eq:gaugedMF}
\end{align}
In \eqref{eq:branefac}, $+$ sign is taken at the right boundary $x^1= \pi L$ and $-$ is taken at the left boundary $x^1= -\pi L$. 

By  introducing  Chan-Paton factors, one  can  change the Neumann boundary condition for  chiral  multiplets to the Dirichlet boundary condition. For example,  see  \cite{Honda:2013uca} for the localization computation of the hemisphere  partition function with the Dirichlet boundary condition. We will see the two methods, i.e., the Neumann boundary condition with a matrix factorization and the Dirichlet boundary condition 
   agree each other in simple examples.

Next we explain the definition of supersymmetric indices on $I \times S^1$.   We assume the same boundary condition is imposed at $x^{1}=\pm \pi L$.
We take the following twisted boundary condition along $S^1$ direction:
\begin{align}
&\Psi (x^1,x^2+2\pi R)=\prod_{i} e^{z_i F_i} \Psi (x^1,x^2)\,.
\label{eq:twistbc2d}
\end{align}
Here $F_i$ is the  generator of a $U(1)$ flavor symmetry and $z_i$ is the fugacity of $F_i$.
Then the supersymmetric index on $I \times S^1$ is defined by  
\begin{align}
Z_{\mathcal{W}_L \mathcal{W}_R}:&= 
\mathrm{Tr}_{\mathcal{H}} (-1)^F e^{-2 \pi   R  H} 
  \prod_{i} e^{ -z_i F_i} \,.
\label{eq:22index}
\end{align}
The localization formula of  the index is given by 
\begin{align}
Z_{\mathcal{W}_L \mathcal{W}_R} &= 
 \frac{1}{|W_{ G }|} \sum_{u_{\ast} \in \mathfrak{M}_{\text{sing}} } 
\mathop{\text{JK-Res}}_{ u = {u}_*} ({Q}_*,{\eta})  \,  
\nonumber \\
& \qquad \times \mathrm{Str}_{\mathcal{V}_L} ( e^{-u})  \mathrm{Str}_{\mathcal{V}_R} ( e^{u}) \,
 Z^{1\text{-loop}}_{I \times S^1}  Z^{1\text{-loop}}_{S^1_L}  Z^{1\text{-loop}}_{S^1_R}
\wedge_{a=1}^{\mathrm{rk}(G)} d u^a  \,.
\label{eq:indexIxS1}
\end{align}
Here 
$u$  is the saddle point value  of $\oint ( {\rm i}  A_{2} - \sigma^{\prime})$.
 $Z^{1\text{-loop}}_{I \times S^1},  Z^{1\text{-loop}}_{S^1_L}$ and $  Z^{1\text{-loop}}_{S^1_R}$
 are the one-loop determinant of the 2d $\mathcal{N}=(2,2)$ theory on $I \times S^1$,  the one-loop determinants of the 1d $\mathcal{N}=2$ theories on $S^1_L$ and  $S^1_R$ defined by
\begin{align}
 Z^{1\text{-loop}}_{I \times S^1}&= Z^{I \times S^1}_{\text{2d.vec}, G} (u) \prod   Z^{I\times S^1}_{\text{chi}, {\sf D}, {\bf R}} (u, z ) \prod   Z^{I\times S^1}_{\text{chi}, {\sf N}, {\bf R}} (u, z ) \,,
\label{eq:loop2dS1xI} \\
 Z^{1\text{-loop}}_{S^1_L}&=  \prod   Z_{\text{1d.chi}, {\bf R}_L} (u,z_{L} ) \prod   Z_{\text{1d.Fermi}, {\bf R}_L} ({ u}, z_L ) \,,\\
 Z^{1\text{-loop}}_{S^1_R}&=  \prod Z_{\text{1d.chi}, {\bf R}_R} (u, z_{R} ) \prod  Z_{\text{1d.Fermi}, {\bf R}_R} ({ u}, z_R ) \,.
\label{eq:loop1dS1}
\end{align}
The 2d $\mathcal{N}=(2,2)$ one-loop determinant \eqref{eq:loop2dS1xI} consists of a 2d $\mathcal{N}=(2,2)$ $G$ vector multiplet $Z^{I \times S^1}_{\text{2d.vec}, G}$, a 2d chiral multiplet $Z^{I \times S^1}_{\text{chi}, {\sf N},{\bf R}}$  with the Neumann  boundary condition and a $Z_{\text{chi}, {\sf D},{\bf R}}$  with the Dirichlet boundary   condition. The 1d $\mathcal{N}=2$ one-loop determinants consist of a 1d  chiral multiplet $Z_{\text{1d.chi}, {\bf R}}$ and a 1d  fermi multiplet $ Z_{\text{1d.Fermi}, {\bf R}}$. 
The products   are taken over all the multiplets.

The one-loop determinants of the supermultiplets  are given by
\begin{align}
Z^{I \times S^1}_{\text{2d.vec}, G} (u) &= \prod_{\alpha \in \text{rt}(\mathfrak{g})} 
 2 \sinh \left( \frac{\alpha (u)}{2} \right) \,,\\
 Z^{I \times S^1}_{\text{chi}, {\sf D}, {\bf R}} ({u} ; z) &=
Z_{\text{1d.Fermi}, {\bf R}} ({ u}, z)=
\prod_{Q  \in \text{wt} (\mathbf{R})} 
  \prod_{Q^F  \in \text{wt} ({\bf F})}
 2 \sinh \left( \frac{Q(u)+Q^F (z)}{2} \right) \,, 
 \\
Z^{I \times S^1}_{\text{chi}, {\sf N}, {\bf R}} (u ; z)
&= Z_{\text{1d.chi}, {\bf R}} (u, z )=\prod_{Q  \in \text{wt} (\mathbf{R})} 
  \prod_{Q^F  \in \text{wt} ({\bf F})}
 \frac{1}{2 \sinh  \left( \frac{Q(u)+Q^F (z)}{2} \right)} \,.
\label{eq:2doneloop}
\end{align}
We find that the one-loop determinants of the 2d $\mathcal{N}=(2,2)$ multiplets on $I \times S^1$ are independent of the length of the interval $I$ and  
agree with  the one-loop determinants of 1d $\mathcal{N}=2$ multiplets  on $S^1$ in \cite{Hori:2014tda, Hwang:2014uwa}.
A  formula without   the Dirichlet boundary condition and the boundary 1d multiplets was briefly mentioned in \cite{Hori:2013ika}.

The derivation of supersymmetric localization formula of the index on $I \times S^1 $ is almost parallel to that for the index on $I \times T^2$. But there is a  difference  coming from non-compactness of the space of $u, \bar{u}$.  On $I \times S^1$, ${u}$ and $\bar{u}$ come from the constant values of $ \oint ({\rm i} A_{2} \mp \sigma^{\prime})$ 
that span a non-compact space $\mathfrak{M}=(S^1 \times \mathbb{R})^{\mathrm{rk}(G)}$. 
In addition to the residues around $\partial \Delta_{\varepsilon}$, the residues around $\partial (\mathfrak{M} \backslash \Delta_{\varepsilon})$ with $\mathrm{Re} (u_i) = \pm \infty$   possibly  contribute
 to the $I \times S^1$ index. For the non-degenerate case, if the 2d FI-parameter $\zeta$ is contained in the charge cones $\sum_i \mathbb{R}_{ > 0 } Q_i $ at all the singular points $u_*$, we do not have to take the residues with $\mathrm{Re} (u_i) = \pm \infty$ into account. In such situations, the index is given by the localization formula \eqref{eq:indexIxS1} by setting $\eta=\zeta$. 
On the other hand,  if $\zeta$ is not contained in the charge cones,   there is possibly an  extra contribution   to the index. 
Here we assume $\zeta$ satisfies the definition  of the JK residue and do not consider the extra contribution to the index.

Next we consider the expectation value of Q-closed operators on $I \times S^1$. 
Each Q-closed operator is a  Wilson loop such that 
 a path $C$ is a circle along the $x^2$-direction with  $x^3$=constant :
\begin{align} 
W_{\bf R}=  \mathrm{Tr}_{\bf R} P \exp \left( \oint_C dx^2 ( {\rm i} A_{2}-\sigma^{\prime} ) \right) \,.
\label{eq:Wilson}
 \end{align}
Here  ${\bf R}$ is a representation of the 2d gauge group $G$. 
 $W_{\bf R}$ does not necessarily lie on the boundaries.
The saddle point value of the Wilson loop \eqref{eq:Wilson} is given by
\begin{align} 
W_{\bf R} |_{\text{saddle point}}=  \mathrm{Tr}_{\bf R}  e^{ u} \,.
 \end{align}
Note that a correlation function  of  Wilson loops is independent of the $x^3$ position. The localization  formula for the 
 correlation functions  of Wilson loops is given by 
\begin{align}
\langle \prod_{i=1}^n W_{{\bf R}_i} \rangle_{ \mathcal{W}_{\mathcal{V}_L}, \mathcal{W}_{\mathcal{V}_R}}&= 
 \frac{1}{|W_{ G }|} \sum_{u_{\ast} \in \mathfrak{M}_{\text{sing}} } 
\mathop{\text{JK-Res}}_{ u = {u}_*} ({Q}_*,{\eta}) \left(\prod_{i=1}^n \mathrm{Tr}_{{\bf R}_i} e^{  u} \right)
\nonumber \\
& \times \mathrm{Str}_{\mathcal{V}_L} ( e^{-u})  \mathrm{Str}_{\mathcal{V}_R} ( e^{u}) Z^{1\text{-loop}}_{I \times S^1}  Z^{1\text{-loop}}_{S^1_L}  Z^{1\text{-loop}}_{S^1_R}
\wedge_{a=1}^{\mathrm{rk}(G)} d u^a  \,.
\label{eq:vevWilson}
\end{align}
One can also insert the  Wilson loops $\mathrm{Tr}_{\bf F} e^{z}$   for the flavor symmetry group in the correlation functions.

\subsection{$I \times S^1$ indices, Wilson loops  and open string Witten indices}
 
Here we briefly study boundary conditions in $I \times S^1$ indices and compare them with the indices  based on the  geometric computation  and the results in the Gepner models in the  CFT computation.  

\subsubsection{Projective space $\mathbb{P}^{M-1}$}
We consider a $U(1)$ GLSM with $M$ chiral multiplets with the gauge charge $+1$ without a superpotential. In the negative FI-parameter region $\zeta <0$, the supersymmetric vacuum does not exist. 
 In a generic point in the   positive FI parameter region $\zeta >0$,  
the moduli space of  the Higgs branch vacua is a complex projective space $\mathbb{P}^{M-1}$.
We impose the Neumann boundary condition for the $M$ chiral multiplets and introduce a Wilson loop 
with a charge $a$. From the localization formula \eqref{eq:vevWilson} the vacuum expectation value (vev) of the Wilson loop is given by 
\begin{align}
\langle 
e^{a {\rm i} \oint (A_2+ {\rm i} \sigma^{\prime})  } \rangle
&= \oint_{u=0}   \frac{d u}{2 \pi {\rm i} }  \frac{e^{a u }}{( e^{u/2} -e^{-u/2})^{M}} \,. 
\label{eq:Pnpair}
\end{align}
Here we have chosen the FI-parameter $\zeta >0$ in the JK residue evaluation.
Note that \eqref{eq:Pnpair} is  invariant under the  global gauge transformation $u \mapsto u + 2 \pi {\rm i}$, only if 
the condition $a +\frac{M}{2} \equiv 0 \, ( \text{mod } 2)$ is satisfied.  
  Especially $M=\text{even}$ is required   for  $a \in \mathbb{Z}$. When $M=\text{even}$ and $a \in \mathbb{Z}$,
 the $I \times S^1$ index directly agrees with an  index with an $\hat{A}$-class:
\begin{align}
\langle e^{a {\rm i} \oint (A_2+ {\rm i} \sigma^{\prime})  } \rangle
&= \oint_{u=0}   \frac{d u}{2 \pi {\rm i} u^M}  e^{a u } \left ( \frac{u/2} {\sinh( u/2) } \right)^M  \nonumber \\
&=\int_{\mathbb{P}^{M-1} }  \mathrm{ch} (\mathcal{O}(a) ) \hat{ A}(T\mathbb{P}^{M-1}) \,.
\end{align}
In order for the $I \times S^1$ index to be well-defined at an arbitrary integer $M$,  we have to insert  a charge $a\pm M/2$ $U(1)$ Wilson loop 
 with $a \in \mathbb{Z}$. If we choose $a + M/2$ with $a \in \mathbb{Z}$ , the vev of the operator has  a  geometric interpretation:
\begin{align}
\langle e^{(a + \frac{M}{2} ){\rm i} \oint (A_2+ {\rm i} \sigma^{\prime})  } \rangle
&= \oint_{u=0}   \frac{d u}{2 \pi {\rm i} u^M}  e^{a u } \left ( \frac{u} {1-e^{-u}} \right)^M  \nonumber \\
&=\int_{\mathbb{P}^{M-1} }  \mathrm{ch} (\mathcal{O}(a) ) \mathrm{Td}(T\mathbb{P}^{M-1}) 
\nonumber \\
&=\left(
    \begin{array}{c}
      M-1 +a   \\
      a 
    \end{array}
  \right)\,.
\end{align}
Here $\text{Td}$ is the Todd class.  If we turn on the flavor fugacities for the flavor group $SU(M)$, 
the vev is modified to   the equivariant index of $\mathbb{P}^{M-1}$. 
The vev of Wilson loops with $e^{-a  {\rm i} \oint (A_2+ {\rm i} \sigma^{\prime})  }, e^{b  {\rm i} \oint (A_2+ {\rm i} \sigma^{\prime})  } $ inserted at the left and the right boundaries 
is interpreted as the Euler pairing of $\mathcal{O}(b)$ and $\mathcal{O}(a)$: 
\begin{align}
\langle e^{-a {\rm i}  \oint (A_2+ {\rm i} \sigma^{\prime})  } e^{b {\rm i}  \oint (A_2+ {\rm i} \sigma^{\prime})  } e^{ \frac{M}{2} {\rm i}  \oint (A_2+ {\rm i} \sigma^{\prime})  }\rangle
&= \oint_{u=0}   \frac{d u}{2 \pi {\rm i} u^M}  e^{(b-a) u } \left ( \frac{u} {1-e^{-u}} \right)^M  \nonumber \\
&=\int_{\mathbb{P}^{M-1} }  \mathrm{ch} (\mathcal{O}(a)^{\vee})  \mathrm{ch} (  \mathcal{O}(b)  ) \mathrm{Td}(T\mathbb{P}^{M-1}) 
\nonumber \\
&=\chi_{\mathbb{P}^{M-1}} (\mathcal{O}(a), \mathcal{O}(b)).
\end{align}

\subsubsection{Geometric phase, Landau--Ginzburg phase and matrix factorization}
\label{sec:section72}
Next we study a 2d $\mathcal{N}=(2,2)$ GLSM  with  a superpotential with two methods. 
 We consider a  $U(1)$ gauge theory with chiral multiplets $\phi_i$  with gauge charges $+1$'s for $i=1,\cdots, M$ 
and a chiral multiplet $P$ with a gauge charge $-k$. We introduce a superpotential term
\begin{align}
W=P f(\phi_1,\cdots,\phi_M) \,,
\end{align}
where $f(\phi_1,\cdots,\phi_M)$ is a degree $k$ homogeneous polynomial of $\phi_1,\cdots, \phi_M$.
At a generic point in the positive FI-parameter region, the moduli space of the Higgs branch vacua is  a degree $k$ hypersurface $X$ defined by  $f(\phi_1,\cdots,\phi_M)=0$ in $\mathbb{P}^{M-1}$. 
The hypersurface $X$ is a Fano $(M-2)$-fold for $k < M$, and a  Calabi-Yau $(M-2)$-fold for $k = M$.

Let us study the boundary condition and its  consequence in physics.
First we consider the geometric phase, i.e., the positive FI-parameter region.
We take the Neumann boundary condition for $\phi_i$'s with $i=1,\cdots,M$ and the Dirichlet boundary condition for $P$. The vev of the Wilson loop is  expressed as
\begin{align}
\langle e^{a {\rm i} \oint (A_2+ {\rm i} \sigma^{\prime})  } \rangle
&= \oint_{u=0}   \frac{d u}{2 \pi {\rm i} }  e^{a u }  \frac{  e^{-k u/2} -e^{k u/2} } { (e^{u/2} -e^{-u/2})^{M} }   \nonumber \\
&=- \int_{X}    \mathrm{ch}\left(\mathcal{O}\left(a-(k-M)/2\right) \right) \text{Td}(TX)\,,
\end{align}
where $a-(k-M)/2$ has to be  an integer. 
On the other hand, if we impose the Neumann boundary condition for $P$ and the Dirichlet boundary 
conditions for $\phi_i$'s, the index is given by  
\begin{align}
\langle e^{a {\rm i} \oint (A_2 + {\rm i} \sigma^{\prime})  } \rangle
&= \oint_{u=0}   \frac{d u}{2 \pi {\rm i} }  e^{a u }  \frac{   (e^{u/2} -e^{-u/2})^{M} } {e^{-k u/2} -e^{k u/2} }   \nonumber \\
&= -\sum_{l=0}^{k-1 }\oint_{u=\frac{2 \pi {\rm i} l}{k} }   \frac{d u}{2 \pi {\rm i} }  e^{\left(a+\frac{M-k}{2}  \right)u }  \frac{  (1-e^{-u})^{M}   } {1 -e^{-k u}}  \,. 
\end{align}

To study a relation with SCFT computation, we consider the case in which the GLSM flows to superconformal field theories, i.e., $k=M$. 
In the large positive FI parameter region, the GLSM flows to  an NLSM with the target space $X$.
In this region, $\phi_i$'s parameterize the target space $X$. Then we impose the Neumann boundary conditions for $\phi_i$'s. The correlation function is written as 
\begin{align}
\langle e^{-a {\rm i} \oint (A_2+ {\rm i} \sigma^{\prime})  } e^{b {\rm i} \oint (A_2+ {\rm i} \sigma^{\prime})  } \rangle
&=\oint_{u=0}   \frac{d u}{2 \pi {\rm i} }  e^{(b-a)u }  \frac{  1 -e^{-M u} } {(1-e^{-u})^{M}  }   \nonumber \\
&=-\int_{X} \mathrm{ch} (\mathcal{O}(-a) ) \otimes  \mathcal{O}(b)) \mathrm{Td} (TX) \,.
\label{eq:Eulerpairing}
\end{align}
Let us reproduce the \eqref{eq:Eulerpairing} from a matrix factorization;  
 the $\phi_i$'s and $P$ with the Neumann boundary conditions and a matrix factorization \eqref{eq:gaugedMF} that makes the superpotential  invariant.
We take a tachyon profile at the left and the right boundaries as 
\begin{align}
{\sf Q}= f(\phi) \bar{\eta}+  P {\eta}  \text{ with } \{\eta, \bar{\eta} \}=1, \,\, \eta^2=\bar{\eta}^2=0  \,.
\end{align}
Here ${\sf Q}$ acts on the Chan--Paton vector spaces  
 $\mathcal{V}_{L/R}=| 0 \rangle_{L/R} \oplus \bar{\eta} | 0 \rangle_{L/R} $ at the left ($L$) and the right ($R$) boundaries. We assign  gauge charges $a+\frac{M}{2}$ for $| 0 \rangle_{L}$ and $b+\frac{M}{2}$ for $| 0 \rangle_{R}$. The saddle point values of the Chan--Paton factors are given by
\begin{align}
\mathrm{Str}_{\mathcal{V}_L} ( e^{-u})&=e^{-a u} (e^{ \frac{M u}{2}}-e^{\frac{-M u}{2}})\,, \nonumber \\ 
  \mathrm{Str}_{\mathcal{V}_R} ( e^{u})&=e^{b u} (e^{-\frac{M u}{2}}-e^{\frac{M u}{2}})\,.
\end{align}
Thus the localization formula for the  $I \times S^1$ index with the Chan--Paton factors is given by 
\begin{align}
Z_{\mathcal{W}_L \mathcal{W}_R}
&=
\oint_{u=0} \frac{d u}{2 \pi {\rm i} } \frac{\mathrm{Str}_{\mathcal{V}_L} ( e^{-u}) \mathrm{Str}_{\mathcal{V}_R} ( e^{u})}
{(e^{-\frac{M u}{2}}-e^{ \frac{M u}{2}})(e^{ \frac{ u}{2}}-e^{- \frac{ u}{2}})^M}  \nonumber \\
&=\oint_{u=0}   \frac{d u}{2 \pi {\rm i} }  e^{(b-a)u }  \frac{  1 -e^{-M u} } {(1-e^{-u})^{M}  }  \,.
\label{eq:MFEuler}
\end{align}
Therefore we obtain the same result as \eqref{eq:Eulerpairing} up to an overall sign.
By comparing \eqref{eq:Eulerpairing} with \eqref{eq:MFEuler}, we find that  the two boundary interactions $\mathcal{W}_{\mathcal{V}_L}$ and $\mathcal{W}_{\mathcal{V}_R}$ turn the Neumann boundary condition for $P$ to the Dirichlet one. 

Next let us consider the Landau-Ginzburg phase, i.e., the negative FI-parameter region. 
We choose the Neumann boundary condition for $P$ and the Dirichlet boundary conditions for $\phi_i$'s. The index with two Wilson loops  is given by
\begin{align}
\chi^{\text{LG}}_{a b}:&=\langle e^{b {\rm i} \oint (A_2+ {\rm i} \sigma^{\prime})  }  e^{-a {\rm i} \oint (A_2+ {\rm i} \sigma^{\prime})  }  \rangle \nonumber \\
&= \sum_{l=0}^{M-1}\oint_{u=\frac{2 \pi {\rm  i} l}{M}}   \frac{d u}{2 \pi {\rm i} }  e^{(b-a)u }  \frac{  (1-e^{-u})^{M} } {1 -e^{-M u}  }   \,.
\label{eq:LGND}
\end{align}
For example $M=k=5$,  the $\chi^{\text{LG}}_{a b}$ for $a,b=0,\cdots,4$
 are given by 
\begin{align}
\chi^{\text{LG}}_{a b} =
\left(
\begin{array}{ccccc}
 0 & 5 & -10 & 10 & -5 \\
 -5 & 0 & 5 & -10 & 10 \\
 10 & -5 & 0 & 5 & -10 \\
 -10 & 10 & -5 & 0 & 5 \\
 5 & -10 & 10 & -5 & 0 \\
\end{array}
\right) \,.
\label{eq:LGopen}
\end{align}
We find that $\chi^{\text{LG}}_{a b}$ $(a, b=0,1, 2, 3, 4)$ correctly reproduce   open string Witten indices  $I_{a b}={\rm Tr}_{{a b},R}(-1)^Fq^{L_0-\frac{c}{24}}$  in the Gepner model for the quintic 3-fold in \cite{Brunner:1999jq}.   From the open/closed string duality, $I_{a b}$'s  are   calculated  by   cylinder amplitudes for  B-type  boundary states in \cite{Recknagel:1997sb}.

Let us derive  \eqref{eq:LGND} from the Neumann boundary conditions with a matrix factorization. We take   a   tachyon profile \cite{Hori:2013ika} as
\begin{align}
{\sf Q}= \sum_{i=1}^{M} \left( \phi_i \bar{\eta}_i+\frac{1}{M} P  \frac{ \partial f(\phi)}{\partial {\phi_i}}  {\eta}_i \right)  \text{ with } \{\eta_i, \bar{\eta}_j \}=\delta_{i j}\,,  
\,\, \{\eta_i, \eta_j \}=\{\bar{\eta}_i, \bar{\eta}_j \}=0 \,,
\end{align}
which acts on the following graded vector space:
\begin{align}
\mathcal{V}= \bigoplus_{i=0}^{M} \bigwedge^{i} E\,,
\end{align}
where $E$ is an $M$-dimensional vector space spanned by
$\{  \bar{\eta}_{i}  |0 \rangle \}_{i=1}^{M}$.
For simplicity, we suppressed the  subscripts $L/R$ for the left and the right boundaries.
Then the saddle point values of the brane factors are given by
\begin{align}
\mathrm{Str}_{\mathcal{V}_L}( e^{-u})&=e^{-(a-\frac{M}{2}) u}\sum_{i=0}^{M} (-1)^i
\left(
    \begin{array}{c}
      M  \\
       i 
    \end{array}
  \right)
e^{-i u} 
 =e^{-a u}(e^{\frac{u}{2}} - e^{-\frac{u}{2}})^M \,,
\nonumber \\
\mathrm{Str}_{\mathcal{V}_R} ( e^{u} ) & =e^{(b -\frac{M}{2}) u} \sum_{i=0}^{M} (-1)^i
\left(
    \begin{array}{c}
      M  \\
       i 
    \end{array}
  \right)
e^{i u} 
 =e^{b  u} (e^{-\frac{u}{2}} - e^{\frac{u}{2}})^M \,.
\end{align}
Here we assign a gauge charge $a-\frac{M}{2}$ (resp. $b-\frac{M}{2}$ ) for  $| 0 \rangle_L$ (resp. $| 0 \rangle_R$). Then the localization formula with two brane factors is given by
\begin{align}
Z_{\mathcal{W}_L \mathcal{W}_R}
&=\sum_{l=0}^{k-1}
\oint_{u=\frac{2 \pi {\rm i} l }{M} } \frac{d u}{2 \pi {\rm i} } \frac{\mathrm{Str}_{\mathcal{V}_L} ( e^{-u}) \mathrm{Str}_{\mathcal{V}_R} ( e^{u})}
{(e^{-\frac{M u}{2}}-e^{ \frac{M u}{2}})(e^{ \frac{u}{2}}-e^{- \frac{u}{2}})^M} \nonumber \\
&=(-1)^{M+1}\sum_{i=0}^{M-1}\oint_{u=\frac{2 \pi {\rm  i} i}{M}}   \frac{d u}{2 \pi {\rm i} }  e^{(b-a)u }  \frac{  (1-e^{-u})^{M} } {1 -e^{-M u}  }  \,.
\label{eq:LGCPfactor}
\end{align}
Thus we obtain the same result as \eqref{eq:LGND} up to an overall sign.
By comparing \eqref{eq:LGND} with \eqref{eq:LGCPfactor}, we find that  the two brane factors $\mathcal{W}_{\mathcal{V}_L}$ and $\mathcal{W}_{\mathcal{V}_R}$ turn the Neumann boundary conditions for $\phi_i$'s for $i=1,\cdots, M$ to the Dirichlet ones.

\section{Summary and future directions}
\label{sec:summary}
We have introduced $I \times T^2$ indices  for  3d $\mathcal{N}=2$ supersymmetric theories coupled to 2d $\mathcal{N}=(0,2)$ boundary theories and 
have studied properties of the indices. We summarize our results and comments on their implications in 
future studies.

In section \ref{sec:section4} we have studied  the 3d $\mathcal{N}=4$ theory which is mirror dual of 
the 3d $\mathcal{N}=8$ super Yang-Mills theory. We find that $I \times T^2$ index for the 3d $\mathcal{N}=4$ theory agrees with the M-string partition function up to a fugacity $y$.
Since the 3d $\mathcal{N}=8$ super Yang-Mills theory  
flows to the ABJM model  with $\kappa=1$,   an $I \times T^2$ index for the ABJM model   with an appropriate boundary condition is expected to   reproduce M-string partition functions.  
  In this direction, the authors of    \cite{Hosomichi:2014rqa} studied the level $\kappa=1$  ABJM model on  the interval in the zero length limit and compared it  with  the M-string partition function. It was found that the partition function of the dimensionally reduced ABJM model partially agrees with  the M-string partition function.  It is interesting to  explore the boundary conditions in the ABJM model and the $I \times T^2$ index relevant to  the M-string partition function.

In section \ref{sec:section5},  we have studied three dimensional dualities with boundaries like the SQED and the XYZ model.   
In typical cases of 3d dualities between two gauge theories on  half spaces $\mathbb{R}_{\le 0} \times \mathbb{R}^2$ or on $S^1 \times D^2$,  
it was conjectured in \cite{Dimofte:2017tpi}  that the Neumann boundary condition for the vector multiplet is mapped to the Dirichlet boundary condition for the vector multiplet 
in the dual model. The Neumann boundary condition for the vector multiplet is same as the boundary condition \eqref{eq:vecdbdy}. 
On the other hand, the Dirichlet boundary condition for the vector multiplet is not treated in this article. 
In the Dirichlet boundary condition, the global gauge symmetries are preserved at the boundary that  are identified with flavor symmetries in the dual model.  
It would be nice  to develop the localization computation on $ I \times T^2$ with   
the Dirichlet boundary condition for the vector multiplet and to study dualities between various combinations of boundary conditions for the supermultiplets.

In section \ref{sec:section6}, we have treated chiral algebras for simple 3d theories on the interval. 
For the half spaces $\mathbb{R}_{\le 0} \times \mathbb{C}$, 
it is known that chiral algebras are associated to the H-twisted 3d $\mathcal{N}=4$ gauge theories \cite{Costello:2018fnz}. 
In their construction, the gauge theory data correspond to the chiral algebras as follows;  the $G$ vector multiplet,  the hypermultiplets and the boundary fermi multiplets are associated to the $\mathfrak{g}$ $bc$-ghost,  the symplectic bosons and the fermions. The complex moment map in the  3d gauge theory  corresponds to   the  current for symplectic bosons that enters  in the definition of the  BRST charge. The  gauge anomaly cancellation between 3d and 2d theories corresponds to the nilpotency of the BRST charge. The $S^1 \times D^2$ index corresponds to the vacuum character of the chiral algebra.
It is interesting to explore the general rules for the chiral algebras associated with  the gauge theories on $I \times M_2$ 
 like the cases for the H-twisted gauge theories on $\mathbb{R}_{\le 0} \times \mathbb{C}$.

In section \ref{sec:section7}, we have studied the localization formula for indices on $I \times S^1$. In explicit computations in several examples we have shown that indices on $I \times S^1$ with loop operators 
agree with open string Witten indices; Euler pairings  in the geometric phase and the cylinder  amplitudes of the B-type RR ground states for the Gepner models in the Landau-Ginzburg phase. 
To the best of our knowledge, the evaluation of open string Witten indices  for the Gepner models based on the  GLSMs on $I \times S^1$ is new.  

The Euler parings appear in the physics associated to the  geometry of D-branes  and they are related to several topics of quantum geometries of target spaces and SUSY cycles. 
Among them, there is an interesting  property between 
 the K\"{a}hler potential for a Calabi--Yau $n$-fold, period integrals and  Euler pairings: 
\begin{align}
\exp \left( - {\rm  i}^{n}  K(z, \bar{z})\right) =\sum_{a,b} \chi^{a b} \int_{A_a} \Omega (z) \int_{A_b} \overline{\Omega} (\bar{z}) \,.
\label{eq:kahler}
\end{align}
Here $K(z, \bar{z})$  is the K\"{a}hler potential for the Calabi--Yau $n$-fold. 
The left and the right hand sides of \eqref{eq:kahler}  are expressed as partition functions of 2d $\mathcal{N}=(2,2)$ theories as follows.  It was conjectured in \cite{Jockers:2012dk} that $e^{- {\rm  i}^{n}  K(z, \bar{z})}$
is given by an $S^2$ partition function in \cite{Benini:2012ui, Doroud:2012xw}.  Period integrals $\int_{A_a} \Omega (z)$ and their conjugates are given by $D^2$ partition functions in \cite{Honda:2013uca, Hori:2013ika}.  $\chi^{a b}$ is the inverse matrix of Euler pairings ($I \times S^1$ indices). 
Since  $I \times T^2$ indices and $S^1 \times D^2$ indices in \cite{Yoshida:2014ssa} are 
 $S^1$-extensions (q-deformations) of Euler parings and  period integrals, 
we expect a factorization similar to \eqref{eq:kahler} holds between partition functions on closed 3-manifolds,  $S^1 \times D^2$ indices and $I \times T^2$ indices.
 Another future direction is as follows. 
In three dimensions, quantum differential equations become q-difference equations \cite{MR1943747} that annihilate $S^1 \times D^2$ indices, more precisely annihilate  $K$-theoretic $I$-functions in $S^1 \times D^2$ indices. 
For q-difference equations, counter parts of monodromy matrices are called connection matrices. The relation between  monodromy matrices of quantum differential equations  and $I \times S^1$ indices  
imply that   connection matrices for  q-difference equations for  the $K$-theoretic $I$-functions are described by $I \times T^2$ indices. The relation between  the connection matrices and $I \times T^2$ indices will be studied elsewhere.

\section*{Acknowledgements}
We would like to thank Kentaro Hori for comments on the draft.
YY  is supported  by JSPS KAKENHI Grant Number JP16H06335 and also by World Premier International Research Center Initiative (WPI), MEXT Japan.



\bibliography{refs}

\end{document}